\documentclass[%
 reprint,
 amsmath,amssymb,
 aps,
pra,
]{revtex4-2}

\usepackage{graphicx}
\usepackage{color}
\usepackage[utf8]{inputenc}
\usepackage{dcolumn}
\usepackage{bm}
\usepackage[mathlines]{lineno}
\usepackage{multirow}
\usepackage{braket}
 
\newcommand{\der}{\mathrm{d}}
\newcommand{\dert}{\mathrm{d}t}

\newcommand{\Jocc}{\hat{J}_{occ}}
\newcommand{\Jvir}{\hat{J}_{vir}}
\newcommand{\J}{\hat{J}}
\newcommand{\CIS}{\Phi_a^p}
\newcommand{\GS}{\Phi_0^\mathrm{DF}}
\newcommand{\el}{\ell}

\newcommand{\Qca}{Q_a}
\newcommand{\Pca}{P_a}
\newcommand{\Qcb}{Q_b}
\newcommand{\Pcb}{P_b}

\def\bf#1{\mathrm{\mathbf{#1}}}

\def\ket#1{\left| #1 \right\rangle}
\def\bra#1{\left\langle #1\right|}
\def\ann#1{\hat{a}_{#1}}
\def\cre#1{\hat{a}_{#1}^\dagger}

\begin{document}

\preprint{APS/123-QED}


\title{Relativistic time-dependent configuration-interaction singles}
\author{Felipe Zapata$^1$}
\author{Jimmy Vinbladh$^{1,2}$}
\author{Anton Ljungdahl$^2$}
\author{Eva Lindroth$^2$}
\author{Jan Marcus Dahlstr\"om$^1$}

\affiliation{\small $^1$Department of Physics, Lund University, 22100 Lund, Sweden.}

\affiliation{\small  $^2$Department of Physics, Stockholm University, AlbaNova University Center, 10691 Stockholm, Sweden.}

\begin{abstract}
\noindent 
In this work, a derivation and implementation of the relativistic time-dependent configuration interaction singles (RTDCIS) method is presented. Various observables for krypton and xenon atoms obtained by RTDCIS are compared with experimental data and alternative relativistic calculations. This includes energies of occupied orbitals in the Dirac-Fock ground state, Rydberg state energies, Fano resonances and  photoionization cross sections. Diagrammatic many-body perturbation theory, based on the relativistic random phase approximation, is used as a benchmark with excellent agreement between RTDCIS reported at the Tamm-Dancoff level. Results from RTDCIS are computed in the length gauage, where the negative energy states can be omitted with acceptable loss of accuracy. A complex absorbing potential, that is used to remove photoelectrons far from the ion, is implemented as a scalar potential and validated for RTDCIS. The RTDCIS methodology presented here opens for future studies of strong-field processes, such as attosecond transient absorption and high-order harmonic generation, with electron and hole spin dynamics and other relativistic effects described by first principle via the Dirac equation. 
\end{abstract}
  
\maketitle

\section{Introduction}
Attosecond physics aims to unravel the electron motion and coherence in atoms and molecules. A major contribution to this field was the study of valance-shell electrons in krypton ions made by Goulielmakis \emph{et al.} in 2010 \cite{Goulielmakis10}. In this pioneering experiment, the motion of electrons was characterized by means of attosecond transient absorption spectroscopy (ATAS) \cite{Beck15}. Since then, ATAS has been widely used in different scenarios. For example, to reconstruct the time-dependent two-electron wave packet of an excited helium atom \cite{Ott14}, to investigate the instantaneous ac Stark shift \cite{Wirth11}, to control the line shapes of Fano resonances \cite{Ott13} and to probe inner-valance transitions in neon \cite{Ding16,Beck14} and in xenon \cite{Kobayashi17}. In addition, autoionizing states of different noble gases have been studied theoretically within the framework of ATAS \cite{Chu13,Petersson17,Chew18}. The theory and derivation of strong-field ATAS can be found in the following review \cite{Wu16}.

Weak-field ATAS calculations including spin–orbit effects have been performed by Baggesen \emph{et al.} \cite{Baggesen12} and by Kolbasova \emph{et al.} \cite{Kolbasova21}. A relativistic many-body approach was used to describe the bound states and the hole transitions, in krypton and in xenon, with dynamics computed using time-dependent perturbation theory. Strong-field ATAS studies so far have been based on \emph{ad hoc} relativistic theory, although the importance of spin-orbit coupling was established already by the first ATAS experiment \cite{Goulielmakis10}. Pabst \emph{et al.} have pioneered this subject by solving the time-dependent Schrödinger equation (TDSE), within the time-dependent configuration-interaction singles (TDCIS) method \cite{Pabst12}, with spin-orbit effects incorporated to the hole orbitals by hand. This was done by performing a recoupling of the hole angular momentum, $\ell$, and spin, $s$, to a total hole angular momentum, $j$, and then adjusting its energy to match experimental values, while no corresponding recoupling was performed for the particle states. Recently, two-component time-dependent R-matrix calculations are possible with the RMT code \cite{Wragg20,Brown20}, which has been used, for example, to resolve the electron spin dynamics in krypton with a combination of parallel- and cross-polarized laser pulses \cite{Wragg19}.

Regarding the lack of a \emph{relativistic} transient absorption theory that handles laser fields beyond the perturbative regime, we have decided to develop a general relativistic ATAS method to study heavy elements in strong fields. The first step in our development was the derivation of the relativistic transient absorption theory based on the time-dependent Dirac equation (TDDE) \cite{Zapata21}. Once the equations of the relativistic transient absorption theory have been validated, the next step is to solve the many-electron TDDE. As this is not a trivial task, in our opinion, it merits its own attention. Thus, the aim of the present article is to discuss the approximations we have applied in order to solve the many-electron TDDE. As a compromise between computational cost and accuracy, a Relativistic formulation of the TDCIS method (i.e. RTDCIS), has been chosen for our purpose. The development of RTDCIS was carried out following the implementation of the TDCIS method done by Rohringer \emph{et al.} \cite{Rohringer06} and by Greenman \emph{et al.} \cite{Greenman10}. The main difference between RTDCIS and TDCIS is that, in the former method, the atomic orbitals are 4-component spinors obtained by solving the relativistic Hartree-Fock equations (also known as the Dirac-Fock equations), while in the later method, the atomic orbitals are obtained from a non-relativistic Hartree-Fock calculation. In consequence, as in TDCIS, the different hole excitation channels are going to be coupled by the presence of the electron-electron interaction term in the relativistic Hamiltonian and by the action of external fields \cite{You2016}. Thus, RTDCIS is not a ``single-active electron model'' because many-body/multi-channel effects are included in the theory. The advantage of using RTDCIS is the possibility to have access to the fine structure of the atomic spectra beyond the perturbative regime without the necessity of including an \emph{ad hoc} Pauli-type potential to the Hamiltonian. For the moment, our implementation of RTDCIS is restricted to closed-shell atoms. The description of the light-matter is restrained to the dipole approximation, which is sufficient for strong, albeit not extreme fields. Discussions about ``beyond-dipole effects'' in the TDDE can be found in Refs. \cite{Aleksander16,Kjellsson17}.  Pair production may be induced by extreme processes, such as heavy ion collisions at relativistic velocities, or in presence of a super-strong laser field that polarizes the vacuum \cite{Greiner00}. Our work on relativistic ATAS is far from these extreme scenarios and the \emph{no-virtual-pair approximation} \cite{Grant06} is applied. The limitations of RTDCIS, which are explored in the present work, comes from the fact that it is a \emph{single reference method} that only includes single excitations. Similar limitations were reported for TDCIS, \emph{see} for example Refs. \cite{Rohringer06,Greenman10,Krebs14}. Generally speaking, the method proposed here can be used to study spin-resolved ATAS experiments in heavy elements beyond the perturbative regime with relativistic effects described by first principle via the Dirac equation. Similarly, it can be applied to other strong-field processes, such as high-order harmonic generation, above-threshold ionization and laser-assisted photoionization.   

The outline of the present article is at follows. In Section \ref{theory}, the theory is presented with a derivation of the RTDCIS equations of motion and observables.  
In Section \ref{bsplines}, details on the chosen B-spline basis set and the numerical propagator are commented. In Section \ref{results}, results are presented and discussed. 
In order to validate the RTDCIS theory, different observables have been calculated for krypton and xenon. First, the ``quality'' of the relativistic configuration-interaction singles (RCIS) space has been investigated. Our results have been compared with experimental spectral data \cite{Saloman04,Saloman07} provided by the National Institute of Standards and Technology (NIST) \cite{NIST:database} and with 4-component configuration-interaction singles calculations performed with the DIRAC19 code \cite{DIRAC19}. 
Second, the implementation of RTDCIS has been verified using  
photoionization cross sections that have been compared with experimental data \cite{Samson02,Shannon77,Becker89,Kammerling89} and with relativistic random phase approximation (RRPA). To the best of our knowledge, such RRPA calculations have not been compared with explicit time propagation of a relativistic method in the past, where we find that the Tamm-Dancoff approximation (RRPA(TD)) provides an efficient benchmark for the implementation. Finally, in Section \ref{conclusion}, the conclusion is given.

\section{Theory}
\label{theory}
In this section, the formulation of RTDCIS is presented. The equations of motion are derived in Sec.~\ref{sec:eom} and we discuss the computation of the required ``source orbitals'' (4-component spinors of the Dirac-Fock equation) in Sec.~\ref{sec:df_with_cap}. Due to the presence of a complex absorbing potential (CAP), the Dirac-Fock equations become non-Hermitian and particular evaluation of the matrix elements is highlighted. Afterwards, the no-virtual-pair approximation is discussed in Sec.~\ref{sec:virtual_non_pair}. Next, the computation of some observables is addressed. First, the computation of the relativistic singly-excited state energy levels is described in Sec.~\ref{sec:single_energy}. Second, the equations to compute the total angular momentum of a relativistic singly-excited  state are derived for a closed-shell atom in Sec.~\ref{sec:total_j}. Finally, the calculation of the photoionization cross sections is presented in Sec.~\ref{sec:photoionization_cross_sections}. Atomic units (a.u.) are used unless otherwise stated, $e=\hbar=m_e=4\pi\epsilon_0=1$.

\subsection{Relativistic formulation of TDCIS}

\subsubsection{Equations of motion}
\label{sec:eom}
The relativistic electron dynamics of a closed-shell atom under the influence of a laser field is encoded in the TDDE, which can be written in the Hamiltonian formulation as follows,
\begin{equation}
\label{TDDE}
    i\frac{\partial}{\partial t}\ket{\Psi(t)}=\left[\hat{H}+\hat{V}(t)\right]\ket{\Psi(t)},
\end{equation}
where $\ket{\Psi(t)}$ is the $N$-electron wave function, $\hat{H}$ is the field-free Hamiltonian and $\hat{V}(t)$ describes the interaction of the atom with the laser field. Eq.(\ref{TDDE}) cannot  be solved exactly and different approximations must be taken into account. Within the framework of the RTDCIS, the time-dependent $N$-electron wave function $\ket{\Psi(t)}$ is going to be expressed as a linear combination of the Dirac-Fock ground state, $\ket{\GS}$, and the single particle-hole excitation states, $\ket{\CIS}$. Thus, the $N$-electron wave function  $\ket{\Psi(t)}$ is given by
\begin{equation}
\label{td-wave}
    \ket{\Psi(t)}=c_0(t)\ket{\GS}+\sum_{a,p}c_a^p(t)\ket{\CIS},
\end{equation}
where $\ket{\CIS}=\cre{p}\ann{a}\ket{\GS}$ and $\ket{\GS}=\cre{N}\dots\cre{c}\cre{b}\cre{a}\ket{0}$
with $\ket{0}$ being the vacuum state. Here and in the following, indices $a,b,c,d,\dots$ are used for one-particle occupied (core) orbitals in $\ket{\GS}$, while indices $p,q,r,s,\dots$ are employed for unoccupied (virtual) orbitals. For general one-particle orbitals (occupied or unoccupied) $i,j,k,l,\dots$ indices are used. The one-particle orbitals, $\ket{i}$, are given by 4-component Dirac-Fock spinors, which are obtained after solving the following eigenvalue problem, 
\begin{equation}
\label{dirac-fock-equation}
  \hat{h}_0^\mathrm{DF}\ket{i}=\varepsilon_i\ket{i},  
\end{equation}
where $\hat{h}_0^\mathrm{DF}$ is the one-particle Dirac-Fock operator and $\varepsilon_i$ are the one-particle Dirac-Fock orbital energies, which are given by  
\begin{eqnarray}
\nonumber
\varepsilon_i&=&\bra{i}\hat{h}^\mathrm{DF}_0\ket{i}\\
\nonumber
             &=&\bra{i}\hat{h}^\mathrm{D}\ket{i}+\bra{i}\hat{v}^\mathrm{DF}\ket{i}\\
             &=&\bra{i}\hat{h}^\mathrm{D}\ket{i}+\sum_b\left[\bra{ib}r_{12}^{-1}\ket{ib}-\bra{ib}r_{12}^{-1}\ket{bi}\right],
\end{eqnarray}
where the Dirac operator $\hat{h}^\mathrm{D}$ is defined as 
\begin{equation}
    \hat{h}^\mathrm{D}=c\;{\bm\alpha}\cdot{\bf p}+\beta m_e c^2-\frac{Z}{r},
\end{equation}
where $c$ is the speed of light, $m_ec^2$ the electron rest mass energy, ${\bf p}=-i{\bm \nabla}$ the electron momentum operator, $r$ the electron position and $Z$ the nuclear charge. The Dirac matrices ${\bm  \alpha}=(\alpha_x,\alpha_y,\alpha_z)$ and $\beta$ are given by
\begin{equation}
\nonumber
    \alpha_\xi=\left(\begin{array}{cc}
         0&\sigma_\xi  \\
         \sigma_\xi&0 
    \end{array}\right);\;\; \mathrm{and}\;\;\beta=\left(\begin{array}{cc}
         I&0  \\
        0 & -I
    \end{array}\right),
\end{equation}
with 
\begin{equation}
\nonumber
    \sigma_x= 
    \left(\begin{array}{cc}
        0 & 1  \\
        1 & 0 
    \end{array}\right);\;\;\sigma_y= 
    \left(\begin{array}{cc}
        0 & -i  \\
        i & 0 
    \end{array}\right);\;\;\mathrm{and}\;\;
    \sigma_z= 
    \left(\begin{array}{cc}
        1 & 0  \\
        0 & -1 
    \end{array}\right),
\end{equation}
where the set of $\sigma_\xi$ is given by the Pauli matrices and $I$ is a $2\times2$ unitary matrix  \cite{Grant06}. The action of the Dirac-Fock potential $\hat{v}^\mathrm{DF}$ on $\ket{i}$ is expressed in terms of the direct $\bra{ia}r_{12}^{-1}\ket{ia}$ and exchange $\bra{ia}r_{12}^{-1}\ket{ai}$ Coulomb two-electron integrals (written  here in the so-called ``physicist's notation'') where $r_{12}=|{\bf r_1}-{\bf r_2}|$. The complete expression for the relativistic two-electron integrals is given below. Finally, the one-particle Dirac-Fock orbitals are expressed in spherical coordinates as follows \cite{Grant70}
\begin{equation}
\label{spinor}
\bra{\bf{r}}i\rangle\equiv\varphi_{n,\kappa,m}(\bf{r})=\frac{1}{r}\binom{P_{n,\kappa}(r)\chi_{\kappa,m}(\Omega)}{iQ_{n,\kappa}(r)\chi_{-\kappa,m}(\Omega)},
\end{equation}
where  the quantum number $\kappa$ relates $\ell$ and $j$ as follows: $\kappa=\ell$ for $j=\ell-1/2$ and $\kappa=-(\ell+1)$ for $j=\ell+1/2$. The spin-angular functions $\chi_{\pm\kappa,m}(\Omega)$ are expressed using the ``$\ell s$-coupling'' as follows
\begin{equation}
    \chi_{\pm\kappa,m}(\Omega)=\frac{1}{\sqrt{2\ell+1}}\left(\begin{array}{cc}
        \pm \sqrt{\ell\pm m +1/2}&Y_\ell^{m-1/2}(\Omega)\\
            \sqrt{\ell\mp m +1/2}&Y_\ell^{m+1/2}(\Omega)
    \end{array}\right),
\end{equation}
where $\Omega$ stands for the angles $\theta,\phi$. The radial functions $P_{n,\kappa}(r)$ and $Q_{n,\kappa}(r)$ are the so-called ``large'' and ``small'' components, respectively. 

In consequence, the field-free Hamiltonian in Eq.(\ref{TDDE}) can be expressed as follows \cite{Szabo96},
\begin{equation}
    \hat{H}=\hat{H}_0^\mathrm{(DF)}+\hat{H}_1-E_0^\mathrm{(DF)},
\end{equation}
where the reference Hamiltonian (also known as the ``zero-order Hamiltonian'') is given by the $N$-electron Dirac-Fock Hamiltonian, i.e.
\begin{equation}
    \hat{H}_0^\mathrm{(DF)}=\sum_{ij}\bra{i}\hat{h}^\mathrm{DF}_0\ket{j}\cre{i}\ann{j},
\end{equation}
and the ``perturbation'' is given by the difference between the exact electron-electron Coulomb interaction and the 
Dirac-Fock potentials, i.e. 
\begin{equation}
    \hat{H}_1=\frac{1}{2}\sum_{ijkl}\bra{ij}r_{12}^{-1}\ket{kl}\cre{i}\cre{j}\ann{l}\ann{k}-\sum_{ij}\bra{i}\hat{v}^\mathrm{DF}\ket{j}\cre{i}\ann{j}.
\end{equation}
In order to have compact equations of motion, the spectrum of the field-free Hamiltonian is shifted by the Dirac-Fock ground state energy, i.e.
\begin{equation}
    E_0^\mathrm{(DF)}=\sum_a\varepsilon_a-\frac{1}{2}\sum_{ab}\left[\bra{ab}r_{12}^{-1}\ket{ab}-\bra{ab}r_{12}^{-1}\ket{ba}\right],
\end{equation}
which is taken as the zero energy reference. 

The interaction with the laser field is going to be treated within the dipole approximation \cite{Sakurai2020}. Neglecting the magnetic laser-field effects, and using a linearly polarized pulse of duration $\tau$ along the $z$-axis, we write the interaction in length gauge as 
\begin{equation}
\label{lengthgauge}
    \hat{V}(t)=\mathcal{E}(t)\sum_{ij}\cre{i}\ann{j}\bra{i}\hat{z}\ket{j},
\end{equation}
where $\hat{z}$ is the $z$-component of the dipole operator and $\mathcal{E}(t)$ is the electric field. The complete expression for the dipole-transition elements $\bra{i}\hat{z}\ket{j}$ is given in our previous work \cite{Zapata21}. 

Inserting the ansatz given by Eq.(\ref{td-wave}) into Eq.(\ref{TDDE}), and projecting onto either $\ket{\GS}$ or $\ket{\CIS}$, we obtain the equations of motion for the time-dependent coefficients $c_0(t)$ and $c_a^p(t)$, respectively. The resulting matrix elements are determined using the anticommutation relations of the creation and the annihilation operators (equivalent to the so-called Slater-Condon rules, \emph{see} Ref. \cite{Szabo96}). Some of the required matrix elements are given here: 
\begin{equation}
\begin{array}{rcl}
\bra{\GS}\hat{H}\ket{\CIS}&= &0 \;\;(\mathrm{Brillouin\; theorem}); \\
\bra{\GS}\hat{V}(t)\ket{\CIS}&=&\mathcal{E}(t)\bra{a}\hat{z}\ket{p};\\
\bra{\CIS}\hat{H}\ket{\Phi_b^q}&=&(\varepsilon_p-\varepsilon_a)\delta_{ab}\delta_{pq}+\bra{bp}r_{12}^{-1}\ket{qa}\\
&&-\bra{bq}r_{12}^{-1}\ket{aq};\\
\bra{\CIS}\hat{V}(t)\ket{\Phi_b^q}&=&\mathcal{E}(t)\left[\bra{p}\hat{z}\ket{q}\delta_{ab}-\bra{b}\hat{z}\ket{a}\delta_{pq}\right].
\end{array}
\end{equation}

Thus, the RTDCIS equations of motions are given by  
\begin{subequations}
\label{EOM}
\begin{align}
         i\dot{c}_0(t)=&\sum_{ap}c_a^p(t)\mathcal{E}(t)\bra{a}\hat{z}\ket{p};  \\
         \nonumber
         i\dot{c}_a^p(t)=&(\varepsilon_p-\varepsilon_a)c_a^p(t)\\
         \nonumber
         &+\sum_{bq}c_b^q(t)[\bra{bp}r_{12}^{-1}\ket{qa}-\bra{bp}r_{12}^{-1}\ket{aq}] \\
         \nonumber
          &+\mathcal{E}(t)\left[c_0(t)\bra{p}\hat{z}\ket{a}+\sum_q c_a^q(t)\bra{p}\hat{z}\ket{q}\right.\\
          &-\left.\sum_b c_b^p(t)\bra{b}\hat{z}\ket{a}\right],
\end{align}
\end{subequations}
where $\varepsilon_a$ and $\varepsilon_p$ are the hole and particle energies of the one-particle Dirac-Fock orbitals $\ket{a}$ and $\ket{p}$, respectively.

As we can see, Eq.(\ref{EOM}) is similar to the non-relativistic equations of motion derived by Rohringer \emph{et al} \cite{Rohringer06} and Greenman \emph{et al} \cite{Greenman10}.  Note that in the non-relativistic TDCIS implementation, factors of two arise due to the use of spin-adapted configurations (i.e. due to the summation of the spin degrees of freedom). However, the main difference between our implementation and Refs. \cite{Rohringer06} and \cite{Greenman10} is the relativistic nature of the one-particle orbitals used in Eq.(\ref{EOM}). This fact will strongly modify the possible number of ``active'' holes and then the total number of ionization channels per simulation.  Finally, in order to avoid unphysical reflections during the numerical propagation of TDCIS, core and virtual orbitals are obtained from a Hartree-Fock calculation in presence of a CAP \cite{Greenman10}. Likewise, in the present relativistic version, core and virtual orbitals are obtained after solving Eq.(\ref{dirac-fock-equation}) (i.e. the Dirac-Fock equations) in presence of a CAP. 

\subsubsection{Dirac-Fock equations with CAP}
\label{sec:df_with_cap}
In the present work, the Dirac-Fock equations have been implemented following the investigations of Grant \cite{Grant70,Grant06,Grant09} and Lindgren and Rosén \cite{Lindgren74}. The electron-electron interaction has been treated using the instantaneous Coulomb potential. The resulting two-electron integrals are given by the following multipolar expansion 
\begin{equation}
\label{slater}
\bra{ij}r_{12}^{-1}\ket{kl} = \sum_{uw}(-1)^w R^u(ijkl)\bra{i}X_w^u\ket{k}\bra{j}X_{-w}^u\ket{l},
\end{equation}
where the angular coefficients, $\bra{i}X_w^u\ket{k}$ and $\bra{j}X_{-w}^u\ket{l}$, are expressed in terms of $3j$-symbols, \emph{see} Eq.(7.9) in Ref. \cite{Grant70}, and the radial part is given by the so-called relativistic Slater integral, 
\begin{eqnarray}
\nonumber
R^u(ijkl)=\int_0^\infty I_u(ik)\left[P_j^*(r_1)P_l(r_1)+Q_j^*(r_1)Q_l(r_1)\right]\der r_1,
\end{eqnarray}
with 
\begin{equation}
\nonumber
    I_u(ik)=\int_0^\infty\frac{r^u_<}{r_>^{u+1}}\left[P_i^*(r_2)P_k(r_2)+Q_i^*(r_2)Q_k(r_2)\right]\der r_2,
\end{equation}
where $r_<=\mathrm{min}(r_1,r_2)$ and $r_>=\mathrm{max}(r_1,r_2)$. Therefore, for an one-particle orbital $\ket{a}$, the Dirac-Fock equations can be written as follows
\begin{widetext}
\begin{subequations}
\label{DFE}
\begin{align}
  -c \;Q'_a+\frac{c\kappa}{r}\Qca+\left[\mathcal{U}_\mathrm{CAP}(r)-\frac{Z}{r}\right]\Pca
  +\sum_{bu}\left[C^0(abu)I_u(bb)\Pca+D^0(abu)I_u(ab)\Pcb\right]&=&\varepsilon_a\Pca;\\
  c\;P'_a+\frac{c\kappa}{r}\Pca-\left[\mathcal{U}_\mathrm{CAP}(r)+2m_ec^2+\frac{Z}{r}\right]\Qca+\sum_{bu}\left[C^0(abu)I_u(bb)\Qca+D^0(abu)I_u(ab)\Qcb\right]&=&\varepsilon_a\Qca,
\end{align}
\end{subequations}
\end{widetext}
where the angular coefficients are given by
\begin{subequations}
\begin{align}
    C^0(abu)&=(2j_b+1)\delta_{u,0};\\
    D^0(abu)&=-(2j_b+1)\left(\begin{array}{ccc}
        j_a & u & j_b \\
        -\frac{1}{2}&0 & \frac{1}{2} 
    \end{array}\right)^2, 
\end{align}
\end{subequations}
for $\ell_a+\ell_b+u$ even, otherwise $D^0(abu)=0$. In Eq.(\ref{DFE}), the zero energy has been defined so that an electron at rest at infinity has zero energy.

As is customary Eq.(\ref{DFE}) is here solved numerically using a $L^2$-basis approximation where Dirichlet boundary conditions are imposed \cite{Grant06}. In order to prevent unphysical reflections during time propagation at the end of the simulation box, a complex absorbing potential $\mathcal{U}_\mathrm{CAP}(r)$ has been incorporated as a scalar potential following  the implementation made by Ackad and Horbatsch in Refs. \cite{Ackad07,Horbatsch07,Horbatsch07b}. The CAP used here is defined as in Ref. \cite{Riss93}, i.e.
\begin{equation}
    \mathcal{U}_\mathrm{CAP}(r)=\left\{\begin{array}{rcr}
        0&\mathrm{if} & r\leq \mathcal{R}_\mathrm{CAP};   \\
        -i\eta(r-\mathcal{R}_\mathrm{CAP})^2&\mathrm{if} & r>\mathcal{R}_\mathrm{CAP},
    \end{array}\right.
\end{equation}
being $\eta$ a positive parameter that determines the strength of the potential. Due to the presence of the CAP, the Dirac-Fock Hamiltonian becomes non-Hermitian. As a consequence, the Hermitian inner product is not satisfied in this basis. Nevertheless, the resulting complex symmetric Dirac-Fock Hamiltonian enables a redefinition of the inner product. This problem was also addressed in the implementation of the non-relativistic TDCIS method in reference \cite{Greenman10}. In practice, one does not take the complex conjugate of the radial function in the left vector when computing matrix elements. In the present work, the adopted CAP strength was $\eta=6\times10^{-4}$. 


\subsubsection{No-virtual-pair approximation}
\label{sec:virtual_non_pair}
The solution of Eq.(\ref{DFE}) is composed by two sets of solutions: the positive-energy states and the negative-energy states \cite{Grant06}. When positron-electron pair creation is energetically out of reach, the possibility of removing the negative-energy states from the basis is very tempting from a computational point of view. With the basis set reduced by a factor of two, the time propagation will be less demanding. The question on how to treat the negative-energy states in the many-electron mean-field problem has been addressed by several authors in the last decades, \emph{see} for example \cite{Sucher80,Sucher84,Heully86,Grant06,Kutzelnigg12,Almoukhalalati16,Liu20,Toulouse21} and references therein. Furthermore, this problem has been attacked by different relativistic atomic and molecular codes, \emph{see} for example Refs. \cite{Fischer19,Saue20,Belpassi20}. 

In essence, it can be shown that a positive-energy solution of the relativistic Hartree-Fock Hamiltonian does not contain any negative component \cite{Heully86}. Therefore, the negative-energy states can be easily rejected by inspection of the energy value. In the so-called \emph{no-virtual-pair approximation}, the sums in Eq.(\ref{DFE}) are restricted to orbital indices that belong to the positive-energy solutions only. Even though this is a very natural approximation, only together the positive- and the negative-energy solutions form a complete basis set. This issue has important consequences when a time-dependent perturbation is added to the Hamiltonian. As Furry \cite{furry:51} showed in 1951, when a spectral method is used to evaluate a time-dependent perturbation beyond the lowest order, both positive- and negative-energy states are needed to expand the intermediate ``virtual'' states correctly. Moreover, if the perturbation is described by a non-diagonal operator with respect to the large and the small components of the wave function, it can be shown that the contribution of the negative-energy states is of the same order of magnitude as the contribution corresponding to the positive. This has been investigated in details by Selst{\o} \emph{et al.} \cite{Selsto09} in connection with contributions beyond the dipole approximation and by Vanne and Saenz \cite{Vanne12} in relation with multi-photon ionization within the dipole approximation. As it was previously explained in Ref. \cite{Zapata21}, the velocity form of the dipole operator is indeed non-diagonal with respect to the large and the small component and multi-photon contributions will in this gauge not be correctly represented with just the positive-energy spectrum. On the other hand, the length form is diagonal and the contribution of negative-energy states is greatly suppressed. In fact, it will be suppressed beyond the leading relativistic contribution with, at least, one order of the fine-structure constant $(\alpha_\mathrm{fs}\approx1/137)$. This was also demonstrated numerically in Ref. \cite{Vanne12}. As a consequence, one is forced to use the length gauge form of the light-mater interaction when the negative-energy states are excluded from the propagation of the TDDE. 

\subsection{\label{observables}Observables}

\subsubsection{Relativistic singly-excited state energy levels}
\label{sec:single_energy}
 In order to compute the relativistic singly-excited state energy levels, one needs to project the field-free Hamiltonian $\hat{H}$ onto the RCIS space. This procedure leads to the following eigenvalue problem, 
\begin{equation}
\label{RCIS_equation}
    \bf A \bf C_n = \omega_n \bf C_n,
\end{equation}
where $\bf A$ represents the field-free Hamiltonian $\hat{H}$ in the RCIS space, $\omega_n$ is the singly-excited state energy level defined as $\omega_n=E_\mathrm{RCIS}-E_0^\mathrm{(DF)}$, and the vector $\bf C_n$ contains the RCIS expansion coefficients. The matrix elements of $\bf A$ are then given by
\begin{equation}
\label{A_matrix}
    A_{ap,bq}=(\varepsilon_p-\varepsilon_a)\delta_{ab}\delta_{pq}+\langle bp|r_{12}^{-1}|qa\rangle-\langle bp|r_{12}^{-1}|aq\rangle,
\end{equation}
where the two-electron integrals are computed using Eq.(\ref{slater}), and the orbital energies $\varepsilon_a$ and $\varepsilon_p$ are obtained after solving Eq.(\ref{DFE}).

\subsubsection{Total angular momentum in closed-shell atoms}
\label{sec:total_j}
In second quantization, the total angular momentum operator is given by 
\begin{equation}
\label{Jtotal}
    \hat{J}_\mathrm{total}=\sum_{ij}\cre{i}\ann{j}\bra{i}\hat{J}\ket{j},
\end{equation}
where the sum runs for all occupied and virtual one-particle orbitals and $\hat{J}=\hat{J}_x+\hat{J}_y+\hat{J}_z$. As the operator $\hat{J}$ does not couple the occupied and the virtual orbitals in a closed-shell atom, the sum in Eq.(\ref{Jtotal}) can be rewritten as $\sum_{ij}=\sum_{ab}^{occ}+\sum_{pq}^{vir}$. This fact allows us to express the total angular momentum operator as the sum of the separate occupied and virtual contributions as follows,
\begin{eqnarray}
\nonumber
 \hat{J}_\mathrm{total}&=&\hat{J}_{occ}+\hat{J}_{vir}\\
                       &=&\sum_{ab}^{occ}\cre{b} \ann{a}\bra{b}\hat{J}\ket{a}+\sum_{pq}^{vir}\cre{q} \ann{p}\bra{q}\hat{J}\ket{p}.
\end{eqnarray}
 Moreover, the total angular momentum operator $\hat{J}_\mathrm{total}^2$ can be defined as 
 \begin{equation}
     \hat{J}_\mathrm{total}^2=\hat{J}_{occ}^2+\hat{J}_{vir}^2+2\hat{J}_{occ}\hat{J}_{vir}.
 \end{equation}
 
As shown in Appendix \ref{appendixA}, the expectation value of the total angular momentum for a given RCIS state is defined as 
 \begin{equation}
 \label{total_J}
    \langle \hat{J}_\mathrm{total}^2 \rangle_n = \mathbf{C}_n^\dagger\;\mathbf{J}^2\;\mathbf{C}_n,
\end{equation}
 where the vector $\mathbf{C}_n$ is obtained after solving Eq.(\ref{RCIS_equation}) and the matrix elements of $\mathbf{J}^2$ are expressed in terms of the occupied and the virtual one-particle orbital quantum numbers, $\{j_a,m_{a}\}$ and $\{j_p,m_{p}\}$ respectively, i.e. 
 \begin{equation}
 \label{j_element}
\begin{array}{lll}
 \mathbf{J}^2_{n',n}&=&\bra{\Phi_b^q}\hat{J}_\mathrm{total}^2\ket{\Phi_a^p}\\
 &=&\{\;k_1\;\delta_{m_{a},m_{b}}\delta_{m_{p},m_{q}}\\
 & &-k_2\;\delta_{m_{a}-1,m_{b}-1}\delta_{m_{p}-1,m_{q}-1}\\
 & &-k_3\;\delta_{m_{a}+1,m_{b}+1}\delta_{m_{p}+1,m_{q}+1}\}\\
 & &\times\;\delta_{j_a,j_b}\delta_{j_p,j_q},
\end{array}
\end{equation}
where the angular coefficients are defined as 
\begin{eqnarray}
\nonumber
k_1&=&j_a(j_a+1)+j_p(j_p+1)-2m_{a}m_{p};\\
\nonumber
k_2&=&[(j_a+m_{a})(j_a-m_{a}+1)]^{1/2}\\
\nonumber
&\times&[(j_p+m_{p})(j_p-m_{p}+1)]^{1/2};\\
\nonumber
k_3&=&[(j_a-m_{a})(j_a+m_{a}+1)]^{1/2}\\
\nonumber
&\times&[(j_p-m_{p})(j_p+m_{p}+1)]^{1/2}.
\end{eqnarray} 
Note that the matrix elements in Eq.(\ref{j_element}) are diagonal with respect to the $j$'s quantum numbers but mix the $m$'s quantum numbers.

\subsubsection{Photoionization cross sections}
\label{sec:photoionization_cross_sections}
Atomic photoionization cross sections can be calculated as follows \cite{Krebs14}
\begin{equation}
\label{cross_section}
    \sigma(\omega)=\frac{4\pi\omega}{c}\mathrm{Re}\left[\int_0^\infty C(t)e^{i\omega t}\dert\right],
\end{equation}
where the time-dependent correlation function $C(t)$ is defined here as the overlap between an initial dipole-perturbed ground state $\ket{\Psi'(0)}=\hat{\mathcal{Q}}\ket{\GS}$ and the field-free propagated state $\ket{\Psi'(t)}$, being $\hat{\mathcal{Q}}$ the total $N$-electron position operator. Within the CIS framework, one can easily find that the correlation function can be written as $C(t)=\sum_{ap}c_a^p(t)\bra{a}\hat{z}\ket{p}$,
where the time-dependent coefficients are found by solving the field-free set of equations of motion (i.e. by solving Eq.(\ref{EOM}) with $\mathcal{E}(t)=0$) with the following initial conditions: $c_0(0)=0$ and $c_a^p(0)=\bra{p}\hat{z}\ket{a}$. Moreover, if a CAP is used to generate the Dirac-Fock orbitals, the dipole transition matrix elements must be evaluated in an inner region which is not affected by the CAP. Details on how to evaluate dipole elements in the inner region can be found in Ref. \cite{Zapata21}. Finally, in order to perform the inverse Fourier transform in Eq.(\ref{cross_section}), a filter function must be used to damp the infinite oscillating behavior of the correlation function $C(t)$. In the present work, the inverse of the cumulative distribution function has been implemented as a filter where 
\begin{equation}
f(t)=\frac{1}{2}\left[1+\mathrm{erf}\left(\frac{t-\mu_1}{\mu_2\sqrt{2}}\right)\right],
\end{equation}
with $\mu_1$ being equal to $75\%$ of the total propagation time and $\mu_2$ around $10\%$, depending on the desired spectral resolution.  

\section{\label{bsplines}Numerical implementation}
This section contains numerical details for our implementation of RTDCIS. In Sec. \ref{bsplinesA}, we present the time-propagation scheme implemented to solve Eq.(\ref{EOM}). In Sec. \ref{bsplinesB}, we give the parameters for the B-spline representation of the radial components of the Dirac-Fock orbitals. 

\subsection{\label{bsplinesA}Time-propagation scheme}
In this work,  Eq.(\ref{EOM}) is propagated numerically using a second-order finite-differencing scheme \cite{Leforestier91}. This propagation scheme was previously used by us to propagate the TDDE in hydrogen \cite{Zapata21}. It is also the same propagation scheme that was used for TDCIS in Ref. \cite{Greenman10}. As a result, the time-dependent coefficients $c_0(t)$ and $c_a^p(t)$ are computed at each time step $\Delta t$ as follows
\begin{subequations}
\label{scheme}
\begin{align}
    c_0(t+\Delta t)=&c_0(t-\Delta t)\\
    \nonumber
    &+2i\Delta t\mathcal{E}(t)\sum_{ap}c_a^p(t)\langle a|\hat{z}|p\rangle; \\
    \nonumber
    c_a^p(t+\Delta t)=&e^{-2i(\varepsilon_p-\varepsilon_a)\Delta t}c_a^p(t-\Delta t)\\
    &-2i\Delta t e^{-i(\varepsilon_p-\varepsilon_a)\Delta t}F_a^p(t),
\end{align}
\end{subequations}
where 
\begin{eqnarray}
\nonumber
F_a^p(t)&=&\sum_{bq}c_b^q(t)[\bra{bp}r_{12}^{-1}|qa\rangle-\bra{bp}r_{12}^{-1}|aq\rangle] \\
\nonumber
&&+\mathcal{E}(t)\left[c_0(t)\bra{p}\hat{z}\ket{a}+\sum_q c_a^q(t)\bra{p}\hat{z}\ket{q}\right.\\
\nonumber
 &&-\left.\sum_b c_b^p(t)\bra{b}\hat{z}\ket{a}\right].
\end{eqnarray}
Note that Eq.(\ref{scheme}) preserves the norm only for Hermitian systems. Numerical stability is reached for $\Delta t<\hbar/|\varepsilon_\mathrm{max}|$, $\varepsilon_\mathrm{max}$ being the largest eigenvalue of the total Hamiltonian operator \cite{Leforestier91}. 
In the present work, the time-dependent correlation function $C(t)$ was recorded during a propagation time of 6000 a.u. and $\Delta t=5\times10^{-3}$ a.u.

\subsection{\label{bsplinesB}B-spline basis set}
The accuracy in describing processes that involve Rydberg and continuum states can be related to the choice of the basis set \cite{Labeye18}. In the case of atomic photoionization, B-spline basis sets are preferred, \emph{see} for example Ref. \cite{Zapata19}. Following our previous work \cite{Zapata21}, the large and the small components are expanded in two different B-spline basis sets. As suggested by Froese-Fischer and Zatsarinny \cite{Charlotte08}, different polynomial orders for the large and small components are used for the purpose of removing the so-called \emph{spurious states} known to appear in the numerical spectrum of the Dirac-Fock Hamiltonian after discretization, i.e. 
\begin{subequations}
\begin{align}
   P_{n,\kappa}(r)&=\sum_{i=1}^{n_s'}\beta_iB_i^{k_s'}(r);\\
 Q_{n,\kappa}(r)&=\sum_{j=1}^{n_s}\beta_jB_j^{k_s}(r), 
\end{align}
\end{subequations}
where the dimensions of the basis are defined by
$n_s'$ and $n_s=n_s'+1$
 and the order of the B-splines by 
$k_s'$ and $k_s=k_s'+1$. In the present work, both B-spline sets have been defined on the same sequence of increasing knot points while the boundary knots have been chosen to be either $k_{s}$- or $k_{s}'$-fold degenerate, e.g. $r_{1}=r_{2}=...=r_{k_{s}}=r_{\mathrm{min}}$ and $r_{n_{s}+1}=r_{n_{s}+2}=...r_{n_{s}+k_{s}}=r_{\mathrm{max}}$. In order to ensure the zero boundary conditions of $P_{n,\kappa}(r)$ and $Q_{n,\kappa}(r)$ at $r=r_{\mathrm{min}}$ and $r=r_{\mathrm{max}}$, the first and the last B-splines, in both sets, were removed from the calculation. Converged results were obtained using an exponential-linear hybrid knot distribution as in Ref. \cite{Qiu99} with $r_\mathrm{max}=100$ a.u. and $\mathcal{R}_\mathrm{CAP}=70$ a.u. The exponential region was described by 12 knot points for krypton and by 18 knot points in the case of xenon. The linear region was represented by 200 points where the last 60 knots described the outer region ($r>\mathcal{R}_\mathrm{CAP}$) where the CAP is non-zero. The order of the B-splines was chosen to be $k_s=8$ and $k_s'=7$. For krypton, the chosen grid generates a total number of $n_s=216$ B-splines for the small component and $n_{s'}=215$ for the large component. For xenon, $n_s=222$ B-splines for the small component and $n_{s'}=221$ for the large component. Given a set of B-spline parameters, one obtains $n_{s'}$ positive-energy states and $n_s$ negative-energy state solutions per spin-angular symmetry. In the present work, $\ell_\mathrm{max}=4$, and in order to speed up the time-propagation, the high energy components of the spectrum were not taken into account. This can be done without compromising the results. The adopted cutoff energy was $\varepsilon_\mathrm{cutoff}=15$ a.u.

\section{\label{results} Results and discussion}
In order to validate the implementation of RTDCIS, krypton and xenon have been chosen as target systems in Sec. \ref{sec:kr} and \ref{sec:xe}, respectively. We have first explored the quality of the RCIS space by reproducing the Rydberg series with configuration $ns^2np^5(^2P^o_{j_a})n'\ell'$, where $n'>n$ with $n=4$ for krypton and $n=5$ for xenon. Our relativistic singly-excited state energy levels and their corresponding total angular momentum have been compared with 4-component CIS calculations performed with the LUCIAREL module of the DIRAC19 code \cite{DIRAC19,Fleig03,Fleig06,Olsen90,Knecht10}, and with experimental data provided by NIST  \cite{Saloman04,Saloman07,NIST:database}. Finally, RTDCIS photoionization cross sections have been compared with experimental data \cite{Samson02,Shannon77,Becker89,Kammerling89} and with RRPA calculations.  Details on RRPA calculations are given in Appendix \ref{appendixB}. Important conversion factors are for the energy, 1 a.u. equals to 27.2114 eV, and for the cross section, 1 a.u. equals 28.0028 Mb.

\subsection{Krypton}
\label{sec:kr}
As a first comparison between DIRAC19 and our RCIS code, the Dirac-Fock orbital energies of krypton are shown in Table \ref{DF_energies_Kr}. Dirac-Fock energies obtained with our code are labeled as ``RCIS''. The calculation performed with DIRAC19 was done using a Gaussian-type orbital (GTO) basis set with parameters given in Appendix \ref{appendixC}. Excellent agreement is found between the two calculations. Over-all, the differences are not above the $0.01\%$. The B-splines parameters chosen here are able to reproduces the one-particle Dirac-Fock orbital energies as good as a GTO-type basis set. The Dirac-Fock orbital energies provide further a good approximation to the ionization energies from the outermost orbitals: NIST~\cite{NIST:database}  gives the energy needed for ionization to the ground state of Kr$^+$ $4p^5(^2P_{3/2})$ to $13.9996$~eV, and  to $14.6654$~eV for ionization to $4p^5(^2P_{1/2})$, in close agreement with the values given in Table~\ref{DF_energies_Kr}. For ionization from deeper core-orbitals, however,  the true ionization energies are typically several eV lower than the Dirac-Fock orbital energies due to the increased importance of orbital relaxation.
\begin{table}[t!]
\begin{center}
\caption{\label{DF_energies_Kr}Dirac-Fock orbital energies of krypton.}
 \begin{tabular}{c c c | r | r |  c} 
 \hline\hline
 \multirow{2}{*}{$n$}& \multirow{2}{*}{$\el$} & \multirow{2}{*}{$j$} & \multicolumn{1}{c|}{RCIS\footnote{Energies computed with Eq.(\ref{DFE}).}} & \multicolumn{1}{c|}{DIRAC19\footnote{Energies computed with  DIRAC19.}} & \multirow{2}{*}{$\Sigma_\mathrm{DIRAC}$\footnote{$\Sigma_\mathrm{DIRAC}=|1-\varepsilon_\mathrm{RCIS}/\varepsilon_\mathrm{DIRAC}|\times100$.}}\\
 \cline{4-5}
 & & & $\varepsilon_{n,\el,j}$ (eV) & $\varepsilon_{n,\el,j}$ (eV)& \\ 
 \hline
1&      0&      1/2&    -14413.49791&   -14413.48729&  0.00007\%\\  
\hline
2&      0&      1/2&    -1961.39410&    -1961.40662 &  0.00064\%\\
2&      1&      1/2&    -1765.30087&    -1765.27230 &  0.00162\%\\
2&      1&      3/2&    -1710.99426&    -1711.04216 &  0.00280\%\\
\hline
3&      0&      1/2&    -305.45367&     -305.44033  &  0.00437\%\\
3&      1&      1/2&    -234.58893&     -234.55246  &  0.01555\%\\
3&      1&      3/2&    -226.21326&     -226.20836  &  0.00217\%\\
3&      2&      3/2&    -102.79160&     -102.80004  &  0.00821\%\\
3&      2&      5/2&    -101.40654&     -101.41688  &  0.01020\%\\
\hline
4&      0&      1/2&    -32.32088&      -32.32143   &  0.00170\%\\
4&      1&      1/2&    -14.73688&      -14.73416   &  0.01846\%\\
4&      1&      3/2&    -13.99564&      -13.99618   &  0.00386\%\\   
\hline\hline
\end{tabular}
\end{center}
\end{table}

In order to investigate the quality of the RCIS space, several singly-excited state energy levels for the series $4s^24p^5(^2P_{1/2}^o)n'\ell'$ and $4s^24p^5(^2P_{3/2}^o)n'\ell'$ are shown in Table \ref{CIS_energies_Kr}. Calculations have been carried out using only the following active holes: $4p_{3/2}$ and $4p_{1/2}$. The RCIS energy levels and their corresponding total angular momenta have been obtained following the prescriptions given in Section \ref{observables}. First, experimental excitation energy levels are shown together with their total angular momentum $J$ and with their corresponding configuration. The energy levels computed with DIRAC19 are shown together with their corresponding parity.  DIRAC19 cannot exploit symmetry at the CIS-level, however, for closed-shell atoms, orbitals have a well-defined parity, i.e. orbitals can be ``gerade'' or ``ungerade'' (g/u). As a result, levels can be easily characterized by the combination of hole and particle parities. Finally, we show the energy levels obtained with our RCIS code. The RCIS energy levels are not shown together with their corresponding total angular momenta as the total angular momenta computed with Eq.(\ref{total_J}) are similar to the experimental values up to the machine accuracy. In addition, for a given level, the columns with ``$\ell_{a,\mathrm{max}}$'', ``$j_{a,\mathrm{max}}$'', ``$\ell_{p,\mathrm{max}}$'' and ``$j_{p,\mathrm{max}}$'' contain the one-particle orbital- and total-angular momenta of the most relevant core and particle orbitals in the coefficient expansion vector $\bf C_n$ in Eq.(\ref{RCIS_equation}). As we can observe, for some levels there is a co-existence of several angular momenta. That means their weights are comparable in magnitude in the coefficient vector $\bf C_n$. In general, ``$\ell_{a,\mathrm{max}}$'' and ``$j_{a,\mathrm{max}}$''  indicate the correct quantum numbers of the remaining hole. On the contrary, the assignment of the particle orbital- and $j$-quantum numbers with ``$\ell_{p,\mathrm{max}}$'' and ``$j_{p,\mathrm{max}}$'' is less conclusive,  since $s$- and $d$-waves, as well as $p_{1/2}$ - and $p_{3/2}$-waves, generally  mix to describe the excited electron. In order to theoretically predict the total angular momentum of a singly-excited state level in the RCIS space, one needs to use Eq.(\ref{total_J}). 

Closer inspection of Table~\ref{CIS_energies_Kr} reveals that the RCIS levels are mostly  above the experimental energy values. As discussed in many references, \emph{see} for example Ref.~\cite{Dreuw05}, the excitation energies computed with a CIS-based method are usually overestimated in comparison with their corresponding experimental values. 
This is related to the fact that the singly-excited determinants $\{\ket{\CIS}\}$, derived from the Dirac-Fock ground state $\ket{\Phi_0}$, can be seen as a first approximation to the true excited states. Higher-order correlations, typically requiring doubly-excited determinants as $\{\ket{\Phi_{ac}^{pr}}\}$, generally lower the energies, and are needed to improve the singly-excited state energies. The comparison of the RCIS energy levels with the DIRAC19 levels allows us to have another benchmark. As we can see, differences between RCIS and DIRAC19 are very small. Nevertheless, at higher excitation energies, the difference between RCIS and DIRAC19 increases dramatically (not shown here). This is a common problem related to the finite dimension of the implemented basis set. The same issue is observed between RCIS and NIST but at even higher excitation energies (not shown here). Generally speaking, the dimension of a GTO-type basis set is going to be limited by the apparition of the linear dependencies when diagonalizing the Dirac-Fock Hamiltonian. In the case of the B-spline representation, the dimension of the basis is limited by the size of the radial box, i.e. $r_\mathrm{max}$, which in principle can be selected to reach any specific degree of convergence. Overall, and within the degree of convergence that we have obtained with DIRAC19 and with our RCIS code, the present result indicates that with our method we are able to reproduce (almost quantitatively) the space of the relativistic singly-excited states. 

\begin{figure}[t!]
\begin{center}
\includegraphics[width=0.47\textwidth]{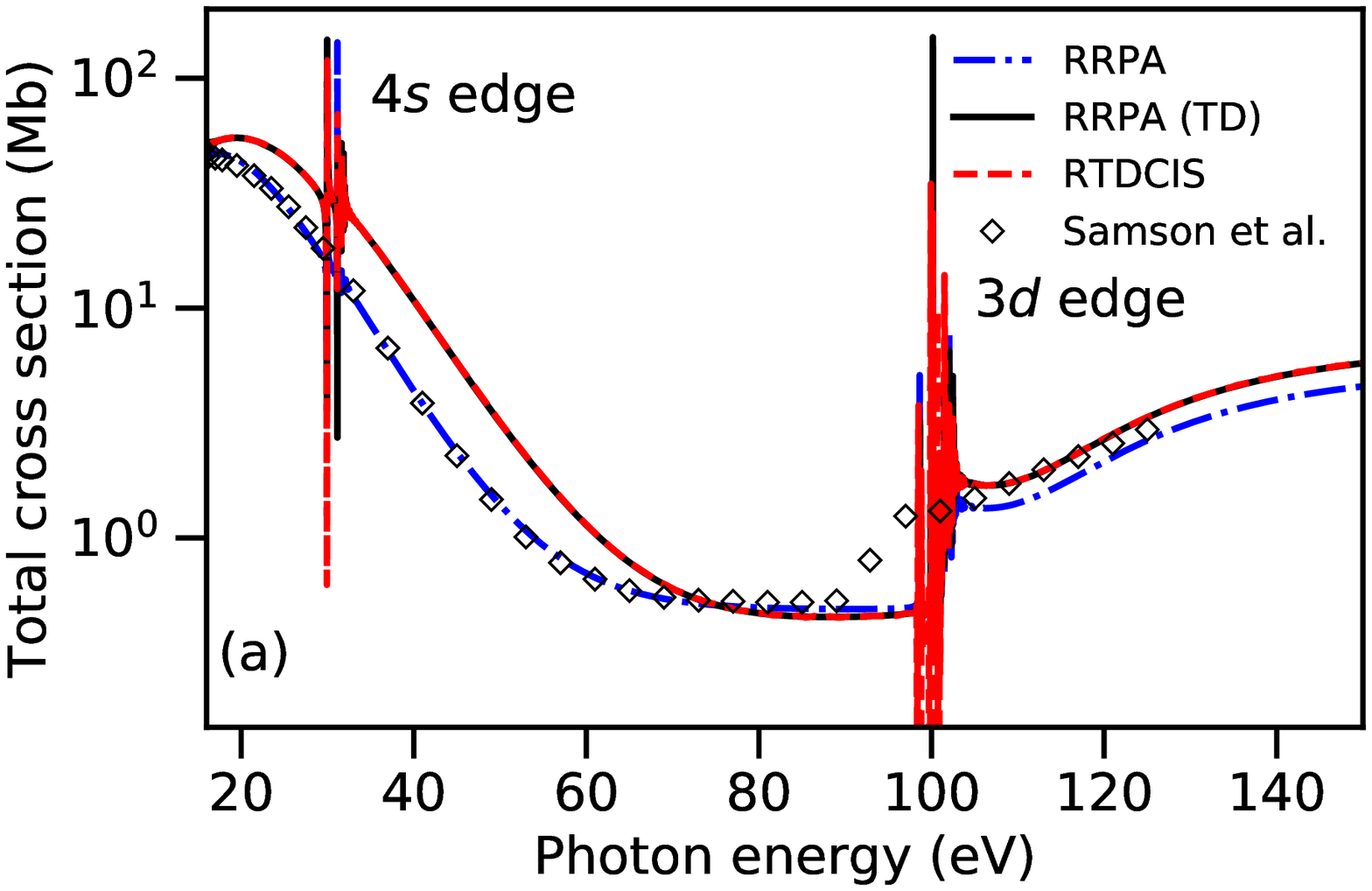}
\includegraphics[width=0.47\textwidth]{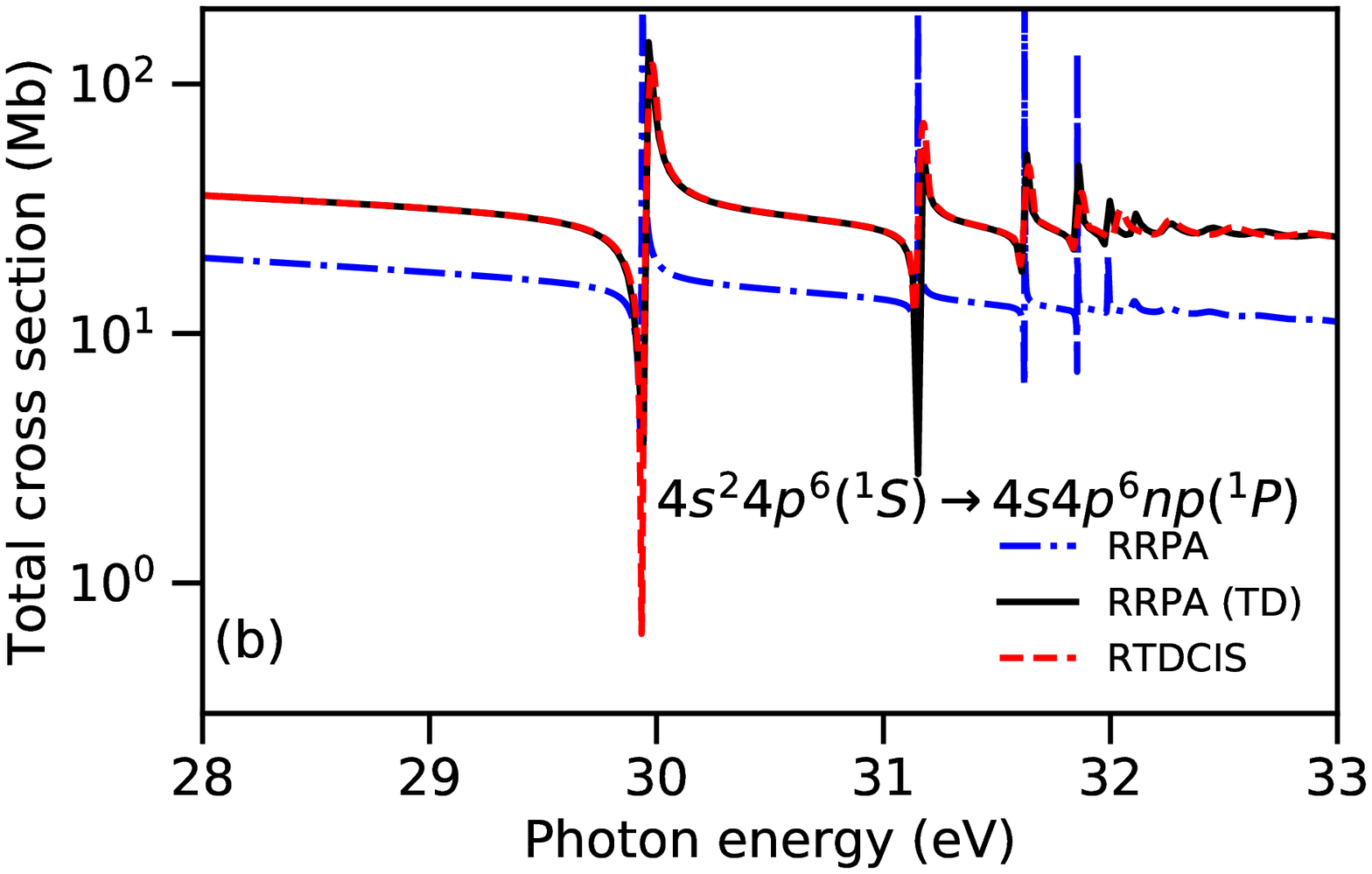}
\caption{\label{CS_Kr}Upper panel (a): theoretical and experimental total photoionization cross section of krypton. Experimental data from Ref. \cite{Samson02}. Lower panel (b): total photoionization cross section of krypton in the $4s^24p^6(^1S)\to 4s4p^6np(^1P)$ autoionization energy range.}
\end{center}
\end{figure}
In order to reproduce the experimental photoionization spectrum of krypton, RTDCIS and RRPA calculations were performed with the following active holes: $4p_{3/2}$, $4p_{1/2}$, $4s_{1/2}$, $3d_{5/2}$ and $3d_{3/2}$. In Figure~\ref{CS_Kr}, the total photoionization cross section of krypton is displayed. In the upper panel (a), a complete profile of the cross section is shown. As one can see, RTDCIS reproduces the important features of the photoionization spectrum of krypton: the quadratic decrease after the $4s$ - edge up to a minimum around 80 eV and the smooth increase after the $3d$-edge. Nevertheless, RTDCIS overestimates the experimental cross section. On the contrary, the full RRPA calculation matches very well the entire experimental profile up to the so-called $3d$-edge, where it exhibits some deviation. This deviation is in fact produced by the $\sim 5$ eV energy shift existing between the Dirac-Fock orbital energies and the binding energies of the $3d$ orbitals, as estimated from experiments and accurate calculations in Ref.~\cite{deslattes:rmp}. In order to understand the overestimation of the total photoionization cross section produced by RTDCIS, we decided to run a RRPA(TD) calculation. As one can observe, RTDCIS matches RRPA(TD) very well. The differences between RTDCIS and RRPA(TD), that can be detected in the shapes of the autoionization resonances, are simply related to the fact that a filter function was used to compute the Fourier Transform of the time-dependent correlation function in Eq.(\ref{cross_section}). Numerically identical results could only be obtained by propagating RTDCIS for an infinite amount of time which is not feasible. Apart from that, RTDCIS and RRPA(TD) results can be considered to be in excellent agreement. Thus, the Tamm-Dancoff approximation reduces RRPA to Eq.(\ref{RCIS_equation}), i.e. to the RCIS level of theory, which is used in RTDCIS. One important disadvantage of the Tamm-Dancoff approximation is that RRPA no longer obeys the Thomas-Reiche-Kuhn sum rule, which states that the sum of the transition dipole moments shall be equal to the number of electrons \cite{Bethe13,Amusia90,Dreuw05}. Therefore, properties such as total photoionization cross sections cannot be expected to be quantitatively accurate with RRPA(TD) or RTDCIS. 

In the lower panel (b), the $4s^24p^6(^1S)\to 4s4p^6np(^1P)$ autoionization energy range is presented. A comparison with the measurements performed by Chan \emph{et al.}~\cite{Chan92} shows that neither RRPA, nor RRPA(TD) or RTDCIS are able to properly reproduce the autoionization process in krypton.  Figure 15 in Ref.~\cite{Chan92}  displays typical window resonances, in sharp contrast to panel (b) in Figure ~\ref{CS_Kr}. The source of this disagreement is the lack of the ``two-electron--two-hole'' excitations. This problem is intrinsic to all RRPA and RRPA(TD) calculations. Amusia and Kheifets \cite{Amusia82}, as well as Carette \emph{et al.} \cite{Carette13}, addressed this question in argon, where these excitations  are also important. In order to obtain the correct $q$-parameter in Fano's theory of autoionization (i.e. the parameter that defines the shape of the autoionization resonances) 
one needs to include  such configurations that are close in energy to the dominating one-hole--one-particle configuration. For the resonances shown in panel (b) in Figure~\ref{CS_Kr}, which are labelled as being due to excitations to $4s4p^6np$, this means that it, in particular, is important to add configurations such as $4s^24p^44dnp$.  To overcome this limitation in RTDCIS, one will need to include doubly-excited configurations (i.e. $\ket{\Phi_{ac}^{pr}}$) in Eq.(\ref{td-wave}). However, in terms of computational time, the addition of doubly-excited states into the expansion of the time-dependent wave function is  very expensive. For the moment, our investigation on ATAS will be restricted to the space of singly excited configurations.

\subsection{Xenon}
\label{sec:xe}
Following the discussion on krypton, a similar investigation has been done for xenon. In Table \ref{DF_energies_Xe}, the Dirac-Fock energies of xenon  are compared with DIRAC19 calculations.  As in the case of krypton, the agreement between the two calculations is very good. The orbital energies for the outermost orbitals agree further well with experimental ionization energies~\cite{NIST:database} (12.1298~eV when the ion is left in  $5p^5\,^2P_{3/2}$ state, and 13.4368~eV when it is left in  $5p^5\,^2P_{1/2}$), while the $5s$ and $4d$ orbital energies are around $4$~eV above the true  positions of the $5s$-edge~\cite{NIST:database} and the $4d$-edges~\cite{svensson:88}, respectively.
\begin{table}[t!]
\begin{center}
\caption{\label{DF_energies_Xe}Dirac-Fock orbital energies of xenon.}
 \begin{tabular}{c c c | r | r |  c} 
 \hline\hline
 \multirow{2}{*}{$n$}& \multirow{2}{*}{$\el$} & \multirow{2}{*}{$j$} & \multicolumn{1}{c|}{RCIS\footnote{Energies computed with Eq.(\ref{DFE}).}} & \multicolumn{1}{c|}{DIRAC19\footnote{Energies computed with DIRAC19}} & \multirow{2}{*}{$\Sigma_\mathrm{DIRAC}$\footnote{$\Sigma_\mathrm{DIRAC}=|1-\varepsilon_\mathrm{RCIS}/\varepsilon_\mathrm{DIRAC}|\times100$.}}\\
 \cline{4-5}
 & & & $\varepsilon_{n,\el,j}$ (eV) & $\varepsilon_{n,\el,j}$ (eV)& \\ 
 \hline
 1 &0 &1/2&-34755.43795 &      -34755.96427  &    0.00151\% \\ 
\hline                                                    
2& 0& 1/2& -5508.96252 &       -5509.38625   &    0.00769\% \\
2& 1& 1/2& -5161.48991 &       -5161.40855   &    0.00158\% \\
2& 1& 3/2& -4835.61037 &       -4835.61570   &    0.00011\% \\
\hline                                                      
3& 0& 1/2& -1170.29682 &       -1170.39998   &    0.00881\% \\
3& 1& 1/2& -1024.78963 &       -1024.78765   &    0.00019\% \\
3& 1& 3/2&  -961.25527 &        -961.27295   &    0.00184\% \\
3& 2& 3/2&  -708.13702 &        -708.15645   &    0.00274\% \\
3& 2& 5/2&  -694.90484 &        -694.92646   &    0.00311\% \\
\hline                                                       
4& 0& 1/2&  -229.37366 &        -229.40692   &    0.01450\% \\
4& 1& 1/2&  -175.58375 &        -175.59037   &    0.00377\% \\
4& 1& 3/2&  -162.80167 &        -162.81276   &    0.00681\% \\
4& 2& 3/2&   -73.78031 &         -73.79646   &    0.02189\% \\
4& 2& 5/2&   -71.66946 &         -71.68587   &    0.02290\% \\
\hline                                                       
5& 0& 1/2&   -27.48507 &         -27.48995   &    0.01776\% \\
5& 1& 1/2&   -13.40393 &         -13.40252   &    0.01052\% \\
5& 1& 3/2&   -11.96796 &         -11.96762   &    0.00284\% \\                                
\hline\hline
\end{tabular}
\end{center}
\end{table}

In Table \ref{CIS_energies_Xe}, several energy levels 
belonging to the $5s^25p^5(^2P_{1/2}^o)n'\ell'$ and $5s^25p^5(^2P_{3/2}^o)n'\ell'$ series
of singly-excited state in xenon 
are given. Calculations have been carried out using only the following active holes: $5p_{3/2}$ and $5p_{1/2}$.  First of all, the experimental energy levels are shown together with their total angular momentum $J$ and with their corresponding electron configuration. The energy levels computed with DIRAC19 are shown together with their corresponding parity. Moreover, an extra column has been added to the DIRAC19 results. This column contains the energy levels when the magnetic part of the electron-electron interaction, in form of the so-called Gaunt term 
\cite{Gaunt29,Saue11},
is accounted for. For the description of the valence-excited states it is typically not
the direct influence of the magnetic interaction which is most important, but the self-consistent treatment of it which changes the central-field potential~\cite{lindroth:89:breit}.
The RCIS energy levels are not shown together with their $J$ values as the total angular momenta computed with Eq.(\ref{total_J}) are equal to the experimental values up to the machine accuracy. In addition, the columns with  ``$\ell_{a,\mathrm{max}}$'', ``$j_{a,\mathrm{max}}$'', ``$\ell_{p,\mathrm{max}}$'' and ``$j_{p,\mathrm{max}}$'' contain the one-particle orbital- and total-angular momenta of the most relevant core and particle orbitals in the coefficient expansion vector $\bf C_n$ in Eq.(\ref{RCIS_equation}). Similarly to krypton, ``$\ell_{a,\mathrm{max}}$'' and ``$j_{a,\mathrm{max}}$'' typically do indicate the dominating quantum numbers of the hole, while several angular momentum channels are needed to describe the excited electron. Finally, in the last two columns of Table \ref{CIS_energies_Xe}, the RCIS energy levels are compared with NIST and DIRAC19. As one can see, the RCIS levels are in good agreement with the levels computed with DIRAC19. We note that both calculations most often underestimate the experimental energies. The deviations from the experimental energy values are related to the lack of higher-order electron correlation, as previously commented on for krypton, but the form of the relativistic electron-electron interaction also starts to come into play. As seen in Table \ref{CIS_energies_Xe}, the energies obtained with DIRAC19 are modified with up to one percent when the Gaunt interaction is added to the Dirac-Fock Hamiltonian. In contrast, only minor modifications of the DIRAC19 energy levels were observed when it was accounted for in the krypton calculations (not shown here). As a  consequence, if heavier elements are to be explored in the future with the RTDCIS code, the Gaunt interaction, or even better, the complete Breit~\cite{Breit32} interaction, should be taken into account.  

\begin{figure}[h!]
\begin{center}
\includegraphics[width=0.47\textwidth]{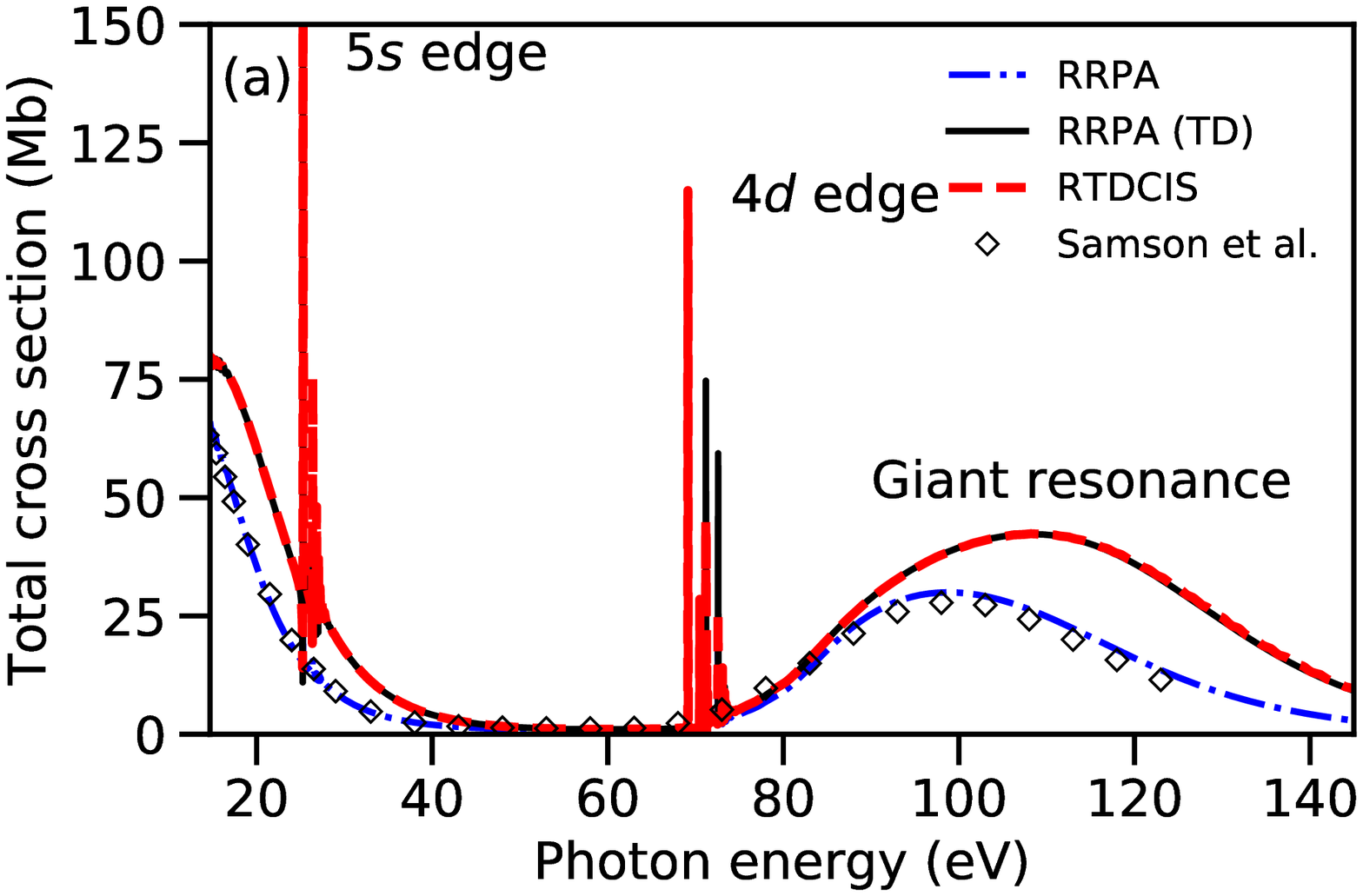}
\includegraphics[width=0.47\textwidth]{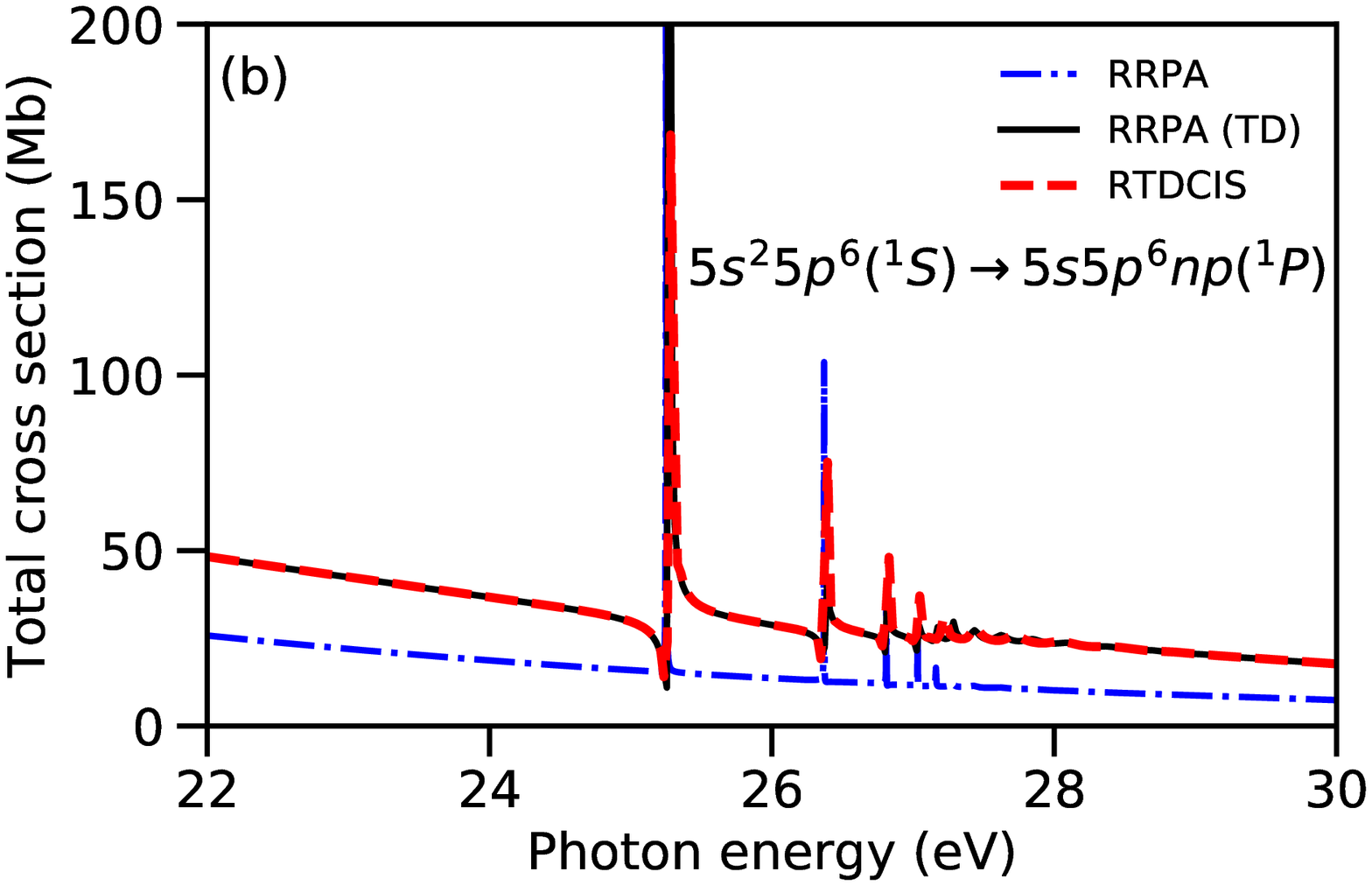}
\includegraphics[width=0.47\textwidth]{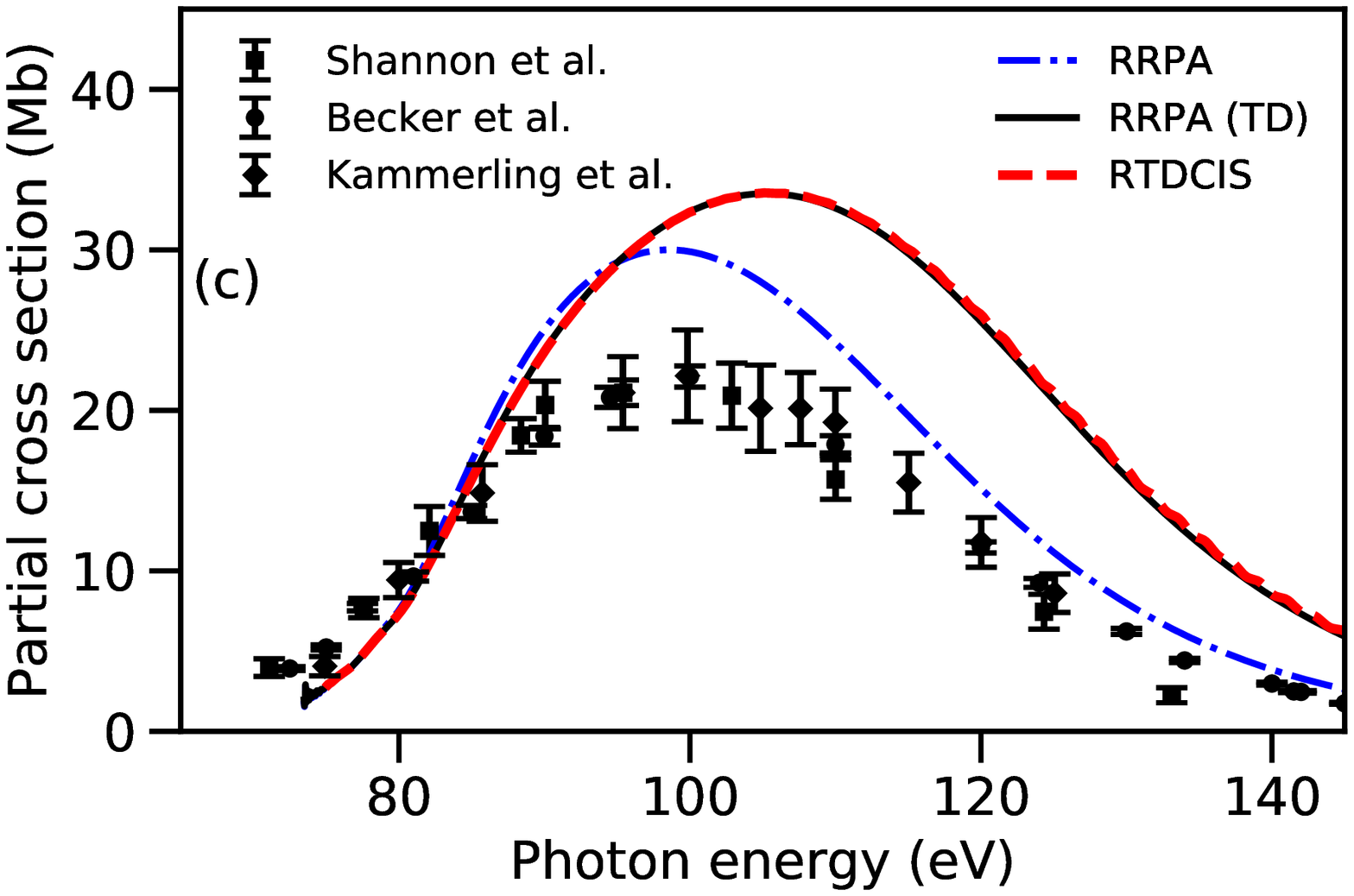}
\caption{\label{CS_Xe}Upper panel (a): theoretical and experimental total photoionization cross section of xenon. Experimental data from Ref. \cite{Samson02}.  Middle panel (b): total photoionization cross section of xenon in the $5s^25p^6(^1S)\to 5s5p^6np(^1P)$ autoionization energy range. Lower panel (c): xenon partial $4d$ photoionization cross section. Experimental data: solid squares from \cite{Shannon77}, solid circles from \cite{Becker89} and solid diamonds from \cite{Kammerling89}.}
\end{center}
\end{figure}

In Figure \ref{CS_Xe}, the total and partial $4d$ photoionization cross sections of xenon are given. In the upper panel, the experimental total photoionization cross section profile is compared with RRPA, RRPA(TD) and RTDCIS calculations. Calculations have been performed using the following active holes: $5p_{3/2}$, $5p_{1/2}$, $5s_{1/2}$, $4d_{5/2}$ and $4d_{3/2}$. As in the case of krypton, RTDCIS overlaps the RRPA(TD) result. These two calculations overestimate the experimental profile but are able to reproduce the important features of the spectrum: the quadratic decreasing up to a minimum around 60 eV and, after the $4d$-edge, the so-called giant resonance. The differences that can be seen between RTDCIS and RRPA(TD) are due to the use of a filter function when computing the Fourier transform in Eq.(\ref{cross_section}) as previously commented for krypton. On the contrary, RRPA matches very well over the whole spectrum. In the middle panel, the total photoionization cross section of xenon is shown in the energy range of the  $5s^25p^6(^1S)\to 5s5p^6np(^1P)$ autoionization resonances. As for krypton, the resonances obtained with RRPA have very narrow energy widths. RTDCIS and RRPA(TD) results present broader resonances. However, a comparison with the measurements by Chan \emph{et al.}\cite{Chan92}, shows as for krypton, clear window resonances, 
\emph{cf} Figure 16 in Ref.~\cite{Chan92}, i.e. with a clearly  different resonance profile than either of the curves in panel (b) in Figure~\ref{CS_Xe}.

Finally, in the lower panel (c), the $4d$ partial photoionization cross section of xenon is presented. RRPA, RRPA(TD) and RTDCIS calculations have been performed following the work of Cheng and Johnson \cite{Cheng83}. Correlation effects from the $5s$ and $5p$ shells, as well as from inner shells, have been neglected and only the $4d_{5/2}$ and $4d_{3/2}$ orbitals are considered as active holes. The origin of the giant resonance in the photoionization spectrum of xenon has been investigated by several authors, \emph{see} for example Refs. \cite{Cooper64,Cheng83,Chen15,Toffoli02}. The existence of this resonance can be explained using a ``single-electron picture'', however, to obtain the correct spectral position and shape one is forced to introduce electron-electron correlation effects beyond the mean-field approximation. As we can see, the RTDCIS method matches the results generated by the RRPA(TD) calculation. However, the partial cross section is overestimated and the spectral position is shifted  in comparison with the experimental data. On the contrary, the RRPA calculation, which includes the so-called ``time-reverse'' diagrams (also called {\em ground state correlation}), does produce the true strength of the resonance and the spectral shift compared to  RRPA(TD). Nevertheless, the still missing relaxation effects should also be included in order to properly describe the giant resonance~\cite{Kutzner89}. Overall, RTDCIS is a simple method that is capable to reproduce single-electron processes almost quantitatively in xenon.

\section{\label{conclusion}{Conclusion}}
RTDCIS is based on the expansion of the $N$-electron time-dependent wave function in the space of single excitations and the first goal  was to investigate the quality of this space. To this end, energy levels of a range of  singly-excited states in krypton and xenon was calculated and compared with experimental data from NIST, as well as with  4-component CIS calculations performed with DIRAC19. The agreement with DIRAC19 is excellent, and the deviations from experiments is within the expected size of contributions  from higher order electron-electron correlation, which are missing in both cases. For xenon there is evidence that contributions from inter-electron  interactions beyond the pure Coulomb contribution play a role. In the future we plan thus to add  a possibility to start from Dirac-Fock-Breit orbitals, which can be done without any major consequences for the time-consumption  during the time-propagation. 

Subsequently, the time-propagation was validated by comparison of our RTDCIS photoionization cross sections, which are extracted after time-propation of the TDDE, with traditionally calculated  cross sections, as well as with experiments. As expected, the RTDCIS cross sections are completely equivalent to the RRPA(TD) results.

In conclusion, our implementation of RTDCIS,  which is now benchmarked with respect to accuracy,  opens the possibility to study one-electron processes in heavy atoms beyond the perturbative regime in the context of attosecond transient absorption spectroscopy, high-order harmonic generation, above-threshold ionization and laser-assisted photoionization.

\section{Acknowledgements} 
The authors acknowledge support from the Knut and Alice Wallenberg Foundation: Grant No. 2017.0104 and 2019.0154, from the Swedish Research Council: Grant No. 2018-03845 and 2020-03315, and from the Olle Engkvist Foundation: Grant No. 194-0734. Dr S.  Carlström is acknowledged for helpful discussions. 

\appendix

\section{\label{appendixA}\\Total angular momentum matrix elements \\ for closed-shell atoms}
When closed-shell atoms are investigated, one can easily find that the elements of the total angular momentum matrix can be expressed in terms of the occupied and virtual one-particle orbital quantum numbers. In this appendix, the matrix elements for the total angular momentum are derived in terms of the occupied and virtual one-particle orbital quantum numbers.

The total angular momentum expectation value is defined as 
\begin{equation}
\langle \hat{J}_\mathrm{total}^2 \rangle_n = \mathbf{C}_n^\dagger\;\mathbf{J}^2\;\mathbf{C}_n,
\end{equation}
where the vector $\bf C_n$ is obtained after diagonalizing the relativistic $N$-electron field-free Hamiltonian in the basis of the relativistic configuration interaction singles (RCIS) states. The matrix elements of $\bf J^2$ are given by
\begin{equation}
\label{j_element1}
    \mathbf{J}^2_{n',n}=\bra{\Phi_b^q}\hat{J}_\mathrm{total}^2\ket{\Phi_a^p},
\end{equation}
where the relativistic singly-excited states are defined in second quantization as
\begin{equation}
\label{RCIS_aa0}
    \ket{\Phi_a^p}=\cre{p} \ann{a}|\Phi_0\rangle. 
\end{equation}
As the operator $\hat{J}$ does not couple the occupied and the virtual one-particle states in a closed-shell atom, the total angular momentum operator can be re-defined as 
\begin{equation}
\label{jtotal2}
    \hat{J}^2_\mathrm{total}=\hat{J}^2_{occ}+\hat{J}^2_{vir}+2\hat{J}_{occ}\hat{J}_{vir}, 
\end{equation}
where 
\begin{subequations}
\label{j_occ_j_vic}
\begin{align}
\Jocc &= \sum_{a,b}^{occ}\cre{b} \ann{a}\bra{b}\hat{J}\ket{a}; \\
\Jvir &= \sum_{p,q}^{vir}\cre{q} \ann{p}\bra{q}\hat{J}\ket{p}.
\end{align}
\end{subequations}
Therefore, in order to compute Eq.(\ref{j_element1}), one needs first to know the action of $\hat{J}^2_{occ}$, $\hat{J}^2_{vir}$ and $\hat{J}_{occ}\hat{J}_{vir}$ on $\ket{\Phi_a^p}$. 

By making use of Eq.(\ref{RCIS_aa0}), Eq.(\ref{j_occ_j_vic}) and the anticommutator relations for the creation and the annihilation operators, given by
\begin{eqnarray}
\label{anticommutator}
\{\cre{i},\cre{j}\}=0; \;\; \{\ann{i},\ann{j}\}= 0;\;\mathrm{and} \;\{\ann{i},\cre{j}\}=\delta _{i,j},
\end{eqnarray}
one obtains the actions for $\Jocc$ and $\Jvir$:
\begin{subequations}
\begin{align}
    \hat{J}_{occ}\ket{\CIS}&=-\sum_{b}\langle a|\J|b\rangle|\Phi_b^p\rangle;\\
    \Jvir|\Phi_a^p\rangle& = \sum_{r}\langle r|\J|p\rangle|\Phi_a^r\rangle.
\end{align}
\end{subequations}
In a second step, one easily gets the actions for $\hat{J}^2_{occ}$ and $\hat{J}^2_{vir}$, which are given by
\begin{subequations}
\begin{align}
\Jocc^2|\Phi_a^p\rangle &= \Jocc(\Jocc|\Phi_a^p\rangle) =  j_a(j_a+1)|\Phi_a^p\rangle;\\
\Jvir^2|\Phi_a^p\rangle &= \Jvir(\Jvir|\Phi_a^p\rangle) = j_p(j_p+1)|\Phi_a^p\rangle.
\end{align}
\end{subequations}

The action for $\Jocc\Jvir$ is computed separately for each $x$-, $y$- and $z$-contribution, i.e.
\begin{equation}
\label{mixterm}
    \Jocc\Jvir=\frac{1}{2}(\Jocc^+\Jvir^-+\Jocc^-\Jvir^+)+\Jocc^z\Jvir^z,  
\end{equation}
 where the terms $\Jocc^x\Jvir^x$ and $\Jocc^y\Jvir^y$ have been rewritten by making use of the definition of the ladder operators, i.e.
\begin{eqnarray}
\label{ladders}
\hat{J}_x=\frac{1}{2}(\hat{J}^++\hat{J}^-);\;\mathrm{and}\;\hat{J}_y=\frac{1}{2i}(\hat{J}^+-\hat{J}^-).
\end{eqnarray}
In this work, the ladder operators have been defined following the phase convention established by Condon and Shortley \cite{Condon53}, i.e.
\begin{subequations}
\label{ladder_action}
\begin{align}
\hat{J}^+\varphi_{j,m}&=&[(j-m)(j+m+1)]^{1/2}\varphi_{j,m+1};\\
\hat{J}^-\varphi_{j,m}&=&[(j+m)(j-m+1)]^{1/2}\varphi_{j,m-1}.
\end{align}
\end{subequations}

Thus, the action of $\Jocc^+\Jvir^-$ on $\ket{\CIS}$ is derived in three steps as follows:
\begin{enumerate}
    \item[\emph{Step 1.}] We make use of the definitions given in Eq.(\ref{ladder_action}). 
          \begin{equation}
          \nonumber
          \begin{array}{rll}
          \Jocc^+\Jvir^-|\Phi_a^p\rangle
           &=& \Jocc^+\sum_{r}\langle r|\J^-|p\rangle|\Phi_a^r\rangle\\
           &=&\Jocc^+\sum_{r}[(j_p+m_{p})(j_p-m_{p}+1)]^{1/2}\\
           & &\times\delta_{j_r,j_p}\delta_{m_{r},m_{p}-1}|\Phi_a^r\rangle.
          \end{array}
          \end{equation} 
    \item[\emph{Step 2.}] For $m_{r}=m_{p}-1$ and using the notation $p^-\equiv\{j_p,m_{p}-1\}$, the action in step 1 is re-written as follows,
          \begin{equation}
          \nonumber
          \begin{array}{lll}
          \Jocc^+\Jvir^-|\Phi_a^p\rangle&=&[(j_p+m_{p})(j_p-m_{p}+1)]^{1/2}\Jocc^+|\Phi_a^{p^-}\rangle\\
          &=&[(j_p+m_{p})(j_p-m_{p}+1)]^{1/2}\\
          & &\times[-\sum_{b}\langle a|\J^+|b\rangle|\Phi_b^{p^-}\rangle]\\
          &=&-[(j_p+m_{p})(j_p-m_{p}+1)]^{1/2}\\
          & &\times \sum_{b}[(j_b-m_{b})(j_b+m_{b}+1)]^{1/2} \\
          & &\times \delta_{j_a,j_b}\delta_{m_{a},m_{b}+1}|\Phi_{b}^{p^-}\rangle.
          \end{array}
          \end{equation}
    \item[\emph{Step 3.}]For $m_{b}=m_{a}-1$ and using the notation $a^-\equiv\{j_a,m_{a}-1\}$, the resulting action is given by, 
          \begin{equation}
          \label{jocc_+_jvir_-}
          \begin{array}{lll}
          \Jocc^+\Jvir^-|\Phi_a^p\rangle&=&-[(j_p+m_{p})(j_p-m_{p}+1)]^{1/2}\\
          &&\times[(j_a-m_{a}+1)(j_a+m_{a})]^{1/2}|\Phi_{a^-}^{p^-}\rangle.   
          \end{array}
          \end{equation}
\end{enumerate}

The action of $\Jocc^-\Jvir^+$ on $\ket{\CIS}$ is derived also in three steps as follows:
\begin{enumerate}
    \item[\emph{Step 1.}] We make use of the definitions given in Eq.(\ref{ladder_action}).
          \begin{equation}
          \nonumber
          \begin{array}{lll}
          \Jocc^-\Jvir^+|\Phi_a^p\rangle
          &=&\Jocc^-\sum_{r}\langle r|\J^+|p\rangle|\Phi_a^r\rangle\\
          &=&\Jocc^-\sum_{r}[(j_p-m_{p})(j_p+m_{p}+1)]^{1/2}\\
          &&\times \delta_{j_r,j_p}\delta_{m_{r},m_{p}+1}|\Phi_a^r\rangle.
          \end{array}
          \end{equation}
    \item[\emph{Step 2.}] For $m_{r}=m_{p}+1$ and using the notation $p^+\equiv\{j_p,m_{p}+1\}$, the action in step 1 is re-written as follows, 
          \begin{equation}
          \nonumber
          \begin{array}{lll}
          \Jocc^-\Jvir^+|\Phi_a^p\rangle&=&[(j_p-m_{p})(j_p+m_{p}+1)]^{1/2}\Jocc^-|\Phi_a^{p^+}\rangle\\
          &=&[(j_p-m_{p})(j_p+m_{p}+1)]^{1/2}\\
          & &\times[-\sum_{b}\langle a |\J^-|b\rangle |\Phi_b^{p^+}\rangle]\\
          &=&-[(j_p-m_{p})(j_p+m_{p}+1)]^{1/2}\\
          & &\times \sum_{b}[(j_b+m_{b})(j_b-m_{b}+1)]^{1/2}\\
          & &\times \delta_{j_a,j_b}\delta_{m_{a},m_{b}-1}|\Phi_{b}^{p^+}\rangle. 
          \end{array}
          \end{equation}
    \item[\emph{Step 3.}] For $m_{b}=m_{a}+1$ and using the notation $a^+\equiv\{j_a,m_{a}+1\}$, the resulting action is given by
          \begin{equation}
          \label{jocc_-_jvir+}
          \begin{array}{lll}
          \Jocc^-\Jvir^+|\Phi_a^p\rangle&=&-[(j_p-m_{p})(j_p+m_{p}+1)]^{1/2}\\
          & &\times[(j_a+m_{a}+1)(j_a-m_{a})]^{1/2}|\Phi_{a^+}^{p^+}\rangle.
          \end{array}
          \end{equation} 
\end{enumerate}

Finally, the last action one needs to compute is the action for $J_{occ}^zJ_{vir}^z$, and it is given by
\begin{equation}
\label{jocc_z_jvir_z}
\begin{array}{lll}
\Jocc^z\Jvir^z|\Phi_a^p\rangle &=&\Jocc^z(\Jvir^z|\Phi_a^p\rangle) \\
     &=&(-m_{a})m_{p}|\Phi_a^p\rangle.  
\end{array}
\end{equation}

 
In conclusion, the action of $\J_\mathrm{total}^2$ on  $\ket{\CIS}$ is given by
\begin{equation}
\label{action}
\begin{array}{lll}
    \J_\mathrm{total}^2\ket{\CIS}&=&\Jocc^2\ket{\CIS}+\Jvir^2\ket{\CIS}\\
    & &+\Jocc^+\Jvir^-\ket{\CIS}+\Jocc^-\Jvir^+\ket{\CIS}\\
    & &+2\Jocc^z\Jvir^z\ket{\CIS}.
\end{array}
\end{equation}

Using Eq.(\ref{j_occ_j_vic}), Eq.(\ref{jocc_+_jvir_-}), Eq.(\ref{jocc_-_jvir+}) and Eq.(\ref{jocc_z_jvir_z}), the action of $\J_\mathrm{total}^2$ on $\ket{\CIS}$  given in Eq.(\ref{action}) is re-written as
\begin{equation}
      \J_\mathrm{total}^2\ket{\CIS}=k_1\ket{\CIS}-k_2|\Phi_{a^-}^{p^-}\rangle-k_3|\Phi_{a^+}^{p^+}\rangle,
\end{equation}
where the angular coefficients are given by
\begin{subequations}
\begin{align}
\nonumber
k_1=&j_a(j_a+1)+j_p(j_p+1)-2m_{a}m_{p};\\
\nonumber
k_2=&[(j_a+m_{a})(j_a-m_{a}+1)]^{1/2}\\
\nonumber
  \times&[(j_p+m_{p})(j_p-m_{p}+1)]^{1/2};\\
\nonumber
k_3=&[(j_a-m_{a})(j_a+m_{a}+1)]^{1/2}\\
\nonumber
        \times&[(j_p-m_{p})(j_p+m_{p}+1)]^{1/2}.
\end{align}
\end{subequations}

Finally, using the fact that the RCIS states are orthonormal, the matrix elements of $\bf J^2$ are given by the following expression: 
\begin{equation}
\begin{array}{lll}
 \mathbf{J}^2_{n',n}&=&\bra{\Phi_b^q}\hat{J}_\mathrm{total}^2\ket{\Phi_a^p}\\
 &=&\{\;k_1\;\delta_{m_{a},m_{b}}\delta_{m_{p},m_{q}}\\
 & &-k_2\;\delta_{m_{a}-1,m_{b}-1}\delta_{m_{p}-1,m_{q}-1}\\
 & &-k_3\;\delta_{m_{a}+1,m_{b}+1}\delta_{m_{p}+1,m_{q}+1}\}\\
 & &\times\;\delta_{j_a,j_b}\delta_{j_p,j_q}.
\end{array}
\end{equation}

\section{\label{appendixB}\\Relativistic random-phase approximation}
The present relativistic implementation of the random-phase approximation (RRPA) follows Refs. \cite{Vinbladh19, MarcusEva14}.  (Note that what we here label RRPA is in the literature  often referred to as  relativistic random phase approximation with exchange (RRPAE) \cite{Amusia90}). The perturbed part of the many-electron wave function is calculated for each photon frequency $\omega$.When the electron released in the photoionization process is coming from the initially occupied orbital $\ket{a}$, this perturbed wave function, $\ket{\rho_a^\pm}$, is given by the following expression,
\begin{eqnarray}
\label{rpae_eqs1}
\nonumber
\ket{\rho_a^{\pm}} &=& \sum_p \dfrac{\bra{p}\hat{z}\ket{a}\ket{p}}{\varepsilon_p-\varepsilon_a\pm\omega}+\dfrac{\ket{p}}{\varepsilon_p - \varepsilon_a\pm\omega}\\
\nonumber
& &\times \sum_b \left[\braket{bp|r_{12}^{-1}|\rho_b^\pm a}-\braket{bp|r_{12}^{-1}|a\rho_b^\pm}\right.\\
& &+\left.\braket{p\rho_b^\mp|r_{12}^{-1}|ab}-\braket{\rho_b^\mp p|r_{12}^{-1}|ab}\right], 
\end{eqnarray}
where $\hat{z}$ is the dipole operator in length gauge, cf.~Eq.\ref{lengthgauge}, and the superscript ``$\pm$'' indicates absorption or emission of a photon of frequency $\omega$. The  third and fourth terms in Eq.~\ref{rpae_eqs1}: $\braket{bp|r_{12}^{-1}|\rho_b^\pm a}$ and $\braket{bp|r_{12}^{-1}|a\rho_b^\pm}$, give the so-called \emph{``time-forward''} contribution. It lets the outgoing electron adjust to the presence of the hole and allow for coupling between different channels. The fifth and sixth terms:  $\braket{p\rho_b^\mp|r_{12}^{-1}|ab}$ and $\braket{\rho_b^\mp p|r_{12}^{-1}|ab}$, are instead labelled the  \emph{``time-reverse''} contribution. These terms  are needed to fully account for  the modifications of the mean-field potential due to the interaction with the electromagnetic field, and they are also needed to ensure that identical results are obtained in length and velocity gauge. The so-called Tamm-Dancoff approximation (RRPA(TD)) is obtained when the time-reverse diagrams are excluded from Eq.(\ref{rpae_eqs1}). The many-body contributions included in RCIS are identical to those in the Tamm-Dancoff approximation. For a deeper discussion on the relationships between different single-reference methods, \emph{see} Ref. \cite{Dreuw05}.

From the perturbed wave function in  in Eq.~\ref{rpae_eqs1}, the partial photoionization cross section 
can be computed from the outgoing electron flux in each channel. This can be done at any 
radius  far enough from the nucleus. The hole and particle  are defined from the Dirac-Fock approximation and expressed in  a B-spline basis set as described in Sec.~\ref{bsplinesB}.   For both krypton and xenon we used B-spline orders: $k_s = 8$ and $k'_s = 7$, for the small and large component, respectively, and a hybrid exponential-linear knot grid distribution. While the RCIS time-propgation was done in the presence of a CAP, the RRPA implementation is using exterior complex scaling (ECS). The scaled region starts at $70$~a.u., where the radial grid is rotated $0.05$~radians into the complex plane, while the full computational box is   $100$~a.u. For krypton, a total of 240 knot points were used, with 9 points in an inner exponential region (close to the nucleus), 155 in the linear (non-rotated) region, and finally 76 knot points in the linear ECS region. For xenon, a total of 243 knot points were used. The difference with respect to the grid for krypton was the addition of 3 extra knot points in the exponential region close to the nucleus.  

\section{\label{appendixC}\\DIRAC19 GTO-basis set}

\subsection*{Krypton}
DIRAC19 calculations for krypton were performed by using a GTO-type basis set consisting in a triple-augmented correlation-consistent triple zeta (t-aug-cc-pVTZ) basis. In addition, in order to increase the accuracy of the high energy Rydberg states, 5 Kaufmann-type functions were added per angular momentum up to $\ell_\mathrm{max}=2$. The contraction coefficients of the Kaufmann-type functions were set to unity and the exponents were computed using Eq.(18) in Ref. \cite{Kaufmann89}, with $Z=1$ and where the fitting parameters $a_\ell$ and $b_\ell$ can be found in Table 2 in the same reference. The resulting GTO basis set was the best one we could design as the addition of more GTO functions generates important linear dependencies problems. To ensure the elimination of linear dependencies, cutoffs of $10^{-9}$ and $10^{-7}$ for the small and the large components were selected in the DIRAC19 input file.  With this set of parameters, the total number of primitives was 769, i.e. 233 for the large components and 536 for the small components.

\subsection*{Xenon}
 DIRAC19 calculations in xenon have been performed with a triple-augmented Dyall-tyoe GTO basis set (t-aug-dyall.v2z) and 5 Kaufmann-type functions per angular momentum up to $\ell_\mathrm{max}=2$. The contraction coefficients of the Kaufmann-type functions were set to unity and the exponents were computed using Eq.(18) in Ref. \cite{Kaufmann89}, with $Z=1$ and where the fitting parameters $a_\ell$ and $b_\ell$ are listed in Table 2 of the same reference. The addition of more GTO functions to the basis set generates important linear dependencies. As in the case of krypton, a cutoff for the large and the small components was established in the DIRAC19 input file. With these parameters, the total number of primitives was 707; i.e. 212 for the large components and 495 for the small components. Note that less primitives were used in the calculation of xenon. In DIRAC19, the option of using a t-aug-cc-pVTZ basis for xenon was not available. This issue will limit the number of accurate singly-excited states that one can obtain for xenon. Note that only few levels are given in Table \ref{CIS_energies_Xe} in comparison with Table \ref{CIS_energies_Kr}. In fact, above the configuration $5s^25p^5(^2P^o_{3/2})7p$, the energy levels computed with DIRAC19 could not be assigned (not shown here). Nevertheless, the degree of convergence reached with the t-aug-dyall.v2z basis is good enough for our purpose. 
 
\section{\label{appendixD}\\Singly-excited state energy levels}

\setcounter{table}{0}
\renewcommand{\thetable}{D\arabic{table}}

\begin{table*}
\centering
\caption{\label{CIS_energies_Kr}Singly-excited state energy levels of krypton for the series $4s^24p^5(^2P^o_{1/2})n'\ell'$ and $4s^24p^5(^2P^o_{3/2})n'\ell'$.}
    \begin{tabular}{r|c|c|c|c|c|c|c|c|c|c|c|c}
  \hline\hline 
\multirow{2}{*}{Configuration}& \multirow{2}{*}{Term}& \multicolumn{2}{c|}{NIST\footnote{Experimental data from Ref. \cite{Saloman07}.}} & \multicolumn{2}{c|}{DIRAC19\footnote{Energy levels computed with the DIRAC19 code (\emph{see} text).}} & \multicolumn{5}{c|}{RCIS\footnote{Energy levels computed with Eq.(\ref{RCIS_equation}).}} &\multirow{2}{*}{$\Sigma_\mathrm{NIST}$\footnote{$\Sigma_\mathrm{NIST}=|1-\mathrm{Level}_\mathrm{RCIS}/\mathrm{Level}_\mathrm{NIST}|\times100$.} }  & \multirow{2}{*}{$\Sigma_\mathrm{DIRAC19}$\footnote{$\Sigma_\mathrm{DIRAC19}=|1-\mathrm{Level}_\mathrm{RCIS}/\mathrm{Level}_\mathrm{DIRAC19}|\times100$.}}  \\
\cline{3-11}
 &         &$J$& Level (eV)&Sym.& Level (eV) & $\ell_{a,\mathrm{max}}$&$j_{a,\mathrm{max}}$&$\ell_{p,\mathrm{max}}$&$j_{p,\mathrm{max}}$ & Level (eV) &   &   \\
\hline      
 $4s^24p^6$                 & $1S$        &0 &  0.00000 &g  & 0.00000 & -  &   -     & -    &  -      & 0.00000 &  0.00000\% & 0.00000\%  \\  
                            &             &  &          &   &         &    &         &      &         &         &            &            \\
 $4s^24p^5(^2P^o_{3/2})5s$  & $^2[3/2]^o$ &2 &  9.91523 &u  &10.01333 & $p$&   3/2   &$s;d$ &  1/2    &10.00562 &  0.91163\% & 0.07706\%  \\
                            &             &1 & 10.03240 &u  &10.18744 & $p$&   3/2   &$s$   &  1/2    &10.18222 &  1.49336\% & 0.05127\%  \\
                            &             &  &          &   &         &    &         &      &         &         &            &            \\
 $4s^24p^5(^2P^o_{1/2})5s$  & $^2[1/2]^o$ &0 & 10.56241 &u  &10.71064 & $p$&   1/2   & $s;d$&  1/2    &10.70653 &  1.36446\% & 0.03839\%  \\
                            &             &1 & 10.64363 &u  &10.85654 & $p$&   1/2   & $s;d$&1/2;3/2  &10.85480 &  1.98400\% & 0.01603\%  \\
                            &             &  &          &   &         &    &         &      &         &         &            &            \\
 $4s^24p^5(^2P^o_{3/2})5p$  & $^2[1/2]$   &1 & 11.30345 &g  &11.30826 & $p$&   3/2   & $p$  &1/2;3/2  &11.27434 &  0.25753\% & 0.30086\%  \\
                            &             &0 & 11.66602 &g  &11.77865 & $p$& 1/2;3/2 & $p$  &1/2;3/2  &11.76734 &  0.86851\% & 0.09611\%  \\
                            &             &  &          &   &         &    &         &      &         &         &            &            \\
 $4s^24p^5(^2P^o_{3/2})5p$  & $^2[5/2]$   &3 & 11.44304 &g  &11.48680 & $p$&   3/2   & $p$  &   3/2   &11.46641 &  0.20423\% & 0.17782\%  \\
                            &             &2 & 11.44465 &g  &11.51301 & $p$&   3/2   & $p$  &   1/2   &11.49764 &  0.46301\% & 0.13368\%  \\
                            &             &  &          &   &         &    &         &      &         &         &            &            \\
 $4s^24p^5(^2P^o_{3/2})5p$  & $^2[3/2]$   &1 & 11.52611 &g  &11.60424 & $p$&   3/2   & $p$  & 1/2;3/2 &11.59248 &  0.57582\% & 0.10145\%  \\
                            &             &2 & 11.54582 &g  &11.62523 & $p$&   3/2   & $p$  &   3/2   &11.61601 &  0.60793\% & 0.07937\%  \\
                            &             &  &          &   &         & $p$&         &      &         &         &            &            \\
 $4s^24p^5(^2P^o_{3/2})4d$  & $^2[1/2]^o$ &0 & 11.99813 &u  &11.97188 & $p$&   3/2   & $s;d$&   3/2   &11.96277 &  0.29471\% & 0.07615\%  \\
                            &             &1 & 12.03702 &u  &12.01820 & $p$&   3/2   & $s;d$&   3/2   &12.00680 &  0.25106\% & 0.09495\%  \\
                            &             &  &          &   &         &    &         &      &         &         &            &            \\
 $4s^24p^5(^2P^o_{1/2})5p$  & $^2[3/2]$   &1 & 12.10035 &g  &12.23126 & $p$&  1/2    &  $p$ &   1/2   &12.21439 &  0.94245\% & 0.13812\%  \\
                            &             &2 & 12.14365 &g  &12.26743 & $p$&  1/2    &  $p$ &   3/2   &12.26442 &  0.99451\% & 0.02454\%  \\
                            &             &  &          &   &         &    &         &      &         &         &            &            \\
 $4s^24p^5(^2P^o_{3/2})4d$  & $^2[3/2]^o$ &2 & 12.11174 &u  &12.10917 & $p$&  3/2    & $s;d$& 3/2;5/2 &12.09343 &  0.15118\% & 0.13015\%  \\
                            &             &1 & 12.35455 &u  &12.50949 & $p$& 1/2;3/2 & $d$  & 3/2;5/2 &12.44189 &  0.70695\% & 0.54333\%  \\
                            &             &  &          &   &         &    &         &      &         &         &            &            \\
 $4s^24p^5(^2P^o_{3/2})4d$  & $^2[7/2]^o$ &4 & 12.12531 &u  &12.15620 & $p$&   3/2   & $d$  &   5/2   &12.13470 &  0.07744\% & 0.17718\%  \\
                            &             &3 & 12.17850 &u  &12.24345 & $p$&   3/2   & $d$  & 3/2;5/2 &12.21649 &  0.31194\% & 0.22069\%  \\
                            &             &  &          &   &         &    &         &      &         &         &            &            \\
 $4s^24p^5(^2P^o_{1/2})5p$  & $^2[1/2]$   &1 & 12.14042 &g  &12.27713 & $p$&   1/2   & $p$  &   3/2   &12.25450 &  0.93967\% & 0.18467\%  \\
                            &             &0 & 12.25646 &g  &12.42570 & $p$& 1/2;3/2 & $p$  & 1/2;3/2 &12.41762 &  1.31490\% & 0.06507\%  \\
                            &             &  &          &   &         &    &         &      &         &         &            &            \\
 $4s^24p^5(^2P^o_{3/2})4d$  & $^2[5/2]^o$ &2 & 12.25799 &u  &12.35396 & $p$&   3/2   & $s;d$& 3/2;5/2 &12.31549 &  0.46908\% & 0.31237\%  \\
                            &             &3 & 12.28427 &u  &12.38822 & $p$&   3/2   & $d$  &   5/2   &12.34991 &  0.53434\% & 0.31020\%  \\
                            &             &  &          &   &         &    &         &      &         &         &            &            \\
 $4s^24p^5(^2P^o_{3/2})6s$  & $^2[3/2]^o$ &2 & 12.35215 &u  &12.39755 & $p$&   3/2   & $s;d$&   3/2   &12.37827 &  0.21146\% & 0.15576\%  \\
                            &             &1 & 12.38528 &u  &12.41790 & $p$&   3/2   & $s;d$&   1/2   &12.40585 &  0.16608\% & 0.09713\%  \\
                            &             &  &          &   &         &    &         &      &         &         &            &            \\
 $4s^24p^5(^2P^o_{3/2})6p$  & $^2[1/2]$   &1 & 12.75638 &g  &12.76813 & $p$&   3/2   & $p$  & 1/2;3/2 &12.75895 &  0.02015\% & 0.07195\%  \\
                            &             &0 & 12.86480 &g  &12.90986 & $p$& 1/2;3/2 & $p$  & 1/2;3/2 &12.90011 &  0.27447\% & 0.07558\%  \\
                            &             &  &          &   &         &    &         &      &         &         &            &            \\
 $4s^24p^5(^2P^o_{3/2})6p$  & $^2[5/2]$   &3 & 12.78470 &g  &12.79665 & $p$&   3/2   & $p$  &   3/2   &12.78856 &  0.03019\% & 0.06326\%  \\
                            &             &2 & 12.78539 &g  &12.80480 & $p$&   3/2   & $p$  &   1/2   &12.79860 &  0.10332\% & 0.04844\%  \\
                            &             &  &          &   &         &    &         &      &         &         &            &            \\
 $4s^24p^5(^2P^o_{1/2})4d$  & $^2[3/2]^o$ &2 & 12.80339 &u  &12.95024 & $p$& 1/2;3/2 & $d$  &   5/2   &12.83762 &  0.26735\% & 0.87727\%  \\
                            &             &1 & 13.00436 &u  &12.95908 & $p$& 1/2;3/2 & $s;d$&  3/2;5/2&12.87001 &  1.03312\% & 0.69207\%  \\
                            &             &  &          &   &         &    &         &      &         &         &            &            \\
 $4s^24p^5(^2P^o_{3/2})6p$  & $^2[3/2]$   &1 & 12.80923 &g  &12.83218 & $p$&   3/2   & $p$  &  1/2;3/2&12.82661 &  0.13568\% & 0.04343\%  \\
                            &             &2 & 12.81533 &g  &12.83778 & $p$&   3/2   & $p$  &  1/2;1/2&12.83260 &  0.13476\% & 0.04037\%  \\  
\hline\hline
\end{tabular}
\end{table*}

\begin{table*}
\centering
\caption{\label{CIS_energies_Xe}Singly-excited state energy levels of xenon for the series $5s^25p^5(^2P^o_{1/2})n'\ell'$ and $5s^25p^5(^2P^o_{3/2})n'\ell'$.}
\begin{tabular}{r|c|c|c|c|c|c|c|c|c|c|c|c|c}
\hline\hline 
\multirow{2}{*}{Configuration}& \multirow{2}{*}{Term}& \multicolumn{2}{c|}{NIST\footnote{Experimental data from Ref. \cite{Saloman07}.}} & \multicolumn{3}{c|}{DIRAC19\footnote{Energy levels computed with the DIRAC19 code (\emph{see} text).}} & \multicolumn{5}{c|}{RCIS\footnote{Energy levels computed with Eq.(\ref{RCIS_equation}).}} &\multirow{2}{*}{$\Sigma_\mathrm{NIST}$\footnote{$\Sigma_\mathrm{NIST}=|1-\mathrm{Level}_\mathrm{RCIS}/\mathrm{Level}_\mathrm{NIST}|\times100$.} }  & \multirow{2}{*}{$\Sigma_\mathrm{DIRAC19}$\footnote{$\Sigma_\mathrm{DIRAC19}=|1-\mathrm{Level}_\mathrm{RCIS}/\mathrm{Level}_\mathrm{DIRAC19}|\times100$.}}  \\
\cline{3-12}
 &         &$J$& Level (eV)&Sym.& Level\footnote{Energy levels computed with Gaunt interaction (\emph{see text}).} (eV)&Level (eV)&$\ell_{a,\mathrm{max}}$ &$j_{a,\mathrm{max}}$ & $\ell_{p,\mathrm{max}}$ & $j_{p,\mathrm{max}}$& Level (eV) &   &   \\
\hline    
$5s^25p^6$                  & $^1S$       & 0  & 0.00000 &g & 0.0000  & 0.00000&-   &    -  & -   & -       &  0.00000&0.00000\%&0.00000\%  \\                                                              
                            &             &    &         &  &         &        &    &       &     &         &         &         &           \\
$5s^25p^5(^2P^o_{3/2})6s$   & $^2[3/2]^o$ & 2  & 8.31531 &u & 8.33595 & 8.29748&$p$ &  3/2  &$s;d$&  1/2    &  8.29633&0.22825\%&0.01386\%  \\  
                            &             & 1  & 8.43652 &u & 8.52396 & 8.48452&$p$ &  3/2  &$s;d$&  1/2    &  8.48382&0.56066\%&0.00825\%  \\
                            &             &    &         &  &         &        &    &       &     &         &         &         &           \\
$5s^25p^5(^2P^o_{1/2})6s$   & $^2[1/2]^o$ & 0  & 9.44719 &u & 9.34829 & 9.31319&$p$ &1/2;3/2&$s;d$&  1/2;3/2&  9.31240&1.42677\%&0.00848\%  \\
                            &             & 1  & 9.56972 &u & 9.53871 & 9.43782&$p$ &  3/2  &$s;d$&  1/2;3/2&  9.43823&1.37402\%&0.00434\%  \\
                            &             &    &         &  &         &        &    &       &     &         &         &         &           \\
$5s^25p^5(^2P^o_{3/2})6p$   & $^2[1/2]$   & 1  & 9.58015 &g & 9.53871 & 9.50225&$p$ &1/2;3/2&$p$  &  1/2;3/2&  9.50187&0.81711\%&0.00400\%  \\
                            &             & 0  & 9.93348 &g & 9.93898 & 9.94571&$p$ &1/2;3/2&$p$  &         &  9.94603&0.12634\%&0.00322\%  \\
                            &             &    &         &  &         &        &    &       &     &         &         &         &           \\
$5s^25p^5(^2P^o_{3/2})6p$   & $^2[5/2]$   & 2  & 9.68562 &g & 9.67043 & 9.62628&$p$ & 3/2   &$p$  &  1/2;3/2&  9.62005&0.67698\%&0.06476\%  \\
                            &             & 3  & 9.72074 &g & 9.66841 & 9.62816&$p$ & 3/2   &$p$  &  3/2    &  9.61578&1.07975\%&0.12875\%  \\
                            &             &    &         &  &         &        &    &       &     &         &         &         &           \\
$5s^25p^5(^2P^o_{3/2})6p$   & $^2[3/2]$   & 1  & 9.78930 &g & 9.77860 & 9.73590&$p$ & 3/2   &$p$  &  1/2;3/2&  9.73353&0.56970\%&0.02435\%  \\
                            &             & 2  & 9.82109 &g & 9.80897 & 9.76607&$p$ & 3/2   &$p$  &   3/2   &  9.76415&0.57977\%&0.01966\%  \\
                            &             &    &         &  &         &        &    &       &     &         &         &         &           \\
$5s^25p^5(^2P^o_{3/2})5d$   & $^2[1/2]^o$ & 0  & 9.89037 &u & 9.98467 & 9.89473&$p$ & 3/2   &$s;d$&   3/2   &  9.89458&0.04257\%&0.00152\%  \\
                            &             & 1  & 9.91707 &u & 10.02071& 9.97563&$p$ &1/2;3/2&$s;d$&   3/2   &  9.97568&0.59100\%&0.00050\%  \\
                            &             &    &         &  &         &        &    &       &     &         &         &         &           \\
$5s^25p^5(^2P^o_{3/2})5d$   & $^2[7/2]^o$ & 4  & 9.94311 &u & 9.86972 & 9.83221&$p$ & 3/2   &$d$  &   5/2   &  9.80440&1.39504\%&0.28365\%  \\
                            &             & 3  &10.03905 &u & 10.01518& 9.98221&$p$ & 3/2   &$d$  &   3/2   &  9.96132&0.77428\%&0.20971\%  \\
                            &             &    &         &  &         &        &    &       &     &         &         &         &           \\
$5s^25p^5(^2P^o_{3/2})5d$   & $^2[3/2]^o$ & 2  & 9.95875 &u & 9.87174 & 9.83455&$p$ & 3/2   &$s;d$&  3/2;5/2&  9.83305&1.26221\%&0.01525\%  \\
                            &             & 1  &10.40103 &u & 10.44642&10.40286&$p$ & 3/2   &$s;d$&  3/2;5/2& 10.37664&0.23450\%&0.25268\%  \\
                            &             &    &         &  &         &        &    &       &     &         &         &         &           \\
$5s^25p^5(^2P^o_{3/2})5d$   & $^2[5/2]^o$ & 2  &10.15746 &u &10.17994 &10.14016&$p$ &3/2    &$d$  &  3/2;5/2& 10.12552&0.31445\%&0.14459\%  \\
                            &             & 3  &10.22004 &u &10.24666 &10.20526&$p$ &3/2    &$d$  &    5/2  & 10.19511&0.24393\%&0.09956\%  \\
                            &             &    &         &  &         &        &    &       &     &         &         &         &           \\
$5s^25p^5(^2P^o_{3/2})7s$   & $^2[3/2]^o$ & 2  &10.56206 &u &10.51817 &10.47475&$p$ &3/2    &$s;d$&   1/2   & 10.44129&1.14343\%&0.32046\%  \\
                            &             & 1  &10.59321 &u &10.57395 &10.53028&$p$ &3/2    &$s;d$&   1/2   & 10.48867&0.98686\%&0.39671\%  \\
                            &             &    &         &  &         &        &    &       &     &         &         &         &           \\
$5s^25p^5(^2P^o_{3/2})7p$   & $^2[1/2]$   & 1  &10.90157 &g &10.79777 &10.75406&$p$ &3/2    &$p$  & 1/2;3/2 & 10.75394&1.35421\%&0.00112\%  \\
                            &             & 0  &11.01503 &g &10.95639 &10.91197&$p$ &3/2    &$p$  & 1/2;3/2 & 10.90575&0.99210\%&0.05703\%  \\
                            &             &    &         &  &         &        &    &       &     &         &         &         &           \\
$5s^25p^5(^2P^o_{3/2})7p$   & $^2[5/2]$   & 2  &10.95421 &g &10.89935 &10.82338&$p$ &3/2    &$p$  &   1/2   & 10.81939&1.23076\%&0.03688\%  \\
                            &             & 3  &10.96878 &g &10.91483 &10.82791&$p$ &3/2    &$p$  &   3/2   & 10.82352&1.32430\%&0.04056\%  \\

\hline\hline
\end{tabular}
\end{table*}

\bibliography{references}

\providecommand{\noopsort}[1]{}\providecommand{\singleletter}[1]{#1}%
\begin{thebibliography}{87}%
\makeatletter
\providecommand \@ifxundefined [1]{%
 \@ifx{#1\undefined}
}%
\providecommand \@ifnum [1]{%
 \ifnum #1\expandafter \@firstoftwo
 \else \expandafter \@secondoftwo
 \fi
}%
\providecommand \@ifx [1]{%
 \ifx #1\expandafter \@firstoftwo
 \else \expandafter \@secondoftwo
 \fi
}%
\providecommand \natexlab [1]{#1}%
\providecommand \enquote  [1]{``#1''}%
\providecommand \bibnamefont  [1]{#1}%
\providecommand \bibfnamefont [1]{#1}%
\providecommand \citenamefont [1]{#1}%
\providecommand \href@noop [0]{\@secondoftwo}%
\providecommand \href [0]{\begingroup \@sanitize@url \@href}%
\providecommand \@href[1]{\@@startlink{#1}\@@href}%
\providecommand \@@href[1]{\endgroup#1\@@endlink}%
\providecommand \@sanitize@url [0]{\catcode `\\12\catcode `\$12\catcode
  `\&12\catcode `\#12\catcode `\^12\catcode `\_12\catcode `\%12\relax}%
\providecommand \@@startlink[1]{}%
\providecommand \@@endlink[0]{}%
\providecommand \url  [0]{\begingroup\@sanitize@url \@url }%
\providecommand \@url [1]{\endgroup\@href {#1}{\urlprefix }}%
\providecommand \urlprefix  [0]{URL }%
\providecommand \Eprint [0]{\href }%
\providecommand \doibase [0]{https://doi.org/}%
\providecommand \selectlanguage [0]{\@gobble}%
\providecommand \bibinfo  [0]{\@secondoftwo}%
\providecommand \bibfield  [0]{\@secondoftwo}%
\providecommand \translation [1]{[#1]}%
\providecommand \BibitemOpen [0]{}%
\providecommand \bibitemStop [0]{}%
\providecommand \bibitemNoStop [0]{.\EOS\space}%
\providecommand \EOS [0]{\spacefactor3000\relax}%
\providecommand \BibitemShut  [1]{\csname bibitem#1\endcsname}%
\let\auto@bib@innerbib\@empty
\bibitem [{\citenamefont {Goulielmakis}\ \emph {et~al.}(2010)\citenamefont
  {Goulielmakis}, \citenamefont {Loh}, \citenamefont {Wirth}, \citenamefont
  {Santra}, \citenamefont {Rohringer}, \citenamefont {Yakovlev}, \citenamefont
  {Zherebtsov}, \citenamefont {Pfeifer}, \citenamefont {Azzeer}, \citenamefont
  {Kling}, \citenamefont {Leone},\ and\ \citenamefont
  {Krausz}}]{Goulielmakis10}%
  \BibitemOpen
  \bibfield  {author} {\bibinfo {author} {\bibfnamefont {E.}~\bibnamefont
  {Goulielmakis}}, \bibinfo {author} {\bibfnamefont {Z.-H.}\ \bibnamefont
  {Loh}}, \bibinfo {author} {\bibfnamefont {A.}~\bibnamefont {Wirth}}, \bibinfo
  {author} {\bibfnamefont {R.}~\bibnamefont {Santra}}, \bibinfo {author}
  {\bibfnamefont {N.}~\bibnamefont {Rohringer}}, \bibinfo {author}
  {\bibfnamefont {V.~S.}\ \bibnamefont {Yakovlev}}, \bibinfo {author}
  {\bibfnamefont {S.}~\bibnamefont {Zherebtsov}}, \bibinfo {author}
  {\bibfnamefont {T.}~\bibnamefont {Pfeifer}}, \bibinfo {author} {\bibfnamefont
  {A.~M.}\ \bibnamefont {Azzeer}}, \bibinfo {author} {\bibfnamefont {M.~F.}\
  \bibnamefont {Kling}}, \bibinfo {author} {\bibfnamefont {S.~R.}\ \bibnamefont
  {Leone}},\ and\ \bibinfo {author} {\bibfnamefont {F.}~\bibnamefont
  {Krausz}},\ }\bibfield  {title} {\bibinfo {title} {Real-time observation of
  valence electron motion},\ }\href {https://doi.org/10.1038/nature09212}
  {\bibfield  {journal} {\bibinfo  {journal} {Nature}\ }\textbf {\bibinfo
  {volume} {466}},\ \bibinfo {pages} {739} (\bibinfo {year}
  {2010})}\BibitemShut {NoStop}%
\bibitem [{\citenamefont {Beck}\ \emph {et~al.}(2015)\citenamefont {Beck},
  \citenamefont {Neumark},\ and\ \citenamefont {Leone}}]{Beck15}%
  \BibitemOpen
  \bibfield  {author} {\bibinfo {author} {\bibfnamefont {A.~R.}\ \bibnamefont
  {Beck}}, \bibinfo {author} {\bibfnamefont {D.~M.}\ \bibnamefont {Neumark}},\
  and\ \bibinfo {author} {\bibfnamefont {S.~R.}\ \bibnamefont {Leone}},\
  }\bibfield  {title} {\bibinfo {title} {Probing ultrafast dynamics with
  attosecond transient absorption},\ }\href
  {https://doi.org/https://doi.org/10.1016/j.cplett.2014.12.048} {\bibfield
  {journal} {\bibinfo  {journal} {Chemical Physics Letters}\ }\textbf {\bibinfo
  {volume} {624}},\ \bibinfo {pages} {119 } (\bibinfo {year}
  {2015})}\BibitemShut {NoStop}%
\bibitem [{\citenamefont {Ott}\ \emph {et~al.}(2014)\citenamefont {Ott},
  \citenamefont {Kaldun}, \citenamefont {Argenti}, \citenamefont {Raith},
  \citenamefont {Meyer}, \citenamefont {Laux}, \citenamefont {Zhang},
  \citenamefont {Bl{\"a}ttermann}, \citenamefont {Hagstotz}, \citenamefont
  {Ding}, \citenamefont {Heck}, \citenamefont {Madro{\~{n}}ero}, \citenamefont
  {Mart{\'i}n},\ and\ \citenamefont {Pfeifer}}]{Ott14}%
  \BibitemOpen
  \bibfield  {author} {\bibinfo {author} {\bibfnamefont {C.}~\bibnamefont
  {Ott}}, \bibinfo {author} {\bibfnamefont {A.}~\bibnamefont {Kaldun}},
  \bibinfo {author} {\bibfnamefont {L.}~\bibnamefont {Argenti}}, \bibinfo
  {author} {\bibfnamefont {P.}~\bibnamefont {Raith}}, \bibinfo {author}
  {\bibfnamefont {K.}~\bibnamefont {Meyer}}, \bibinfo {author} {\bibfnamefont
  {M.}~\bibnamefont {Laux}}, \bibinfo {author} {\bibfnamefont {Y.}~\bibnamefont
  {Zhang}}, \bibinfo {author} {\bibfnamefont {A.}~\bibnamefont
  {Bl{\"a}ttermann}}, \bibinfo {author} {\bibfnamefont {S.}~\bibnamefont
  {Hagstotz}}, \bibinfo {author} {\bibfnamefont {T.}~\bibnamefont {Ding}},
  \bibinfo {author} {\bibfnamefont {R.}~\bibnamefont {Heck}}, \bibinfo {author}
  {\bibfnamefont {J.}~\bibnamefont {Madro{\~{n}}ero}}, \bibinfo {author}
  {\bibfnamefont {F.}~\bibnamefont {Mart{\'i}n}},\ and\ \bibinfo {author}
  {\bibfnamefont {T.}~\bibnamefont {Pfeifer}},\ }\bibfield  {title} {\bibinfo
  {title} {Reconstruction and control of a time-dependent two-electron wave
  packet},\ }\href {https://doi.org/10.1038/nature14026} {\bibfield  {journal}
  {\bibinfo  {journal} {Nature}\ }\textbf {\bibinfo {volume} {516}},\ \bibinfo
  {pages} {374} (\bibinfo {year} {2014})}\BibitemShut {NoStop}%
\bibitem [{\citenamefont {Wirth}\ \emph {et~al.}(2011)\citenamefont {Wirth},
  \citenamefont {Hassan}, \citenamefont {Grguras}, \citenamefont {Gagnon},
  \citenamefont {Moulet}, \citenamefont {Luu}, \citenamefont {Pabst},
  \citenamefont {Santra}, \citenamefont {Alahmed}, \citenamefont {Azzeer},
  \citenamefont {Yakovlev}, \citenamefont {Pervak}, \citenamefont {Krausz},\
  and\ \citenamefont {Goulielmakis}}]{Wirth11}%
  \BibitemOpen
  \bibfield  {author} {\bibinfo {author} {\bibfnamefont {A.}~\bibnamefont
  {Wirth}}, \bibinfo {author} {\bibfnamefont {M.~T.}\ \bibnamefont {Hassan}},
  \bibinfo {author} {\bibfnamefont {I.}~\bibnamefont {Grguras}}, \bibinfo
  {author} {\bibfnamefont {J.}~\bibnamefont {Gagnon}}, \bibinfo {author}
  {\bibfnamefont {A.}~\bibnamefont {Moulet}}, \bibinfo {author} {\bibfnamefont
  {T.~T.}\ \bibnamefont {Luu}}, \bibinfo {author} {\bibfnamefont
  {S.}~\bibnamefont {Pabst}}, \bibinfo {author} {\bibfnamefont
  {R.}~\bibnamefont {Santra}}, \bibinfo {author} {\bibfnamefont {Z.~A.}\
  \bibnamefont {Alahmed}}, \bibinfo {author} {\bibfnamefont {A.~M.}\
  \bibnamefont {Azzeer}}, \bibinfo {author} {\bibfnamefont {V.~S.}\
  \bibnamefont {Yakovlev}}, \bibinfo {author} {\bibfnamefont {V.}~\bibnamefont
  {Pervak}}, \bibinfo {author} {\bibfnamefont {F.}~\bibnamefont {Krausz}},\
  and\ \bibinfo {author} {\bibfnamefont {E.}~\bibnamefont {Goulielmakis}},\
  }\bibfield  {title} {\bibinfo {title} {Synthesized light transients},\ }\href
  {https://doi.org/10.1126/science.1210268} {\bibfield  {journal} {\bibinfo
  {journal} {Science}\ }\textbf {\bibinfo {volume} {334}},\ \bibinfo {pages}
  {195} (\bibinfo {year} {2011})}\BibitemShut {NoStop}%
\bibitem [{\citenamefont {Ott}\ \emph {et~al.}(2013)\citenamefont {Ott},
  \citenamefont {Kaldun}, \citenamefont {Raith}, \citenamefont {Meyer},
  \citenamefont {Laux}, \citenamefont {Evers}, \citenamefont {Keitel},
  \citenamefont {Greene},\ and\ \citenamefont {Pfeifer}}]{Ott13}%
  \BibitemOpen
  \bibfield  {author} {\bibinfo {author} {\bibfnamefont {C.}~\bibnamefont
  {Ott}}, \bibinfo {author} {\bibfnamefont {A.}~\bibnamefont {Kaldun}},
  \bibinfo {author} {\bibfnamefont {P.}~\bibnamefont {Raith}}, \bibinfo
  {author} {\bibfnamefont {K.}~\bibnamefont {Meyer}}, \bibinfo {author}
  {\bibfnamefont {M.}~\bibnamefont {Laux}}, \bibinfo {author} {\bibfnamefont
  {J.}~\bibnamefont {Evers}}, \bibinfo {author} {\bibfnamefont {C.~H.}\
  \bibnamefont {Keitel}}, \bibinfo {author} {\bibfnamefont {C.~H.}\
  \bibnamefont {Greene}},\ and\ \bibinfo {author} {\bibfnamefont
  {T.}~\bibnamefont {Pfeifer}},\ }\bibfield  {title} {\bibinfo {title} {Lorentz
  meets fano in spectral line shapes: A universal phase and its laser
  control},\ }\href {https://doi.org/10.1126/science.1234407} {\bibfield
  {journal} {\bibinfo  {journal} {Science}\ }\textbf {\bibinfo {volume}
  {340}},\ \bibinfo {pages} {716} (\bibinfo {year} {2013})}\BibitemShut
  {NoStop}%
\bibitem [{\citenamefont {Ding}\ \emph {et~al.}(2016)\citenamefont {Ding},
  \citenamefont {Ott}, \citenamefont {Kaldun}, \citenamefont {Bl\"{a}ttermann},
  \citenamefont {Meyer}, \citenamefont {Stooss}, \citenamefont {Rebholz},
  \citenamefont {Birk}, \citenamefont {Hartmann}, \citenamefont {Brown},
  \citenamefont {Hart},\ and\ \citenamefont {Pfeifer}}]{Ding16}%
  \BibitemOpen
  \bibfield  {author} {\bibinfo {author} {\bibfnamefont {T.}~\bibnamefont
  {Ding}}, \bibinfo {author} {\bibfnamefont {C.}~\bibnamefont {Ott}}, \bibinfo
  {author} {\bibfnamefont {A.}~\bibnamefont {Kaldun}}, \bibinfo {author}
  {\bibfnamefont {A.}~\bibnamefont {Bl\"{a}ttermann}}, \bibinfo {author}
  {\bibfnamefont {K.}~\bibnamefont {Meyer}}, \bibinfo {author} {\bibfnamefont
  {V.}~\bibnamefont {Stooss}}, \bibinfo {author} {\bibfnamefont
  {M.}~\bibnamefont {Rebholz}}, \bibinfo {author} {\bibfnamefont
  {P.}~\bibnamefont {Birk}}, \bibinfo {author} {\bibfnamefont {M.}~\bibnamefont
  {Hartmann}}, \bibinfo {author} {\bibfnamefont {A.}~\bibnamefont {Brown}},
  \bibinfo {author} {\bibfnamefont {H.~V.~D.}\ \bibnamefont {Hart}},\ and\
  \bibinfo {author} {\bibfnamefont {T.}~\bibnamefont {Pfeifer}},\ }\bibfield
  {title} {\bibinfo {title} {Time-resolved four-wave-mixing spectroscopy for
  inner-valence transitions},\ }\href {https://doi.org/10.1364/OL.41.000709}
  {\bibfield  {journal} {\bibinfo  {journal} {Opt. Lett.}\ }\textbf {\bibinfo
  {volume} {41}},\ \bibinfo {pages} {709} (\bibinfo {year} {2016})}\BibitemShut
  {NoStop}%
\bibitem [{\citenamefont {Beck}\ \emph {et~al.}(2014)\citenamefont {Beck},
  \citenamefont {Bernhardt}, \citenamefont {Warrick}, \citenamefont {Wu},
  \citenamefont {Chen}, \citenamefont {Gaarde}, \citenamefont {Schafer},
  \citenamefont {Neumark},\ and\ \citenamefont {Leone}}]{Beck14}%
  \BibitemOpen
  \bibfield  {author} {\bibinfo {author} {\bibfnamefont {A.~R.}\ \bibnamefont
  {Beck}}, \bibinfo {author} {\bibfnamefont {B.}~\bibnamefont {Bernhardt}},
  \bibinfo {author} {\bibfnamefont {E.~R.}\ \bibnamefont {Warrick}}, \bibinfo
  {author} {\bibfnamefont {M.}~\bibnamefont {Wu}}, \bibinfo {author}
  {\bibfnamefont {S.}~\bibnamefont {Chen}}, \bibinfo {author} {\bibfnamefont
  {M.~B.}\ \bibnamefont {Gaarde}}, \bibinfo {author} {\bibfnamefont {K.~J.}\
  \bibnamefont {Schafer}}, \bibinfo {author} {\bibfnamefont {D.~M.}\
  \bibnamefont {Neumark}},\ and\ \bibinfo {author} {\bibfnamefont {S.~R.}\
  \bibnamefont {Leone}},\ }\bibfield  {title} {\bibinfo {title} {Attosecond
  transient absorption probing of electronic superpositions of bound states in
  neon: detection of quantum beats},\ }\href
  {https://doi.org/10.1088/1367-2630/16/11/113016} {\bibfield  {journal}
  {\bibinfo  {journal} {New Journal of Physics}\ }\textbf {\bibinfo {volume}
  {16}},\ \bibinfo {pages} {113016} (\bibinfo {year} {2014})}\BibitemShut
  {NoStop}%
\bibitem [{\citenamefont {Kobayashi}\ \emph {et~al.}(2017)\citenamefont
  {Kobayashi}, \citenamefont {Timmers}, \citenamefont {Sabbar}, \citenamefont
  {Leone},\ and\ \citenamefont {Neumark}}]{Kobayashi17}%
  \BibitemOpen
  \bibfield  {author} {\bibinfo {author} {\bibfnamefont {Y.}~\bibnamefont
  {Kobayashi}}, \bibinfo {author} {\bibfnamefont {H.}~\bibnamefont {Timmers}},
  \bibinfo {author} {\bibfnamefont {M.}~\bibnamefont {Sabbar}}, \bibinfo
  {author} {\bibfnamefont {S.~R.}\ \bibnamefont {Leone}},\ and\ \bibinfo
  {author} {\bibfnamefont {D.~M.}\ \bibnamefont {Neumark}},\ }\bibfield
  {title} {\bibinfo {title} {Attosecond transient-absorption dynamics of xenon
  core-excited states in a strong driving field},\ }\href
  {https://doi.org/10.1103/PhysRevA.95.031401} {\bibfield  {journal} {\bibinfo
  {journal} {Phys. Rev. A}\ }\textbf {\bibinfo {volume} {95}},\ \bibinfo
  {pages} {031401} (\bibinfo {year} {2017})}\BibitemShut {NoStop}%
\bibitem [{\citenamefont {Chu}\ and\ \citenamefont {Lin}(2013)}]{Chu13}%
  \BibitemOpen
  \bibfield  {author} {\bibinfo {author} {\bibfnamefont {W.~C.}\ \bibnamefont
  {Chu}}\ and\ \bibinfo {author} {\bibfnamefont {C.~D.}\ \bibnamefont {Lin}},\
  }\bibfield  {title} {\bibinfo {title} {Absorption and emission of single
  attosecond light pulses in an autoionizing gaseous medium dressed by a
  time-delayed control field},\ }\href
  {https://doi.org/10.1103/PhysRevA.87.013415} {\bibfield  {journal} {\bibinfo
  {journal} {Physical Review A}\ }\textbf {\bibinfo {volume} {87}},\ \bibinfo
  {pages} {013415} (\bibinfo {year} {2013})}\BibitemShut {NoStop}%
\bibitem [{\citenamefont {Petersson}\ \emph {et~al.}(2017)\citenamefont
  {Petersson}, \citenamefont {Argenti},\ and\ \citenamefont
  {Martín}}]{Petersson17}%
  \BibitemOpen
  \bibfield  {author} {\bibinfo {author} {\bibfnamefont {C.~L.~M.}\
  \bibnamefont {Petersson}}, \bibinfo {author} {\bibfnamefont {L.}~\bibnamefont
  {Argenti}},\ and\ \bibinfo {author} {\bibfnamefont {F.}~\bibnamefont
  {Martín}},\ }\bibfield  {title} {\bibinfo {title} {Attosecond transient
  absorption spectroscopy of helium above the {N} = 2 ionization threshold},\
  }\href {https://doi.org/10.1103/PhysRevA.96.013403} {\bibfield  {journal}
  {\bibinfo  {journal} {Physical Review A}\ }\textbf {\bibinfo {volume} {96}},\
  \bibinfo {pages} {013403} (\bibinfo {year} {2017})}\BibitemShut {NoStop}%
\bibitem [{\citenamefont {Chew}\ \emph {et~al.}(2018)\citenamefont {Chew},
  \citenamefont {Douguet}, \citenamefont {Cariker}, \citenamefont {Li},
  \citenamefont {Lindroth}, \citenamefont {Ren}, \citenamefont {Yin},
  \citenamefont {Argenti}, \citenamefont {Hill},\ and\ \citenamefont
  {Chang}}]{Chew18}%
  \BibitemOpen
  \bibfield  {author} {\bibinfo {author} {\bibfnamefont {A.}~\bibnamefont
  {Chew}}, \bibinfo {author} {\bibfnamefont {N.}~\bibnamefont {Douguet}},
  \bibinfo {author} {\bibfnamefont {C.}~\bibnamefont {Cariker}}, \bibinfo
  {author} {\bibfnamefont {J.}~\bibnamefont {Li}}, \bibinfo {author}
  {\bibfnamefont {E.}~\bibnamefont {Lindroth}}, \bibinfo {author}
  {\bibfnamefont {X.}~\bibnamefont {Ren}}, \bibinfo {author} {\bibfnamefont
  {Y.}~\bibnamefont {Yin}}, \bibinfo {author} {\bibfnamefont {L.}~\bibnamefont
  {Argenti}}, \bibinfo {author} {\bibfnamefont {W.~T.}\ \bibnamefont {Hill}},\
  and\ \bibinfo {author} {\bibfnamefont {Z.}~\bibnamefont {Chang}},\ }\bibfield
   {title} {\bibinfo {title} {Attosecond transient absorption spectrum of argon
  at the {L} 2 , 3 edge},\ }\href {https://doi.org/10.1103/PhysRevA.97.031407}
  {\bibfield  {journal} {\bibinfo  {journal} {Physical Review A}\ }\textbf
  {\bibinfo {volume} {97}},\ \bibinfo {pages} {031407} (\bibinfo {year}
  {2018})}\BibitemShut {NoStop}%
\bibitem [{\citenamefont {Wu}\ \emph {et~al.}(2016)\citenamefont {Wu},
  \citenamefont {Chen}, \citenamefont {Camp}, \citenamefont {Schafer},\ and\
  \citenamefont {Gaarde}}]{Wu16}%
  \BibitemOpen
  \bibfield  {author} {\bibinfo {author} {\bibfnamefont {M.}~\bibnamefont
  {Wu}}, \bibinfo {author} {\bibfnamefont {S.}~\bibnamefont {Chen}}, \bibinfo
  {author} {\bibfnamefont {S.}~\bibnamefont {Camp}}, \bibinfo {author}
  {\bibfnamefont {K.~J.}\ \bibnamefont {Schafer}},\ and\ \bibinfo {author}
  {\bibfnamefont {M.~B.}\ \bibnamefont {Gaarde}},\ }\bibfield  {title}
  {\bibinfo {title} {Theory of strong-field attosecond transient absorption},\
  }\href {https://doi.org/10.1088/0953-4075/49/6/062003} {\bibfield  {journal}
  {\bibinfo  {journal} {Journal of Physics B: Atomic, Molecular and Optical
  Physics}\ }\textbf {\bibinfo {volume} {49}},\ \bibinfo {pages} {062003}
  (\bibinfo {year} {2016})}\BibitemShut {NoStop}%
\bibitem [{\citenamefont {Baggesen}\ \emph {et~al.}(2012)\citenamefont
  {Baggesen}, \citenamefont {Lindroth},\ and\ \citenamefont
  {Madsen}}]{Baggesen12}%
  \BibitemOpen
  \bibfield  {author} {\bibinfo {author} {\bibfnamefont {J.~C.}\ \bibnamefont
  {Baggesen}}, \bibinfo {author} {\bibfnamefont {E.}~\bibnamefont {Lindroth}},\
  and\ \bibinfo {author} {\bibfnamefont {L.~B.}\ \bibnamefont {Madsen}},\
  }\bibfield  {title} {\bibinfo {title} {Theory of attosecond absorption
  spectroscopy in krypton},\ }\href
  {https://doi.org/10.1103/PhysRevA.85.013415} {\bibfield  {journal} {\bibinfo
  {journal} {Physical Review A}\ }\textbf {\bibinfo {volume} {85}},\ \bibinfo
  {pages} {013415} (\bibinfo {year} {2012})}\BibitemShut {NoStop}%
\bibitem [{\citenamefont {Kolbasova}\ \emph {et~al.}(2021)\citenamefont
  {Kolbasova}, \citenamefont {Hartmann}, \citenamefont {Jin}, \citenamefont
  {Bl\"attermann}, \citenamefont {Ott}, \citenamefont {Son}, \citenamefont
  {Pfeifer},\ and\ \citenamefont {Santra}}]{Kolbasova21}%
  \BibitemOpen
  \bibfield  {author} {\bibinfo {author} {\bibfnamefont {D.}~\bibnamefont
  {Kolbasova}}, \bibinfo {author} {\bibfnamefont {M.}~\bibnamefont {Hartmann}},
  \bibinfo {author} {\bibfnamefont {R.}~\bibnamefont {Jin}}, \bibinfo {author}
  {\bibfnamefont {A.}~\bibnamefont {Bl\"attermann}}, \bibinfo {author}
  {\bibfnamefont {C.}~\bibnamefont {Ott}}, \bibinfo {author} {\bibfnamefont
  {S.-K.}\ \bibnamefont {Son}}, \bibinfo {author} {\bibfnamefont
  {T.}~\bibnamefont {Pfeifer}},\ and\ \bibinfo {author} {\bibfnamefont
  {R.}~\bibnamefont {Santra}},\ }\bibfield  {title} {\bibinfo {title} {Probing
  ultrafast coherent dynamics in core-excited xenon by using attosecond xuv-nir
  transient absorption spectroscopy},\ }\href
  {https://doi.org/10.1103/PhysRevA.103.043102} {\bibfield  {journal} {\bibinfo
   {journal} {Phys. Rev. A}\ }\textbf {\bibinfo {volume} {103}},\ \bibinfo
  {pages} {043102} (\bibinfo {year} {2021})}\BibitemShut {NoStop}%
\bibitem [{\citenamefont {Pabst}\ \emph {et~al.}(2012)\citenamefont {Pabst},
  \citenamefont {Sytcheva}, \citenamefont {Moulet}, \citenamefont {Wirth},
  \citenamefont {Goulielmakis},\ and\ \citenamefont {Santra}}]{Pabst12}%
  \BibitemOpen
  \bibfield  {author} {\bibinfo {author} {\bibfnamefont {S.}~\bibnamefont
  {Pabst}}, \bibinfo {author} {\bibfnamefont {A.}~\bibnamefont {Sytcheva}},
  \bibinfo {author} {\bibfnamefont {A.}~\bibnamefont {Moulet}}, \bibinfo
  {author} {\bibfnamefont {A.}~\bibnamefont {Wirth}}, \bibinfo {author}
  {\bibfnamefont {E.}~\bibnamefont {Goulielmakis}},\ and\ \bibinfo {author}
  {\bibfnamefont {R.}~\bibnamefont {Santra}},\ }\bibfield  {title} {\bibinfo
  {title} {Theory of attosecond transient-absorption spectroscopy of krypton
  for overlapping pump and probe pulses},\ }\href
  {https://doi.org/10.1103/PhysRevA.86.063411} {\bibfield  {journal} {\bibinfo
  {journal} {Phys. Rev. A}\ }\textbf {\bibinfo {volume} {86}},\ \bibinfo
  {pages} {063411} (\bibinfo {year} {2012})}\BibitemShut {NoStop}%
\bibitem [{\citenamefont {Wragg}\ \emph {et~al.}(2020)\citenamefont {Wragg},
  \citenamefont {Ballance},\ and\ \citenamefont {{van der Hart}}}]{Wragg20}%
  \BibitemOpen
  \bibfield  {author} {\bibinfo {author} {\bibfnamefont {J.}~\bibnamefont
  {Wragg}}, \bibinfo {author} {\bibfnamefont {C.}~\bibnamefont {Ballance}},\
  and\ \bibinfo {author} {\bibfnamefont {H.}~\bibnamefont {{van der Hart}}},\
  }\bibfield  {title} {\bibinfo {title} {Breit–pauli r-matrix approach for
  the time-dependent investigation of ultrafast processes},\ }\href
  {https://doi.org/https://doi.org/10.1016/j.cpc.2020.107274} {\bibfield
  {journal} {\bibinfo  {journal} {Computer Physics Communications}\ }\textbf
  {\bibinfo {volume} {254}},\ \bibinfo {pages} {107274} (\bibinfo {year}
  {2020})}\BibitemShut {NoStop}%
\bibitem [{\citenamefont {Brown}\ \emph {et~al.}(2020)\citenamefont {Brown},
  \citenamefont {Armstrong}, \citenamefont {Benda}, \citenamefont {Clarke},
  \citenamefont {Wragg}, \citenamefont {Hamilton}, \citenamefont {Mašín},
  \citenamefont {Gorfinkiel},\ and\ \citenamefont {{van der Hart}}}]{Brown20}%
  \BibitemOpen
  \bibfield  {author} {\bibinfo {author} {\bibfnamefont {A.~C.}\ \bibnamefont
  {Brown}}, \bibinfo {author} {\bibfnamefont {G.~S.}\ \bibnamefont
  {Armstrong}}, \bibinfo {author} {\bibfnamefont {J.}~\bibnamefont {Benda}},
  \bibinfo {author} {\bibfnamefont {D.~D.}\ \bibnamefont {Clarke}}, \bibinfo
  {author} {\bibfnamefont {J.}~\bibnamefont {Wragg}}, \bibinfo {author}
  {\bibfnamefont {K.~R.}\ \bibnamefont {Hamilton}}, \bibinfo {author}
  {\bibfnamefont {Z.}~\bibnamefont {Mašín}}, \bibinfo {author} {\bibfnamefont
  {J.~D.}\ \bibnamefont {Gorfinkiel}},\ and\ \bibinfo {author} {\bibfnamefont
  {H.~W.}\ \bibnamefont {{van der Hart}}},\ }\bibfield  {title} {\bibinfo
  {title} {Rmt: R-matrix with time-dependence. solving the semi-relativistic,
  time-dependent schrödinger equation for general, multielectron atoms and
  molecules in intense, ultrashort, arbitrarily polarized laser pulses},\
  }\href {https://doi.org/https://doi.org/10.1016/j.cpc.2019.107062} {\bibfield
   {journal} {\bibinfo  {journal} {Computer Physics Communications}\ }\textbf
  {\bibinfo {volume} {250}},\ \bibinfo {pages} {107062} (\bibinfo {year}
  {2020})}\BibitemShut {NoStop}%
\bibitem [{\citenamefont {Wragg}\ \emph {et~al.}(2019)\citenamefont {Wragg},
  \citenamefont {Clarke}, \citenamefont {Armstrong}, \citenamefont {Brown},
  \citenamefont {Ballance},\ and\ \citenamefont {van~der Hart}}]{Wragg19}%
  \BibitemOpen
  \bibfield  {author} {\bibinfo {author} {\bibfnamefont {J.}~\bibnamefont
  {Wragg}}, \bibinfo {author} {\bibfnamefont {D.~D.~A.}\ \bibnamefont
  {Clarke}}, \bibinfo {author} {\bibfnamefont {G.~S.~J.}\ \bibnamefont
  {Armstrong}}, \bibinfo {author} {\bibfnamefont {A.~C.}\ \bibnamefont
  {Brown}}, \bibinfo {author} {\bibfnamefont {C.~P.}\ \bibnamefont
  {Ballance}},\ and\ \bibinfo {author} {\bibfnamefont {H.~W.}\ \bibnamefont
  {van~der Hart}},\ }\bibfield  {title} {\bibinfo {title} {Resolving ultrafast
  spin-orbit dynamics in heavy many-electron atoms},\ }\href
  {https://doi.org/10.1103/PhysRevLett.123.163001} {\bibfield  {journal}
  {\bibinfo  {journal} {Phys. Rev. Lett.}\ }\textbf {\bibinfo {volume} {123}},\
  \bibinfo {pages} {163001} (\bibinfo {year} {2019})}\BibitemShut {NoStop}%
\bibitem [{\citenamefont {Zapata}\ \emph {et~al.}(2021)\citenamefont {Zapata},
  \citenamefont {Vinbladh}, \citenamefont {Lindroth},\ and\ \citenamefont
  {Dahlström}}]{Zapata21}%
  \BibitemOpen
  \bibfield  {author} {\bibinfo {author} {\bibfnamefont {F.}~\bibnamefont
  {Zapata}}, \bibinfo {author} {\bibfnamefont {J.}~\bibnamefont {Vinbladh}},
  \bibinfo {author} {\bibfnamefont {E.}~\bibnamefont {Lindroth}},\ and\
  \bibinfo {author} {\bibfnamefont {J.~M.}\ \bibnamefont {Dahlström}},\
  }\bibfield  {title} {\bibinfo {title} {Implementation and validation of the
  relativistic transient absorption theory within the dipole approximation},\
  }\href {https://doi.org/10.1088/2516-1075/abe191} {\bibfield  {journal}
  {\bibinfo  {journal} {Electronic Structure}\ }\textbf {\bibinfo {volume}
  {3}},\ \bibinfo {pages} {014002} (\bibinfo {year} {2021})}\BibitemShut
  {NoStop}%
\bibitem [{\citenamefont {Rohringer}\ \emph {et~al.}(2006)\citenamefont
  {Rohringer}, \citenamefont {Gordon},\ and\ \citenamefont
  {Santra}}]{Rohringer06}%
  \BibitemOpen
  \bibfield  {author} {\bibinfo {author} {\bibfnamefont {N.}~\bibnamefont
  {Rohringer}}, \bibinfo {author} {\bibfnamefont {A.}~\bibnamefont {Gordon}},\
  and\ \bibinfo {author} {\bibfnamefont {R.}~\bibnamefont {Santra}},\
  }\bibfield  {title} {\bibinfo {title} {Configuration-interaction-based
  time-dependent orbital approach for ab initio treatment of electronic
  dynamics in a strong optical laser field},\ }\href
  {https://doi.org/10.1103/PhysRevA.74.043420} {\bibfield  {journal} {\bibinfo
  {journal} {Phys. Rev. A}\ }\textbf {\bibinfo {volume} {74}},\ \bibinfo
  {pages} {043420} (\bibinfo {year} {2006})}\BibitemShut {NoStop}%
\bibitem [{\citenamefont {Greenman}\ \emph {et~al.}(2010)\citenamefont
  {Greenman}, \citenamefont {Ho}, \citenamefont {Pabst}, \citenamefont
  {Kamarchik}, \citenamefont {Mazziotti},\ and\ \citenamefont
  {Santra}}]{Greenman10}%
  \BibitemOpen
  \bibfield  {author} {\bibinfo {author} {\bibfnamefont {L.}~\bibnamefont
  {Greenman}}, \bibinfo {author} {\bibfnamefont {P.~J.}\ \bibnamefont {Ho}},
  \bibinfo {author} {\bibfnamefont {S.}~\bibnamefont {Pabst}}, \bibinfo
  {author} {\bibfnamefont {E.}~\bibnamefont {Kamarchik}}, \bibinfo {author}
  {\bibfnamefont {D.~A.}\ \bibnamefont {Mazziotti}},\ and\ \bibinfo {author}
  {\bibfnamefont {R.}~\bibnamefont {Santra}},\ }\bibfield  {title} {\bibinfo
  {title} {Implementation of the time-dependent configuration-interaction
  singles method for atomic strong-field processes},\ }\href
  {https://doi.org/10.1103/PhysRevA.82.023406} {\bibfield  {journal} {\bibinfo
  {journal} {Physical Review A}\ }\textbf {\bibinfo {volume} {82}},\ \bibinfo
  {pages} {023406} (\bibinfo {year} {2010})}\BibitemShut {NoStop}%
\bibitem [{\citenamefont {You}\ \emph {et~al.}(2016)\citenamefont {You},
  \citenamefont {Rohringer},\ and\ \citenamefont {Dahlstr\"om}}]{You2016}%
  \BibitemOpen
  \bibfield  {author} {\bibinfo {author} {\bibfnamefont {J.-A.}\ \bibnamefont
  {You}}, \bibinfo {author} {\bibfnamefont {N.}~\bibnamefont {Rohringer}},\
  and\ \bibinfo {author} {\bibfnamefont {J.~M.}\ \bibnamefont {Dahlstr\"om}},\
  }\bibfield  {title} {\bibinfo {title} {Attosecond photoionization dynamics
  with stimulated core-valence transitions},\ }\href
  {https://doi.org/10.1103/PhysRevA.93.033413} {\bibfield  {journal} {\bibinfo
  {journal} {Phys. Rev. A}\ }\textbf {\bibinfo {volume} {93}},\ \bibinfo
  {pages} {033413} (\bibinfo {year} {2016})}\BibitemShut {NoStop}%
\bibitem [{\citenamefont {Simonsen}\ \emph {et~al.}(2016)\citenamefont
  {Simonsen}, \citenamefont {Kjellsson}, \citenamefont {F\o{}rre},
  \citenamefont {Lindroth},\ and\ \citenamefont {Selst\o{}}}]{Aleksander16}%
  \BibitemOpen
  \bibfield  {author} {\bibinfo {author} {\bibfnamefont {A.~S.}\ \bibnamefont
  {Simonsen}}, \bibinfo {author} {\bibfnamefont {T.}~\bibnamefont {Kjellsson}},
  \bibinfo {author} {\bibfnamefont {M.}~\bibnamefont {F\o{}rre}}, \bibinfo
  {author} {\bibfnamefont {E.}~\bibnamefont {Lindroth}},\ and\ \bibinfo
  {author} {\bibfnamefont {S.}~\bibnamefont {Selst\o{}}},\ }\bibfield  {title}
  {\bibinfo {title} {Ionization dynamics beyond the dipole approximation
  induced by the pulse envelope},\ }\href
  {https://doi.org/10.1103/PhysRevA.93.053411} {\bibfield  {journal} {\bibinfo
  {journal} {Phys. Rev. A}\ }\textbf {\bibinfo {volume} {93}},\ \bibinfo
  {pages} {053411} (\bibinfo {year} {2016})}\BibitemShut {NoStop}%
\bibitem [{\citenamefont {Kjellsson}\ \emph {et~al.}(2017)\citenamefont
  {Kjellsson}, \citenamefont {Selstø},\ and\ \citenamefont
  {Lindroth}}]{Kjellsson17}%
  \BibitemOpen
  \bibfield  {author} {\bibinfo {author} {\bibfnamefont {T.}~\bibnamefont
  {Kjellsson}}, \bibinfo {author} {\bibfnamefont {S.}~\bibnamefont {Selstø}},\
  and\ \bibinfo {author} {\bibfnamefont {E.}~\bibnamefont {Lindroth}},\
  }\bibfield  {title} {\bibinfo {title} {Relativistic ionization dynamics for a
  hydrogen atom exposed to superintense {XUV} laser pulses},\ }\href
  {https://doi.org/10.1103/PhysRevA.95.043403} {\bibfield  {journal} {\bibinfo
  {journal} {Physical Review A}\ }\textbf {\bibinfo {volume} {95}},\ \bibinfo
  {pages} {043403} (\bibinfo {year} {2017})}\BibitemShut {NoStop}%
\bibitem [{\citenamefont {Greiner}(2000)}]{Greiner00}%
  \BibitemOpen
  \bibfield  {author} {\bibinfo {author} {\bibfnamefont {W.}~\bibnamefont
  {Greiner}},\ }\href {https://www.springer.com/gp/book/9783540674573} {\emph
  {\bibinfo {title} {Relativistic Quantum Mechanics. Wave Equations.}}}\
  (\bibinfo  {publisher} {Springer-Verlag Berlin Heidelberg},\ \bibinfo
  {address} {Berlin},\ \bibinfo {year} {2000})\BibitemShut {NoStop}%
\bibitem [{\citenamefont {Grant}(2006)}]{Grant06}%
  \BibitemOpen
  \bibfield  {author} {\bibinfo {author} {\bibfnamefont {I.~P.}\ \bibnamefont
  {Grant}},\ }\href@noop {} {\emph {\bibinfo {title} {Relativistic Quantum
  Theory of Atoms and Molecules: Theory and Computation (Springer Series on
  Atomic, Optical, and Plasma Physics)}}}\ (\bibinfo  {publisher}
  {Springer-Verlag},\ \bibinfo {address} {Berlin, Heidelberg},\ \bibinfo {year}
  {2006})\BibitemShut {NoStop}%
\bibitem [{\citenamefont {Krebs}\ \emph {et~al.}(2014)\citenamefont {Krebs},
  \citenamefont {Pabst},\ and\ \citenamefont {Santra}}]{Krebs14}%
  \BibitemOpen
  \bibfield  {author} {\bibinfo {author} {\bibfnamefont {D.}~\bibnamefont
  {Krebs}}, \bibinfo {author} {\bibfnamefont {S.}~\bibnamefont {Pabst}},\ and\
  \bibinfo {author} {\bibfnamefont {R.}~\bibnamefont {Santra}},\ }\bibfield
  {title} {\bibinfo {title} {Introducing many-body physics using atomic
  spectroscopy},\ }\href {https://doi.org/10.1119/1.4827015} {\bibfield
  {journal} {\bibinfo  {journal} {American Journal of Physics}\ }\textbf
  {\bibinfo {volume} {82}},\ \bibinfo {pages} {113} (\bibinfo {year}
  {2014})}\BibitemShut {NoStop}%
\bibitem [{\citenamefont {Saloman}(2004)}]{Saloman04}%
  \BibitemOpen
  \bibfield  {author} {\bibinfo {author} {\bibfnamefont {E.~B.}\ \bibnamefont
  {Saloman}},\ }\bibfield  {title} {\bibinfo {title} {Energy levels and
  observed spectral lines of xenon, xe i through xe liv},\ }\href
  {https://doi.org/10.1063/1.1649348} {\bibfield  {journal} {\bibinfo
  {journal} {Journal of Physical and Chemical Reference Data}\ }\textbf
  {\bibinfo {volume} {33}},\ \bibinfo {pages} {765} (\bibinfo {year}
  {2004})}\BibitemShut {NoStop}%
\bibitem [{\citenamefont {Saloman}(2007)}]{Saloman07}%
  \BibitemOpen
  \bibfield  {author} {\bibinfo {author} {\bibfnamefont {E.~B.}\ \bibnamefont
  {Saloman}},\ }\bibfield  {title} {\bibinfo {title} {Energy levels and
  observed spectral lines of krypton, kr i through kr xxxvi},\ }\href
  {https://doi.org/10.1063/1.2227036} {\bibfield  {journal} {\bibinfo
  {journal} {Journal of Physical and Chemical Reference Data}\ }\textbf
  {\bibinfo {volume} {36}},\ \bibinfo {pages} {215} (\bibinfo {year}
  {2007})}\BibitemShut {NoStop}%
\bibitem [{NIS()}]{NIST:database}%
  \BibitemOpen
  \href@noop {} {}\bibinfo {howpublished} {NIST {\it Atomic Spectra Database}.
  (Available at https://www.nist.gov/pml/atomic-spectra-database)}\BibitemShut
  {NoStop}%
\bibitem [{DIR()}]{DIRAC19}%
  \BibitemOpen
  \href@noop {} {}\bibinfo {note} {{DIRAC}, a relativistic ab initio electronic
  structure program, Release {DIRAC19} (2019), written by A.~S.~P.~Gomes,
  T.~Saue, L.~Visscher, H.~J.~{\relax Aa}.~Jensen, and R.~Bast, with
  contributions from I.~A.~Aucar, V.~Bakken, K.~G.~Dyall, S.~Dubillard,
  U.~Ekstr{\"o}m, E.~Eliav, T.~Enevoldsen, E.~Fa{\ss}hauer, T.~Fleig,
  O.~Fossgaard, L.~Halbert, E.~D.~Hedeg{\aa}rd, B.~Heimlich--Paris,
  T.~Helgaker, J.~Henriksson, M.~Ilia{\v{s}}, Ch.~R.~Jacob, S.~Knecht,
  S.~Komorovsk{\'y}, O.~Kullie, J.~K.~L{\ae}rdahl, C.~V.~Larsen, Y.~S.~Lee,
  H.~S.~Nataraj, M.~K.~Nayak, P.~Norman, G.~Olejniczak, J.~Olsen,
  J.~M.~H.~Olsen, Y.~C.~Park, J.~K.~Pedersen, M.~Pernpointner, R.~di~Remigio,
  K.~Ruud, P.~Sa{\l}ek, B.~Schimmelpfennig, B.~Senjean, A.~Shee, J.~Sikkema,
  A.~J.~Thorvaldsen, J.~Thyssen, J.~van~Stralen, M.~L.~Vidal, S.~Villaume,
  O.~Visser, T.~Winther, and S.~Yamamoto (available at
  \url{http://dx.doi.org/10.5281/zenodo.3572669}, see also
  \url{http://www.diracprogram.org})}\BibitemShut {NoStop}%
\bibitem [{\citenamefont {Samson}\ and\ \citenamefont
  {Stolte}(2002)}]{Samson02}%
  \BibitemOpen
  \bibfield  {author} {\bibinfo {author} {\bibfnamefont {J.}~\bibnamefont
  {Samson}}\ and\ \bibinfo {author} {\bibfnamefont {W.}~\bibnamefont
  {Stolte}},\ }\bibfield  {title} {\bibinfo {title} {Precision measurements of
  the total photoionization cross-sections of he, ne, ar, kr, and xe},\ }\href
  {https://doi.org/https://doi.org/10.1016/S0368-2048(02)00026-9} {\bibfield
  {journal} {\bibinfo  {journal} {Journal of Electron Spectroscopy and Related
  Phenomena}\ }\textbf {\bibinfo {volume} {123}},\ \bibinfo {pages} {265}
  (\bibinfo {year} {2002})},\ \bibinfo {note} {determination of cross-sections
  and momentum profiles of atoms, molecules and condensed matter}\BibitemShut
  {NoStop}%
\bibitem [{\citenamefont {Shannon}\ \emph {et~al.}(1977)\citenamefont
  {Shannon}, \citenamefont {Codling},\ and\ \citenamefont {West}}]{Shannon77}%
  \BibitemOpen
  \bibfield  {author} {\bibinfo {author} {\bibfnamefont {S.~P.}\ \bibnamefont
  {Shannon}}, \bibinfo {author} {\bibfnamefont {K.}~\bibnamefont {Codling}},\
  and\ \bibinfo {author} {\bibfnamefont {J.~B.}\ \bibnamefont {West}},\
  }\bibfield  {title} {\bibinfo {title} {The absolute photoionization cross
  sections of the spin-orbit components of the xenon 4d electron from 70-130
  {eV}},\ }\href {https://doi.org/10.1088/0022-3700/10/5/018} {\bibfield
  {journal} {\bibinfo  {journal} {Journal of Physics B: Atomic and Molecular
  Physics}\ }\textbf {\bibinfo {volume} {10}},\ \bibinfo {pages} {825}
  (\bibinfo {year} {1977})}\BibitemShut {NoStop}%
\bibitem [{\citenamefont {Becker}\ \emph {et~al.}(1989)\citenamefont {Becker},
  \citenamefont {Szostak}, \citenamefont {Kerkhoff}, \citenamefont {Kupsch},
  \citenamefont {Langer}, \citenamefont {Wehlitz}, \citenamefont {Yagishita},\
  and\ \citenamefont {Hayaishi}}]{Becker89}%
  \BibitemOpen
  \bibfield  {author} {\bibinfo {author} {\bibfnamefont {U.}~\bibnamefont
  {Becker}}, \bibinfo {author} {\bibfnamefont {D.}~\bibnamefont {Szostak}},
  \bibinfo {author} {\bibfnamefont {H.~G.}\ \bibnamefont {Kerkhoff}}, \bibinfo
  {author} {\bibfnamefont {M.}~\bibnamefont {Kupsch}}, \bibinfo {author}
  {\bibfnamefont {B.}~\bibnamefont {Langer}}, \bibinfo {author} {\bibfnamefont
  {R.}~\bibnamefont {Wehlitz}}, \bibinfo {author} {\bibfnamefont
  {A.}~\bibnamefont {Yagishita}},\ and\ \bibinfo {author} {\bibfnamefont
  {T.}~\bibnamefont {Hayaishi}},\ }\bibfield  {title} {\bibinfo {title}
  {Subshell photoionization of xe between 40 and 1000 ev},\ }\href
  {https://doi.org/10.1103/PhysRevA.39.3902} {\bibfield  {journal} {\bibinfo
  {journal} {Phys. Rev. A}\ }\textbf {\bibinfo {volume} {39}},\ \bibinfo
  {pages} {3902} (\bibinfo {year} {1989})}\BibitemShut {NoStop}%
\bibitem [{\citenamefont {Kammerling}\ \emph {et~al.}(1989)\citenamefont
  {Kammerling}, \citenamefont {Kossman},\ and\ \citenamefont
  {Schmidt}}]{Kammerling89}%
  \BibitemOpen
  \bibfield  {author} {\bibinfo {author} {\bibfnamefont {B.}~\bibnamefont
  {Kammerling}}, \bibinfo {author} {\bibfnamefont {H.}~\bibnamefont
  {Kossman}},\ and\ \bibinfo {author} {\bibfnamefont {V.}~\bibnamefont
  {Schmidt}},\ }\bibfield  {title} {\bibinfo {title} {4d photoionisation in
  xenon: absolute partial cross section and relative strength of 4d
  many-electron processes},\ }\href
  {https://doi.org/10.1088/0953-4075/22/6/010} {\bibfield  {journal} {\bibinfo
  {journal} {Journal of Physics B: Atomic, Molecular and Optical Physics}\
  }\textbf {\bibinfo {volume} {22}},\ \bibinfo {pages} {841} (\bibinfo {year}
  {1989})}\BibitemShut {NoStop}%
\bibitem [{\citenamefont {Grant}(1970)}]{Grant70}%
  \BibitemOpen
  \bibfield  {author} {\bibinfo {author} {\bibfnamefont {I.}~\bibnamefont
  {Grant}},\ }\bibfield  {title} {\bibinfo {title} {Relativistic calculation of
  atomic structures},\ }\href {https://doi.org/10.1080/00018737000101191}
  {\bibfield  {journal} {\bibinfo  {journal} {Advances in Physics}\ }\textbf
  {\bibinfo {volume} {19}},\ \bibinfo {pages} {747} (\bibinfo {year}
  {1970})}\BibitemShut {NoStop}%
\bibitem [{\citenamefont {Szabo}\ and\ \citenamefont
  {Ostlund}(1996)}]{Szabo96}%
  \BibitemOpen
  \bibfield  {author} {\bibinfo {author} {\bibfnamefont {A.}~\bibnamefont
  {Szabo}}\ and\ \bibinfo {author} {\bibfnamefont {N.}~\bibnamefont
  {Ostlund}},\ }\href {https://books.google.se/books?id=6mV9gYzEkgIC} {\emph
  {\bibinfo {title} {Modern Quantum Chemistry: Introduction to Advanced
  Electronic Structure Theory}}},\ Dover Books on Chemistry\ (\bibinfo
  {publisher} {Dover Publications},\ \bibinfo {year} {1996})\BibitemShut
  {NoStop}%
\bibitem [{\citenamefont {Sakurai}\ and\ \citenamefont
  {Napolitano}(2020)}]{Sakurai2020}%
  \BibitemOpen
  \bibfield  {author} {\bibinfo {author} {\bibfnamefont {J.~J.}\ \bibnamefont
  {Sakurai}}\ and\ \bibinfo {author} {\bibfnamefont {J.}~\bibnamefont
  {Napolitano}},\ }\href@noop {} {\emph {\bibinfo {title} {Modern Quantum
  Mechanics}}},\ \bibinfo {edition} {3rd}\ ed.\ (\bibinfo  {publisher}
  {Cambridge University Press},\ \bibinfo {year} {2020})\BibitemShut {NoStop}%
\bibitem [{\citenamefont {Grant}(2009)}]{Grant09}%
  \BibitemOpen
  \bibfield  {author} {\bibinfo {author} {\bibfnamefont {I.~P.}\ \bibnamefont
  {Grant}},\ }\bibfield  {title} {\bibinfo {title} {B-spline methods for radial
  dirac equations},\ }\href {https://doi.org/10.1088/0953-4075/42/5/055002}
  {\bibfield  {journal} {\bibinfo  {journal} {Journal of Physics B: Atomic,
  Molecular and Optical Physics}\ }\textbf {\bibinfo {volume} {42}},\ \bibinfo
  {pages} {055002} (\bibinfo {year} {2009})}\BibitemShut {NoStop}%
\bibitem [{\citenamefont {Lindgren}\ and\ \citenamefont
  {Rosen}(1974)}]{Lindgren74}%
  \BibitemOpen
  \bibfield  {author} {\bibinfo {author} {\bibfnamefont {I.}~\bibnamefont
  {Lindgren}}\ and\ \bibinfo {author} {\bibfnamefont {A.}~\bibnamefont
  {Rosen}},\ }\bibfield  {title} {\bibinfo {title} {Relativistic
  self-consistent-field calculations with application to atomic hyperfine
  interaction. ii. relativistic theory of atomic hyperfine interaction},\
  }\bibfield  {journal} {\bibinfo  {journal} {Case Stud. At. Phys., v. 4, no.
  3, pp. 150-196}\ }\href {https://www.osti.gov/biblio/4210365} {} (\bibinfo
  {year} {1974})\BibitemShut {NoStop}%
\bibitem [{\citenamefont {Ackad}\ and\ \citenamefont
  {Horbatsch}(2007{\natexlab{a}})}]{Ackad07}%
  \BibitemOpen
  \bibfield  {author} {\bibinfo {author} {\bibfnamefont {E.}~\bibnamefont
  {Ackad}}\ and\ \bibinfo {author} {\bibfnamefont {M.}~\bibnamefont
  {Horbatsch}},\ }\bibfield  {title} {\bibinfo {title} {New calculations for
  heavy-ion collisions with super-critical fields},\ }\href
  {https://doi.org/10.1088/1742-6596/88/1/012017} {\bibfield  {journal}
  {\bibinfo  {journal} {Journal of Physics: Conference Series}\ }\textbf
  {\bibinfo {volume} {88}},\ \bibinfo {pages} {012017} (\bibinfo {year}
  {2007}{\natexlab{a}})}\BibitemShut {NoStop}%
\bibitem [{\citenamefont {Ackad}\ and\ \citenamefont
  {Horbatsch}(2007{\natexlab{b}})}]{Horbatsch07}%
  \BibitemOpen
  \bibfield  {author} {\bibinfo {author} {\bibfnamefont {E.}~\bibnamefont
  {Ackad}}\ and\ \bibinfo {author} {\bibfnamefont {M.}~\bibnamefont
  {Horbatsch}},\ }\bibfield  {title} {\bibinfo {title} {Numerical calculation
  of supercritical dirac resonance parameters by analytic continuation
  methods},\ }\href {https://doi.org/10.1103/PhysRevA.75.022508} {\bibfield
  {journal} {\bibinfo  {journal} {Phys. Rev. A}\ }\textbf {\bibinfo {volume}
  {75}},\ \bibinfo {pages} {022508} (\bibinfo {year}
  {2007}{\natexlab{b}})}\BibitemShut {NoStop}%
\bibitem [{\citenamefont {Ackad}\ and\ \citenamefont
  {Horbatsch}(2007{\natexlab{c}})}]{Horbatsch07b}%
  \BibitemOpen
  \bibfield  {author} {\bibinfo {author} {\bibfnamefont {E.}~\bibnamefont
  {Ackad}}\ and\ \bibinfo {author} {\bibfnamefont {M.}~\bibnamefont
  {Horbatsch}},\ }\bibfield  {title} {\bibinfo {title} {Supercritical dirac
  resonance parameters from extrapolated analytic continuation methods},\
  }\href {https://doi.org/10.1103/PhysRevA.76.022503} {\bibfield  {journal}
  {\bibinfo  {journal} {Phys. Rev. A}\ }\textbf {\bibinfo {volume} {76}},\
  \bibinfo {pages} {022503} (\bibinfo {year} {2007}{\natexlab{c}})}\BibitemShut
  {NoStop}%
\bibitem [{\citenamefont {Riss}\ and\ \citenamefont {Meyer}(1993)}]{Riss93}%
  \BibitemOpen
  \bibfield  {author} {\bibinfo {author} {\bibfnamefont {U.~V.}\ \bibnamefont
  {Riss}}\ and\ \bibinfo {author} {\bibfnamefont {H.~D.}\ \bibnamefont
  {Meyer}},\ }\bibfield  {title} {\bibinfo {title} {Calculation of resonance
  energies and widths using the complex absorbing potential method},\ }\href
  {https://doi.org/10.1088/0953-4075/26/23/021} {\bibfield  {journal} {\bibinfo
   {journal} {Journal of Physics B: Atomic, Molecular and Optical Physics}\
  }\textbf {\bibinfo {volume} {26}},\ \bibinfo {pages} {4503} (\bibinfo {year}
  {1993})}\BibitemShut {NoStop}%
\bibitem [{\citenamefont {Sucher}(1980)}]{Sucher80}%
  \BibitemOpen
  \bibfield  {author} {\bibinfo {author} {\bibfnamefont {J.}~\bibnamefont
  {Sucher}},\ }\bibfield  {title} {\bibinfo {title} {Foundations of the
  relativistic theory of many-electron atoms},\ }\href
  {https://doi.org/10.1103/PhysRevA.22.348} {\bibfield  {journal} {\bibinfo
  {journal} {Phys. Rev. A}\ }\textbf {\bibinfo {volume} {22}},\ \bibinfo
  {pages} {348} (\bibinfo {year} {1980})}\BibitemShut {NoStop}%
\bibitem [{\citenamefont {Sucher}(1984)}]{Sucher84}%
  \BibitemOpen
  \bibfield  {author} {\bibinfo {author} {\bibfnamefont {J.}~\bibnamefont
  {Sucher}},\ }\bibfield  {title} {\bibinfo {title} {Foundations of the
  relativistic theory of many-electron bound states},\ }\href
  {https://doi.org/https://doi.org/10.1002/qua.560250103} {\bibfield  {journal}
  {\bibinfo  {journal} {International Journal of Quantum Chemistry}\ }\textbf
  {\bibinfo {volume} {25}},\ \bibinfo {pages} {3} (\bibinfo {year}
  {1984})}\BibitemShut {NoStop}%
\bibitem [{\citenamefont {Heully}\ \emph {et~al.}(1986)\citenamefont {Heully},
  \citenamefont {Lindgren}, \citenamefont {Lindroth},\ and\ \citenamefont
  {Mrtensson-Pendrill}}]{Heully86}%
  \BibitemOpen
  \bibfield  {author} {\bibinfo {author} {\bibfnamefont {J.-L.}\ \bibnamefont
  {Heully}}, \bibinfo {author} {\bibfnamefont {I.}~\bibnamefont {Lindgren}},
  \bibinfo {author} {\bibfnamefont {E.}~\bibnamefont {Lindroth}},\ and\
  \bibinfo {author} {\bibfnamefont {A.-M.}\ \bibnamefont
  {Mrtensson-Pendrill}},\ }\bibfield  {title} {\bibinfo {title} {Comment on
  relativistic wave equations and negative-energy states},\ }\href
  {https://doi.org/10.1103/PhysRevA.33.4426} {\bibfield  {journal} {\bibinfo
  {journal} {Phys. Rev. A}\ }\textbf {\bibinfo {volume} {33}},\ \bibinfo
  {pages} {4426} (\bibinfo {year} {1986})}\BibitemShut {NoStop}%
\bibitem [{\citenamefont {Kutzelnigg}(2012)}]{Kutzelnigg12}%
  \BibitemOpen
  \bibfield  {author} {\bibinfo {author} {\bibfnamefont {W.}~\bibnamefont
  {Kutzelnigg}},\ }\bibfield  {title} {\bibinfo {title} {Solved and unsolved
  problems in relativistic quantum chemistry},\ }\href
  {https://doi.org/https://doi.org/10.1016/j.chemphys.2011.06.001} {\bibfield
  {journal} {\bibinfo  {journal} {Chemical Physics}\ }\textbf {\bibinfo
  {volume} {395}},\ \bibinfo {pages} {16} (\bibinfo {year} {2012})},\ \bibinfo
  {note} {recent Advances and Applications of Relativistic Quantum
  Chemistry}\BibitemShut {NoStop}%
\bibitem [{\citenamefont {Almoukhalalati}\ \emph {et~al.}(2016)\citenamefont
  {Almoukhalalati}, \citenamefont {Knecht}, \citenamefont {Jensen},
  \citenamefont {Dyall},\ and\ \citenamefont {Saue}}]{Almoukhalalati16}%
  \BibitemOpen
  \bibfield  {author} {\bibinfo {author} {\bibfnamefont {A.}~\bibnamefont
  {Almoukhalalati}}, \bibinfo {author} {\bibfnamefont {S.}~\bibnamefont
  {Knecht}}, \bibinfo {author} {\bibfnamefont {H.~J.~A.}\ \bibnamefont
  {Jensen}}, \bibinfo {author} {\bibfnamefont {K.~G.}\ \bibnamefont {Dyall}},\
  and\ \bibinfo {author} {\bibfnamefont {T.}~\bibnamefont {Saue}},\ }\bibfield
  {title} {\bibinfo {title} {Electron correlation within the relativistic
  no-pair approximation},\ }\href {https://doi.org/10.1063/1.4959452}
  {\bibfield  {journal} {\bibinfo  {journal} {The Journal of Chemical Physics}\
  }\textbf {\bibinfo {volume} {145}},\ \bibinfo {pages} {074104} (\bibinfo
  {year} {2016})}\BibitemShut {NoStop}%
\bibitem [{\citenamefont {Liu}(2020)}]{Liu20}%
  \BibitemOpen
  \bibfield  {author} {\bibinfo {author} {\bibfnamefont {W.}~\bibnamefont
  {Liu}},\ }\bibfield  {title} {\bibinfo {title} {Essentials of relativistic
  quantum chemistry},\ }\href {https://doi.org/10.1063/5.0008432} {\bibfield
  {journal} {\bibinfo  {journal} {The Journal of Chemical Physics}\ }\textbf
  {\bibinfo {volume} {152}},\ \bibinfo {pages} {180901} (\bibinfo {year}
  {2020})}\BibitemShut {NoStop}%
\bibitem [{\citenamefont {Toulouse}(2021)}]{Toulouse21}%
  \BibitemOpen
  \bibfield  {author} {\bibinfo {author} {\bibfnamefont {J.}~\bibnamefont
  {Toulouse}},\ }\bibfield  {title} {\bibinfo {title} {{Relativistic
  density-functional theory based on effective quantum electrodynamics}},\
  }\href {https://doi.org/10.21468/SciPostChem.1.1.002} {\bibfield  {journal}
  {\bibinfo  {journal} {SciPost Chem.}\ }\textbf {\bibinfo {volume} {1}},\
  \bibinfo {pages} {2} (\bibinfo {year} {2021})}\BibitemShut {NoStop}%
\bibitem [{\citenamefont {{Froese Fischer}}\ \emph {et~al.}(2019)\citenamefont
  {{Froese Fischer}}, \citenamefont {Gaigalas}, \citenamefont {Jönsson},\ and\
  \citenamefont {Bieroń}}]{Fischer19}%
  \BibitemOpen
  \bibfield  {author} {\bibinfo {author} {\bibfnamefont {C.}~\bibnamefont
  {{Froese Fischer}}}, \bibinfo {author} {\bibfnamefont {G.}~\bibnamefont
  {Gaigalas}}, \bibinfo {author} {\bibfnamefont {P.}~\bibnamefont {Jönsson}},\
  and\ \bibinfo {author} {\bibfnamefont {J.}~\bibnamefont {Bieroń}},\
  }\bibfield  {title} {\bibinfo {title} {Grasp2018—a fortran 95 version of
  the general relativistic atomic structure package},\ }\href
  {https://doi.org/https://doi.org/10.1016/j.cpc.2018.10.032} {\bibfield
  {journal} {\bibinfo  {journal} {Computer Physics Communications}\ }\textbf
  {\bibinfo {volume} {237}},\ \bibinfo {pages} {184} (\bibinfo {year}
  {2019})}\BibitemShut {NoStop}%
\bibitem [{\citenamefont {Saue}\ \emph {et~al.}(2020)\citenamefont {Saue},
  \citenamefont {Bast}, \citenamefont {Gomes}, \citenamefont {Jensen},
  \citenamefont {Visscher}, \citenamefont {Aucar}, \citenamefont {Di~Remigio},
  \citenamefont {Dyall}, \citenamefont {Eliav}, \citenamefont {Fasshauer},
  \citenamefont {Fleig}, \citenamefont {Halbert}, \citenamefont {Hedegård},
  \citenamefont {Helmich-Paris}, \citenamefont {Iliaš}, \citenamefont {Jacob},
  \citenamefont {Knecht}, \citenamefont {Laerdahl}, \citenamefont {Vidal},
  \citenamefont {Nayak}, \citenamefont {Olejniczak}, \citenamefont {Olsen},
  \citenamefont {Pernpointner}, \citenamefont {Senjean}, \citenamefont {Shee},
  \citenamefont {Sunaga},\ and\ \citenamefont {van Stralen}}]{Saue20}%
  \BibitemOpen
  \bibfield  {author} {\bibinfo {author} {\bibfnamefont {T.}~\bibnamefont
  {Saue}}, \bibinfo {author} {\bibfnamefont {R.}~\bibnamefont {Bast}}, \bibinfo
  {author} {\bibfnamefont {A.~S.~P.}\ \bibnamefont {Gomes}}, \bibinfo {author}
  {\bibfnamefont {H.~J.~A.}\ \bibnamefont {Jensen}}, \bibinfo {author}
  {\bibfnamefont {L.}~\bibnamefont {Visscher}}, \bibinfo {author}
  {\bibfnamefont {I.~A.}\ \bibnamefont {Aucar}}, \bibinfo {author}
  {\bibfnamefont {R.}~\bibnamefont {Di~Remigio}}, \bibinfo {author}
  {\bibfnamefont {K.~G.}\ \bibnamefont {Dyall}}, \bibinfo {author}
  {\bibfnamefont {E.}~\bibnamefont {Eliav}}, \bibinfo {author} {\bibfnamefont
  {E.}~\bibnamefont {Fasshauer}}, \bibinfo {author} {\bibfnamefont
  {T.}~\bibnamefont {Fleig}}, \bibinfo {author} {\bibfnamefont
  {L.}~\bibnamefont {Halbert}}, \bibinfo {author} {\bibfnamefont {E.~D.}\
  \bibnamefont {Hedegård}}, \bibinfo {author} {\bibfnamefont {B.}~\bibnamefont
  {Helmich-Paris}}, \bibinfo {author} {\bibfnamefont {M.}~\bibnamefont
  {Iliaš}}, \bibinfo {author} {\bibfnamefont {C.~R.}\ \bibnamefont {Jacob}},
  \bibinfo {author} {\bibfnamefont {S.}~\bibnamefont {Knecht}}, \bibinfo
  {author} {\bibfnamefont {J.~K.}\ \bibnamefont {Laerdahl}}, \bibinfo {author}
  {\bibfnamefont {M.~L.}\ \bibnamefont {Vidal}}, \bibinfo {author}
  {\bibfnamefont {M.~K.}\ \bibnamefont {Nayak}}, \bibinfo {author}
  {\bibfnamefont {M.}~\bibnamefont {Olejniczak}}, \bibinfo {author}
  {\bibfnamefont {J.~M.~H.}\ \bibnamefont {Olsen}}, \bibinfo {author}
  {\bibfnamefont {M.}~\bibnamefont {Pernpointner}}, \bibinfo {author}
  {\bibfnamefont {B.}~\bibnamefont {Senjean}}, \bibinfo {author} {\bibfnamefont
  {A.}~\bibnamefont {Shee}}, \bibinfo {author} {\bibfnamefont {A.}~\bibnamefont
  {Sunaga}},\ and\ \bibinfo {author} {\bibfnamefont {J.~N.~P.}\ \bibnamefont
  {van Stralen}},\ }\bibfield  {title} {\bibinfo {title} {The dirac code for
  relativistic molecular calculations},\ }\href
  {https://doi.org/10.1063/5.0004844} {\bibfield  {journal} {\bibinfo
  {journal} {The Journal of Chemical Physics}\ }\textbf {\bibinfo {volume}
  {152}},\ \bibinfo {pages} {204104} (\bibinfo {year} {2020})}\BibitemShut
  {NoStop}%
\bibitem [{\citenamefont {Belpassi}\ \emph {et~al.}(2020)\citenamefont
  {Belpassi}, \citenamefont {De~Santis}, \citenamefont {Quiney}, \citenamefont
  {Tarantelli},\ and\ \citenamefont {Storchi}}]{Belpassi20}%
  \BibitemOpen
  \bibfield  {author} {\bibinfo {author} {\bibfnamefont {L.}~\bibnamefont
  {Belpassi}}, \bibinfo {author} {\bibfnamefont {M.}~\bibnamefont {De~Santis}},
  \bibinfo {author} {\bibfnamefont {H.~M.}\ \bibnamefont {Quiney}}, \bibinfo
  {author} {\bibfnamefont {F.}~\bibnamefont {Tarantelli}},\ and\ \bibinfo
  {author} {\bibfnamefont {L.}~\bibnamefont {Storchi}},\ }\bibfield  {title}
  {\bibinfo {title} {Bertha: Implementation of a four-component
  dirac–kohn–sham relativistic framework},\ }\href
  {https://doi.org/10.1063/5.0002831} {\bibfield  {journal} {\bibinfo
  {journal} {The Journal of Chemical Physics}\ }\textbf {\bibinfo {volume}
  {152}},\ \bibinfo {pages} {164118} (\bibinfo {year} {2020})},\ \Eprint
  {https://arxiv.org/abs/https://doi.org/10.1063/5.0002831}
  {https://doi.org/10.1063/5.0002831} \BibitemShut {NoStop}%
\bibitem [{\citenamefont {Furry}(1951)}]{furry:51}%
  \BibitemOpen
  \bibfield  {author} {\bibinfo {author} {\bibfnamefont {W.~H.}\ \bibnamefont
  {Furry}},\ }\bibfield  {title} {\bibinfo {title} {On bound states and
  scattering in positron theory},\ }\href
  {https://doi.org/10.1103/PhysRev.81.115} {\bibfield  {journal} {\bibinfo
  {journal} {Phys. Rev.}\ }\textbf {\bibinfo {volume} {81}},\ \bibinfo {pages}
  {115} (\bibinfo {year} {1951})}\BibitemShut {NoStop}%
\bibitem [{\citenamefont {Selst\o{}}\ \emph {et~al.}(2009)\citenamefont
  {Selst\o{}}, \citenamefont {Lindroth},\ and\ \citenamefont
  {Bengtsson}}]{Selsto09}%
  \BibitemOpen
  \bibfield  {author} {\bibinfo {author} {\bibfnamefont {S.}~\bibnamefont
  {Selst\o{}}}, \bibinfo {author} {\bibfnamefont {E.}~\bibnamefont
  {Lindroth}},\ and\ \bibinfo {author} {\bibfnamefont {J.}~\bibnamefont
  {Bengtsson}},\ }\bibfield  {title} {\bibinfo {title} {Solution of the dirac
  equation for hydrogenlike systems exposed to intense electromagnetic
  pulses},\ }\href {https://doi.org/10.1103/PhysRevA.79.043418} {\bibfield
  {journal} {\bibinfo  {journal} {Phys. Rev. A}\ }\textbf {\bibinfo {volume}
  {79}},\ \bibinfo {pages} {043418} (\bibinfo {year} {2009})}\BibitemShut
  {NoStop}%
\bibitem [{\citenamefont {Vanne}\ and\ \citenamefont {Saenz}(2012)}]{Vanne12}%
  \BibitemOpen
  \bibfield  {author} {\bibinfo {author} {\bibfnamefont {Y.~V.}\ \bibnamefont
  {Vanne}}\ and\ \bibinfo {author} {\bibfnamefont {A.}~\bibnamefont {Saenz}},\
  }\bibfield  {title} {\bibinfo {title} {Solution of the time-dependent {Dirac}
  equation for multiphoton ionization of highly charged hydrogenlike ions},\
  }\href {https://doi.org/10.1103/PhysRevA.85.033411} {\bibfield  {journal}
  {\bibinfo  {journal} {Physical Review A}\ }\textbf {\bibinfo {volume} {85}},\
  \bibinfo {pages} {033411} (\bibinfo {year} {2012})}\BibitemShut {NoStop}%
\bibitem [{\citenamefont {Leforestier}\ \emph {et~al.}(1991)\citenamefont
  {Leforestier}, \citenamefont {Bisseling}, \citenamefont {Cerjan},
  \citenamefont {Feit}, \citenamefont {Friesner}, \citenamefont {Guldberg},
  \citenamefont {Hammerich}, \citenamefont {Jolicard}, \citenamefont
  {Karrlein}, \citenamefont {Meyer}, \citenamefont {Lipkin}, \citenamefont
  {Roncero},\ and\ \citenamefont {Kosloff}}]{Leforestier91}%
  \BibitemOpen
  \bibfield  {author} {\bibinfo {author} {\bibfnamefont {C.}~\bibnamefont
  {Leforestier}}, \bibinfo {author} {\bibfnamefont {R.}~\bibnamefont
  {Bisseling}}, \bibinfo {author} {\bibfnamefont {C.}~\bibnamefont {Cerjan}},
  \bibinfo {author} {\bibfnamefont {M.}~\bibnamefont {Feit}}, \bibinfo {author}
  {\bibfnamefont {R.}~\bibnamefont {Friesner}}, \bibinfo {author}
  {\bibfnamefont {A.}~\bibnamefont {Guldberg}}, \bibinfo {author}
  {\bibfnamefont {A.}~\bibnamefont {Hammerich}}, \bibinfo {author}
  {\bibfnamefont {G.}~\bibnamefont {Jolicard}}, \bibinfo {author}
  {\bibfnamefont {W.}~\bibnamefont {Karrlein}}, \bibinfo {author}
  {\bibfnamefont {H.-D.}\ \bibnamefont {Meyer}}, \bibinfo {author}
  {\bibfnamefont {N.}~\bibnamefont {Lipkin}}, \bibinfo {author} {\bibfnamefont
  {O.}~\bibnamefont {Roncero}},\ and\ \bibinfo {author} {\bibfnamefont
  {R.}~\bibnamefont {Kosloff}},\ }\bibfield  {title} {\bibinfo {title} {A
  comparison of different propagation schemes for the time dependent
  schrödinger equation},\ }\href
  {https://doi.org/https://doi.org/10.1016/0021-9991(91)90137-A} {\bibfield
  {journal} {\bibinfo  {journal} {Journal of Computational Physics}\ }\textbf
  {\bibinfo {volume} {94}},\ \bibinfo {pages} {59 } (\bibinfo {year}
  {1991})}\BibitemShut {NoStop}%
\bibitem [{\citenamefont {Labeye}\ \emph {et~al.}(2018)\citenamefont {Labeye},
  \citenamefont {Zapata}, \citenamefont {Coccia}, \citenamefont {Véniard},
  \citenamefont {Toulouse}, \citenamefont {Caillat}, \citenamefont {Taïeb},\
  and\ \citenamefont {Luppi}}]{Labeye18}%
  \BibitemOpen
  \bibfield  {author} {\bibinfo {author} {\bibfnamefont {M.}~\bibnamefont
  {Labeye}}, \bibinfo {author} {\bibfnamefont {F.}~\bibnamefont {Zapata}},
  \bibinfo {author} {\bibfnamefont {E.}~\bibnamefont {Coccia}}, \bibinfo
  {author} {\bibfnamefont {V.}~\bibnamefont {Véniard}}, \bibinfo {author}
  {\bibfnamefont {J.}~\bibnamefont {Toulouse}}, \bibinfo {author}
  {\bibfnamefont {J.}~\bibnamefont {Caillat}}, \bibinfo {author} {\bibfnamefont
  {R.}~\bibnamefont {Taïeb}},\ and\ \bibinfo {author} {\bibfnamefont
  {E.}~\bibnamefont {Luppi}},\ }\bibfield  {title} {\bibinfo {title} {Optimal
  basis set for electron dynamics in strong laser fields: The case of molecular
  ion h2+},\ }\href {https://doi.org/10.1021/acs.jctc.8b00656} {\bibfield
  {journal} {\bibinfo  {journal} {Journal of Chemical Theory and Computation}\
  }\textbf {\bibinfo {volume} {14}},\ \bibinfo {pages} {5846} (\bibinfo {year}
  {2018})}\BibitemShut {NoStop}%
\bibitem [{\citenamefont {Zapata}\ \emph {et~al.}(2019)\citenamefont {Zapata},
  \citenamefont {Luppi},\ and\ \citenamefont {Toulouse}}]{Zapata19}%
  \BibitemOpen
  \bibfield  {author} {\bibinfo {author} {\bibfnamefont {F.}~\bibnamefont
  {Zapata}}, \bibinfo {author} {\bibfnamefont {E.}~\bibnamefont {Luppi}},\ and\
  \bibinfo {author} {\bibfnamefont {J.}~\bibnamefont {Toulouse}},\ }\bibfield
  {title} {\bibinfo {title} {Linear-response range-separated density-functional
  theory for atomic photoexcitation and photoionization spectra},\ }\href
  {https://doi.org/10.1063/1.5096037} {\bibfield  {journal} {\bibinfo
  {journal} {The Journal of Chemical Physics}\ }\textbf {\bibinfo {volume}
  {150}},\ \bibinfo {pages} {234104} (\bibinfo {year} {2019})}\BibitemShut
  {NoStop}%
\bibitem [{\citenamefont {{Froese Fischer}}\ and\ \citenamefont
  {Zatsarinny}(2009)}]{Charlotte08}%
  \BibitemOpen
  \bibfield  {author} {\bibinfo {author} {\bibfnamefont {C.}~\bibnamefont
  {{Froese Fischer}}}\ and\ \bibinfo {author} {\bibfnamefont {O.}~\bibnamefont
  {Zatsarinny}},\ }\bibfield  {title} {\bibinfo {title} {A b-spline galerkin
  method for the dirac equation},\ }\href
  {https://doi.org/https://doi.org/10.1016/j.cpc.2008.12.010} {\bibfield
  {journal} {\bibinfo  {journal} {Computer Physics Communications}\ }\textbf
  {\bibinfo {volume} {180}},\ \bibinfo {pages} {879 } (\bibinfo {year}
  {2009})}\BibitemShut {NoStop}%
\bibitem [{\citenamefont {Qiu}\ and\ \citenamefont {Fischer}(1999)}]{Qiu99}%
  \BibitemOpen
  \bibfield  {author} {\bibinfo {author} {\bibfnamefont {Y.}~\bibnamefont
  {Qiu}}\ and\ \bibinfo {author} {\bibfnamefont {C.~F.}\ \bibnamefont
  {Fischer}},\ }\bibfield  {title} {\bibinfo {title} {Integration by cell
  algorithm for slater integrals in a spline basis},\ }\href
  {https://doi.org/https://doi.org/10.1006/jcph.1999.6361} {\bibfield
  {journal} {\bibinfo  {journal} {Journal of Computational Physics}\ }\textbf
  {\bibinfo {volume} {156}},\ \bibinfo {pages} {257} (\bibinfo {year}
  {1999})}\BibitemShut {NoStop}%
\bibitem [{\citenamefont {Fleig}\ \emph {et~al.}(2003)\citenamefont {Fleig},
  \citenamefont {Olsen},\ and\ \citenamefont {Visscher}}]{Fleig03}%
  \BibitemOpen
  \bibfield  {author} {\bibinfo {author} {\bibfnamefont {T.}~\bibnamefont
  {Fleig}}, \bibinfo {author} {\bibfnamefont {J.}~\bibnamefont {Olsen}},\ and\
  \bibinfo {author} {\bibfnamefont {L.}~\bibnamefont {Visscher}},\ }\bibfield
  {title} {\bibinfo {title} {The generalized active space concept for the
  relativistic treatment of electron correlation. ii. large-scale configuration
  interaction implementation based on relativistic 2- and 4-spinors and its
  application},\ }\href {https://doi.org/10.1063/1.1590636} {\bibfield
  {journal} {\bibinfo  {journal} {The Journal of Chemical Physics}\ }\textbf
  {\bibinfo {volume} {119}},\ \bibinfo {pages} {2963} (\bibinfo {year}
  {2003})}\BibitemShut {NoStop}%
\bibitem [{\citenamefont {Fleig}\ \emph {et~al.}(2006)\citenamefont {Fleig},
  \citenamefont {Jensen}, \citenamefont {Olsen},\ and\ \citenamefont
  {Visscher}}]{Fleig06}%
  \BibitemOpen
  \bibfield  {author} {\bibinfo {author} {\bibfnamefont {T.}~\bibnamefont
  {Fleig}}, \bibinfo {author} {\bibfnamefont {H.~J.~A.}\ \bibnamefont
  {Jensen}}, \bibinfo {author} {\bibfnamefont {J.}~\bibnamefont {Olsen}},\ and\
  \bibinfo {author} {\bibfnamefont {L.}~\bibnamefont {Visscher}},\ }\bibfield
  {title} {\bibinfo {title} {The generalized active space concept for the
  relativistic treatment of electron correlation. iii. large-scale
  configuration interaction and multiconfiguration self-consistent-field
  four-component methods with application to uo2},\ }\href
  {https://doi.org/10.1063/1.2176609} {\bibfield  {journal} {\bibinfo
  {journal} {The Journal of Chemical Physics}\ }\textbf {\bibinfo {volume}
  {124}},\ \bibinfo {pages} {104106} (\bibinfo {year} {2006})}\BibitemShut
  {NoStop}%
\bibitem [{\citenamefont {Olsen}\ \emph {et~al.}(1990)\citenamefont {Olsen},
  \citenamefont {Jørgensen},\ and\ \citenamefont {Simons}}]{Olsen90}%
  \BibitemOpen
  \bibfield  {author} {\bibinfo {author} {\bibfnamefont {J.}~\bibnamefont
  {Olsen}}, \bibinfo {author} {\bibfnamefont {P.}~\bibnamefont {Jørgensen}},\
  and\ \bibinfo {author} {\bibfnamefont {J.}~\bibnamefont {Simons}},\
  }\bibfield  {title} {\bibinfo {title} {Passing the one-billion limit in full
  configuration-interaction (fci) calculations},\ }\href
  {https://doi.org/https://doi.org/10.1016/0009-2614(90)85633-N} {\bibfield
  {journal} {\bibinfo  {journal} {Chemical Physics Letters}\ }\textbf {\bibinfo
  {volume} {169}},\ \bibinfo {pages} {463} (\bibinfo {year}
  {1990})}\BibitemShut {NoStop}%
\bibitem [{\citenamefont {Knecht}\ \emph {et~al.}(2010)\citenamefont {Knecht},
  \citenamefont {Jensen},\ and\ \citenamefont {Fleig}}]{Knecht10}%
  \BibitemOpen
  \bibfield  {author} {\bibinfo {author} {\bibfnamefont {S.}~\bibnamefont
  {Knecht}}, \bibinfo {author} {\bibfnamefont {H.~J.~A.}\ \bibnamefont
  {Jensen}},\ and\ \bibinfo {author} {\bibfnamefont {T.}~\bibnamefont
  {Fleig}},\ }\bibfield  {title} {\bibinfo {title} {Large-scale parallel
  configuration interaction. ii. two- and four-component double-group general
  active space implementation with application to bih},\ }\href
  {https://doi.org/10.1063/1.3276157} {\bibfield  {journal} {\bibinfo
  {journal} {The Journal of Chemical Physics}\ }\textbf {\bibinfo {volume}
  {132}},\ \bibinfo {pages} {014108} (\bibinfo {year} {2010})}\BibitemShut
  {NoStop}%
\bibitem [{\citenamefont {Dreuw}\ and\ \citenamefont
  {Head-Gordon}(2005)}]{Dreuw05}%
  \BibitemOpen
  \bibfield  {author} {\bibinfo {author} {\bibfnamefont {A.}~\bibnamefont
  {Dreuw}}\ and\ \bibinfo {author} {\bibfnamefont {M.}~\bibnamefont
  {Head-Gordon}},\ }\bibfield  {title} {\bibinfo {title} {Single-reference ab
  initio methods for the calculation of excited states of large molecules},\
  }\href {https://doi.org/10.1021/cr0505627} {\bibfield  {journal} {\bibinfo
  {journal} {Chemical Reviews}\ }\textbf {\bibinfo {volume} {105}},\ \bibinfo
  {pages} {4009} (\bibinfo {year} {2005})}\BibitemShut {NoStop}%
\bibitem [{\citenamefont {Deslattes}\ \emph {et~al.}(2003)\citenamefont
  {Deslattes}, \citenamefont {Kessler}, \citenamefont {Indelicato},
  \citenamefont {{de Billy}}, \citenamefont {Lindroth},\ and\ \citenamefont
  {Anton}}]{deslattes:rmp}%
  \BibitemOpen
  \bibfield  {author} {\bibinfo {author} {\bibfnamefont {R.~D.}\ \bibnamefont
  {Deslattes}}, \bibinfo {author} {\bibfnamefont {E.~G.}\ \bibnamefont
  {Kessler}}, \bibinfo {author} {\bibfnamefont {P.}~\bibnamefont {Indelicato}},
  \bibinfo {author} {\bibfnamefont {L.}~\bibnamefont {{de Billy}}}, \bibinfo
  {author} {\bibfnamefont {E.}~\bibnamefont {Lindroth}},\ and\ \bibinfo
  {author} {\bibfnamefont {J.}~\bibnamefont {Anton}},\ }\bibfield  {title}
  {\bibinfo {title} {X-ray transition energies: new approach to a comprehensive
  evaluation},\ }\href@noop {} {\bibfield  {journal} {\bibinfo  {journal} {Rev.
  Mod. Phys.}\ }\textbf {\bibinfo {volume} {75}},\ \bibinfo {pages} {35}
  (\bibinfo {year} {2003})}\BibitemShut {NoStop}%
\bibitem [{\citenamefont {Bethe}\ and\ \citenamefont
  {Salpeter}(2013)}]{Bethe13}%
  \BibitemOpen
  \bibfield  {author} {\bibinfo {author} {\bibfnamefont {H.}~\bibnamefont
  {Bethe}}\ and\ \bibinfo {author} {\bibfnamefont {E.}~\bibnamefont
  {Salpeter}},\ }\href {https://books.google.se/books?id=nxz2CAAAQBAJ} {\emph
  {\bibinfo {title} {Quantum Mechanics of One- and Two-Electron Atoms}}}\
  (\bibinfo  {publisher} {Springer Berlin Heidelberg},\ \bibinfo {year}
  {2013})\BibitemShut {NoStop}%
\bibitem [{\citenamefont {Amusia}(2013)}]{Amusia90}%
  \BibitemOpen
  \bibfield  {author} {\bibinfo {author} {\bibfnamefont {M.}~\bibnamefont
  {Amusia}},\ }\href@noop {} {\emph {\bibinfo {title} {Atomic Photoeffect}}},\
  Physics of Atoms and Molecules\ (\bibinfo  {publisher} {Springer US},\
  \bibinfo {year} {2013})\BibitemShut {NoStop}%
\bibitem [{\citenamefont {Chan}\ \emph {et~al.}(1992)\citenamefont {Chan},
  \citenamefont {Cooper}, \citenamefont {Guo}, \citenamefont {Burton},\ and\
  \citenamefont {Brion}}]{Chan92}%
  \BibitemOpen
  \bibfield  {author} {\bibinfo {author} {\bibfnamefont {W.~F.}\ \bibnamefont
  {Chan}}, \bibinfo {author} {\bibfnamefont {G.}~\bibnamefont {Cooper}},
  \bibinfo {author} {\bibfnamefont {X.}~\bibnamefont {Guo}}, \bibinfo {author}
  {\bibfnamefont {G.~R.}\ \bibnamefont {Burton}},\ and\ \bibinfo {author}
  {\bibfnamefont {C.~E.}\ \bibnamefont {Brion}},\ }\bibfield  {title} {\bibinfo
  {title} {Absolute optical oscillator strengths for the electronic excitation
  of atoms at high resolution. iii. the photoabsorption of argon, krypton, and
  xenon},\ }\href {https://doi.org/10.1103/PhysRevA.46.149} {\bibfield
  {journal} {\bibinfo  {journal} {Phys. Rev. A}\ }\textbf {\bibinfo {volume}
  {46}},\ \bibinfo {pages} {149} (\bibinfo {year} {1992})}\BibitemShut
  {NoStop}%
\bibitem [{\citenamefont {Amusia}\ and\ \citenamefont
  {Kheifets}(1982)}]{Amusia82}%
  \BibitemOpen
  \bibfield  {author} {\bibinfo {author} {\bibfnamefont {M.~Y.}\ \bibnamefont
  {Amusia}}\ and\ \bibinfo {author} {\bibfnamefont {A.~S.}\ \bibnamefont
  {Kheifets}},\ }\bibfield  {title} {\bibinfo {title} {The influence of
  two-electron--two-hole excitations on the 3s$^{-1}$4p autoionization profile
  in ar atoms},\ }\href@noop {} {\bibfield  {journal} {\bibinfo  {journal}
  {Phys. Lett.}\ }\textbf {\bibinfo {volume} {82A}},\ \bibinfo {pages} {407}
  (\bibinfo {year} {1982})}\BibitemShut {NoStop}%
\bibitem [{\citenamefont {Carette}\ \emph {et~al.}(2013)\citenamefont
  {Carette}, \citenamefont {Dahlstr\"om}, \citenamefont {Argenti},\ and\
  \citenamefont {Lindroth}}]{Carette13}%
  \BibitemOpen
  \bibfield  {author} {\bibinfo {author} {\bibfnamefont {T.}~\bibnamefont
  {Carette}}, \bibinfo {author} {\bibfnamefont {J.~M.}\ \bibnamefont
  {Dahlstr\"om}}, \bibinfo {author} {\bibfnamefont {L.}~\bibnamefont
  {Argenti}},\ and\ \bibinfo {author} {\bibfnamefont {E.}~\bibnamefont
  {Lindroth}},\ }\bibfield  {title} {\bibinfo {title} {Multiconfigurational
  hartree-fock close-coupling ansatz: Application to the argon photoionization
  cross section and delays},\ }\href
  {https://doi.org/10.1103/PhysRevA.87.023420} {\bibfield  {journal} {\bibinfo
  {journal} {Phys. Rev. A}\ }\textbf {\bibinfo {volume} {87}},\ \bibinfo
  {pages} {023420} (\bibinfo {year} {2013})}\BibitemShut {NoStop}%
\bibitem [{\citenamefont {Svensson}\ \emph {et~al.}(1988)\citenamefont
  {Svensson}, \citenamefont {Eriksson}, \citenamefont {Martensson},
  \citenamefont {Wendin},\ and\ \citenamefont {Gelius}}]{svensson:88}%
  \BibitemOpen
  \bibfield  {author} {\bibinfo {author} {\bibfnamefont {S.}~\bibnamefont
  {Svensson}}, \bibinfo {author} {\bibfnamefont {B.}~\bibnamefont {Eriksson}},
  \bibinfo {author} {\bibfnamefont {N.}~\bibnamefont {Martensson}}, \bibinfo
  {author} {\bibfnamefont {G.}~\bibnamefont {Wendin}},\ and\ \bibinfo {author}
  {\bibfnamefont {U.}~\bibnamefont {Gelius}},\ }\bibfield  {title} {\bibinfo
  {title} {Electron shake-up and correlation satellites and continuum shake-off
  distributions in x-ray photoelectron spectra of the rare gas atoms},\ }\href
  {https://doi.org/10.1016/0368-2048(88)85020-5} {\bibfield  {journal}
  {\bibinfo  {journal} {Journal of Electron Spectroscopy and Related
  Phenomena}\ }\textbf {\bibinfo {volume} {47}},\ \bibinfo {pages} {327}
  (\bibinfo {year} {1988})}\BibitemShut {NoStop}%
\bibitem [{\citenamefont {Gaunt}\ and\ \citenamefont {Fowler}(1929)}]{Gaunt29}%
  \BibitemOpen
  \bibfield  {author} {\bibinfo {author} {\bibfnamefont {J.~A.}\ \bibnamefont
  {Gaunt}}\ and\ \bibinfo {author} {\bibfnamefont {R.~H.}\ \bibnamefont
  {Fowler}},\ }\bibfield  {title} {\bibinfo {title} {The triplets of helium},\
  }\href {https://doi.org/10.1098/rspa.1929.0037} {\bibfield  {journal}
  {\bibinfo  {journal} {Proceedings of the Royal Society of London. Series A,
  Containing Papers of a Mathematical and Physical Character}\ }\textbf
  {\bibinfo {volume} {122}},\ \bibinfo {pages} {513} (\bibinfo {year}
  {1929})}\BibitemShut {NoStop}%
\bibitem [{\citenamefont {Saue}(2011)}]{Saue11}%
  \BibitemOpen
  \bibfield  {author} {\bibinfo {author} {\bibfnamefont {T.}~\bibnamefont
  {Saue}},\ }\bibfield  {title} {\bibinfo {title} {Relativistic hamiltonians
  for chemistry: A primer},\ }\href
  {https://doi.org/https://doi.org/10.1002/cphc.201100682} {\bibfield
  {journal} {\bibinfo  {journal} {ChemPhysChem}\ }\textbf {\bibinfo {volume}
  {12}},\ \bibinfo {pages} {3077} (\bibinfo {year} {2011})}\BibitemShut
  {NoStop}%
\bibitem [{\citenamefont {Lindroth}\ \emph {et~al.}(1989)\citenamefont
  {Lindroth}, \citenamefont {M{\aa}rtensson-Pendrill}, \citenamefont
  {Ynnerman},\ and\ \citenamefont {{\"O}ster}}]{lindroth:89:breit}%
  \BibitemOpen
  \bibfield  {author} {\bibinfo {author} {\bibfnamefont {E.}~\bibnamefont
  {Lindroth}}, \bibinfo {author} {\bibfnamefont {A.-M.}\ \bibnamefont
  {M{\aa}rtensson-Pendrill}}, \bibinfo {author} {\bibfnamefont
  {A.}~\bibnamefont {Ynnerman}},\ and\ \bibinfo {author} {\bibfnamefont
  {P.}~\bibnamefont {{\"O}ster}},\ }\bibfield  {title} {\bibinfo {title}
  {Self-consistent treatment of the {B}reit interaction, with application to
  the electric dipole moment in thallium},\ }\href@noop {} {\bibfield
  {journal} {\bibinfo  {journal} {J. Phys. B}\ }\textbf {\bibinfo {volume}
  {22}},\ \bibinfo {pages} {2447} (\bibinfo {year} {1989})}\BibitemShut
  {NoStop}%
\bibitem [{\citenamefont {Breit}(1932)}]{Breit32}%
  \BibitemOpen
  \bibfield  {author} {\bibinfo {author} {\bibfnamefont {G.}~\bibnamefont
  {Breit}},\ }\bibfield  {title} {\bibinfo {title} {Dirac's equation and the
  spin-spin interactions of two electrons},\ }\href@noop {} {\bibfield
  {journal} {\bibinfo  {journal} {Phys. Rev.}\ }\textbf {\bibinfo {volume}
  {39}},\ \bibinfo {pages} {616} (\bibinfo {year} {1932})}\BibitemShut
  {NoStop}%
\bibitem [{\citenamefont {Cheng}\ and\ \citenamefont
  {Johnson}(1983)}]{Cheng83}%
  \BibitemOpen
  \bibfield  {author} {\bibinfo {author} {\bibfnamefont {K.~T.}\ \bibnamefont
  {Cheng}}\ and\ \bibinfo {author} {\bibfnamefont {W.~R.}\ \bibnamefont
  {Johnson}},\ }\bibfield  {title} {\bibinfo {title} {Orbital collapse and the
  photoionization of the inner $4d$ shells for xe-like ions},\ }\href
  {https://doi.org/10.1103/PhysRevA.28.2820} {\bibfield  {journal} {\bibinfo
  {journal} {Phys. Rev. A}\ }\textbf {\bibinfo {volume} {28}},\ \bibinfo
  {pages} {2820} (\bibinfo {year} {1983})}\BibitemShut {NoStop}%
\bibitem [{\citenamefont {Cooper}(1964)}]{Cooper64}%
  \BibitemOpen
  \bibfield  {author} {\bibinfo {author} {\bibfnamefont {J.~W.}\ \bibnamefont
  {Cooper}},\ }\bibfield  {title} {\bibinfo {title} {Interaction of maxima in
  the absorption of soft x rays},\ }\href
  {https://doi.org/10.1103/PhysRevLett.13.762} {\bibfield  {journal} {\bibinfo
  {journal} {Phys. Rev. Lett.}\ }\textbf {\bibinfo {volume} {13}},\ \bibinfo
  {pages} {762} (\bibinfo {year} {1964})}\BibitemShut {NoStop}%
\bibitem [{\citenamefont {Chen}\ \emph {et~al.}(2015)\citenamefont {Chen},
  \citenamefont {Pabst}, \citenamefont {Karamatskou},\ and\ \citenamefont
  {Santra}}]{Chen15}%
  \BibitemOpen
  \bibfield  {author} {\bibinfo {author} {\bibfnamefont {Y.-J.}\ \bibnamefont
  {Chen}}, \bibinfo {author} {\bibfnamefont {S.}~\bibnamefont {Pabst}},
  \bibinfo {author} {\bibfnamefont {A.}~\bibnamefont {Karamatskou}},\ and\
  \bibinfo {author} {\bibfnamefont {R.}~\bibnamefont {Santra}},\ }\bibfield
  {title} {\bibinfo {title} {Theoretical characterization of the collective
  resonance states underlying the xenon giant dipole resonance},\ }\href
  {https://doi.org/10.1103/PhysRevA.91.032503} {\bibfield  {journal} {\bibinfo
  {journal} {Phys. Rev. A}\ }\textbf {\bibinfo {volume} {91}},\ \bibinfo
  {pages} {032503} (\bibinfo {year} {2015})}\BibitemShut {NoStop}%
\bibitem [{\citenamefont {Toffoli}\ \emph {et~al.}(2002)\citenamefont
  {Toffoli}, \citenamefont {Stener},\ and\ \citenamefont
  {Decleva}}]{Toffoli02}%
  \BibitemOpen
  \bibfield  {author} {\bibinfo {author} {\bibfnamefont {D.}~\bibnamefont
  {Toffoli}}, \bibinfo {author} {\bibfnamefont {M.}~\bibnamefont {Stener}},\
  and\ \bibinfo {author} {\bibfnamefont {P.}~\bibnamefont {Decleva}},\
  }\bibfield  {title} {\bibinfo {title} {Application of the relativistic
  time-dependent density functional theory to the photoionization of xenon},\
  }\href@noop {} {\bibfield  {journal} {\bibinfo  {journal} {Journal of Physics
  B: Atomic, Molecular and Optical Physics}\ }\textbf {\bibinfo {volume}
  {35}},\ \bibinfo {pages} {1275–1305} (\bibinfo {year} {2002})}\BibitemShut
  {NoStop}%
\bibitem [{\citenamefont {Kutzner}\ \emph {et~al.}(1989)\citenamefont
  {Kutzner}, \citenamefont {Radojevi\ifmmode~\acute{c}\else \'{c}\fi{}},\ and\
  \citenamefont {Kelly}}]{Kutzner89}%
  \BibitemOpen
  \bibfield  {author} {\bibinfo {author} {\bibfnamefont {M.}~\bibnamefont
  {Kutzner}}, \bibinfo {author} {\bibfnamefont {V.}~\bibnamefont
  {Radojevi\ifmmode~\acute{c}\else \'{c}\fi{}}},\ and\ \bibinfo {author}
  {\bibfnamefont {H.~P.}\ \bibnamefont {Kelly}},\ }\bibfield  {title} {\bibinfo
  {title} {Extended photoionization calculations for xenon},\ }\href
  {https://doi.org/10.1103/PhysRevA.40.5052} {\bibfield  {journal} {\bibinfo
  {journal} {Phys. Rev. A}\ }\textbf {\bibinfo {volume} {40}},\ \bibinfo
  {pages} {5052} (\bibinfo {year} {1989})}\BibitemShut {NoStop}%
\bibitem [{\citenamefont {Condon}\ and\ \citenamefont
  {Shortley}(1953)}]{Condon53}%
  \BibitemOpen
  \bibfield  {author} {\bibinfo {author} {\bibfnamefont {E.}~\bibnamefont
  {Condon}}\ and\ \bibinfo {author} {\bibfnamefont {G.}~\bibnamefont
  {Shortley}},\ }\href {https://books.google.se/books?id=CbvvAAAAMAAJ} {\emph
  {\bibinfo {title} {The Theory of Atomic Spectra}}}\ (\bibinfo  {publisher}
  {Cambridge University Press},\ \bibinfo {year} {1953})\BibitemShut {NoStop}%
\bibitem [{\citenamefont {Vinbladh}\ \emph {et~al.}(2019)\citenamefont
  {Vinbladh}, \citenamefont {Dahlstr\"om},\ and\ \citenamefont
  {Lindroth}}]{Vinbladh19}%
  \BibitemOpen
  \bibfield  {author} {\bibinfo {author} {\bibfnamefont {J.}~\bibnamefont
  {Vinbladh}}, \bibinfo {author} {\bibfnamefont {J.~M.}\ \bibnamefont
  {Dahlstr\"om}},\ and\ \bibinfo {author} {\bibfnamefont {E.}~\bibnamefont
  {Lindroth}},\ }\bibfield  {title} {\bibinfo {title} {Many-body calculations
  of two-photon, two-color matrix elements for attosecond delays},\ }\href
  {https://doi.org/10.1103/PhysRevA.100.043424} {\bibfield  {journal} {\bibinfo
   {journal} {Phys. Rev. A}\ }\textbf {\bibinfo {volume} {100}},\ \bibinfo
  {pages} {043424} (\bibinfo {year} {2019})}\BibitemShut {NoStop}%
\bibitem [{\citenamefont {Dahlström}\ and\ \citenamefont
  {Lindroth}(2014)}]{MarcusEva14}%
  \BibitemOpen
  \bibfield  {author} {\bibinfo {author} {\bibfnamefont {J.~M.}\ \bibnamefont
  {Dahlström}}\ and\ \bibinfo {author} {\bibfnamefont {E.}~\bibnamefont
  {Lindroth}},\ }\bibfield  {title} {\bibinfo {title} {Study of attosecond
  delays using perturbation diagrams and exterior complex scaling},\ }\href
  {https://doi.org/10.1088/0953-4075/47/12/124012} {\bibfield  {journal}
  {\bibinfo  {journal} {Journal of Physics B: Atomic, Molecular and Optical
  Physics}\ }\textbf {\bibinfo {volume} {47}},\ \bibinfo {pages} {124012}
  (\bibinfo {year} {2014})}\BibitemShut {NoStop}%
\bibitem [{\citenamefont {Kaufmann}\ \emph {et~al.}(1989)\citenamefont
  {Kaufmann}, \citenamefont {Baumeister},\ and\ \citenamefont
  {Jungen}}]{Kaufmann89}%
  \BibitemOpen
  \bibfield  {author} {\bibinfo {author} {\bibfnamefont {K.}~\bibnamefont
  {Kaufmann}}, \bibinfo {author} {\bibfnamefont {W.}~\bibnamefont
  {Baumeister}},\ and\ \bibinfo {author} {\bibfnamefont {M.}~\bibnamefont
  {Jungen}},\ }\bibfield  {title} {\bibinfo {title} {Universal gaussian basis
  sets for an optimum representation of rydberg and continuum wavefunctions},\
  }\href {https://doi.org/10.1088/0953-4075/22/14/007} {\bibfield  {journal}
  {\bibinfo  {journal} {Journal of Physics B: Atomic, Molecular and Optical
  Physics}\ }\textbf {\bibinfo {volume} {22}},\ \bibinfo {pages} {2223}
  (\bibinfo {year} {1989})}\BibitemShut {NoStop}%
\end{thebibliography}%

\end{document}